\documentclass[fleqn,usenatbib]{mnras}

\usepackage{newtxtext,newtxmath}
\usepackage{longtable}

\usepackage[T1]{fontenc}

\DeclareRobustCommand{\VAN}[3]{#2}
\let\VANthebibliography\thebibliography
\def\thebibliography{\DeclareRobustCommand{\VAN}[3]{##3}\VANthebibliography}

\usepackage{graphicx}	
\usepackage{amsmath}	
\usepackage{booktabs}

\usepackage{pdflscape}

\newcommand{\hi}{H\,{\footnotesize I}}
\newcommand{\cothree}{CO\,(3-2)}
\newcommand{\ideight}{ID83492}

\title[Large molecular gas fraction of post-SBs at $z > 1$]{The large molecular gas fraction of post-starburst galaxies at $z > 1$}

\author[A. Zanella et al.]{
A. Zanella$^{1}$\thanks{E-mail: anita.zanella@inaf.it}, F. Valentino$^{2, 3}$, A. Gallazzi$^{4}$, S. Belli$^{5}$, G. Magdis$^{3,6,7}$, A. Bolamperti$^{2,8}$
\\
$^1$Istituto Nazionale di Astrofisica, Osservatorio di Padova, Vicolo dell'Osservatorio 5, 35122 Padova, Italy\\
$^2$ European Southern Observatory, Karl Schwarzschild Strasse 2, 85748 Garching, Germany\\
$^3$ Cosmic Dawn Center (DAWN), Copenhagen, Denmark\\ 
$^4$ Istituto Nazionale di Astrofisica, Osservatorio Astrofisico di Arcetri, Largo Enrico Fermi 5, I-50125 Firenze, Italy\\
$^5$ Dipartimento di Fisica e Astronomia "Augusto Righi", Università di Bologna, via Gobetti 93/2, 40129 Bologna, Italy\\
$^6$ DTU-Space, Technical University of Denmark, Elektrovej 327, 2800, Kgs. Lyngby, Denmark\\
$^7$ Niels Bohr Institute, University of Copenhagen, Lyngbyvej 2, 2100, Copenhagen Ø, Denmark\\
$^8$ Dipartimento di Fisica ed Astronomia, Università degli Studi di Padova, Vicolo dell'Osservatorio 3, I-35122 Padova, Italy
}

\date{Accepted XXX. Received YYY; in original form ZZZ}

\pubyear{2022}

\begin{document}
\label{firstpage}
\pagerange{\pageref{firstpage}--\pageref{lastpage}}
\maketitle

\begin{abstract}
Post-starburst galaxies are sources that had the last major episode of star formation about 1 Gyr before the epoch of the observations and are on their way to quiescence. It is important to study such galaxies at redshift $z > 1$, during their main quenching phase, and estimate their molecular gas content to constrain the processes responsible for the cessation of star formation. We present CO(3-2) ALMA observations of two massive (M$_\star \sim 5\times10^{10}$ M$_\odot$) post-starburst galaxies at $z > 1$. We measure their molecular gas fraction to be $f_\mathrm{H2} = \mathrm{M_{H2}/M_\star} \sim$ 8\% - 16\%, consistent with $z < 1$ post-starburst galaxies from the literature. The star formation efficiency of our targets is $\sim 10\times$ lower than that of star-forming galaxies at similar redshift, and they are outliers of the $f_\mathrm{H2}$ - specific star formation rate (sSFR) relation of star-forming galaxies, as they have larger $f_\mathrm{H2}$ than expected given their sSFR. The gas fraction of post-starbursts from our sample and the literature correlates with the D$_n$4000 spectral index, a proxy of the stellar population age. This suggests that their gas content decreases after the last major burst of star formation. Finally, one of our targets is undergoing a major merger phase with two highly star-forming companions. This hints at a picture where a perturber event (e.g., major merger) quenches star formation without completely removing the molecular gas.
\end{abstract}

\begin{keywords}
galaxies: evolution $<$ Galaxies, galaxies: formation $<$ Galaxies, galaxies: high-redshift $<$ Galaxies, (galaxies:) intergalactic medium $<$ Galaxies,
galaxies: ISM $<$ Galaxies
\end{keywords}



\section{Introduction}
\label{sec:introduction}

The existence of a well defined separation among massive, red, early-type, quiescent galaxies and blue, late-type, actively star-forming objects is a fundamental instrument for the current modeling of galaxy formation and evolution. Nowadays, the majority of star-forming galaxies are thought to assemble their mass through a secular process of star formation in a relatively steady state, forming a "Main Sequence" up to high redshift (\citealt{Daddi2007}; \citealt{Speagle2014}) and following a tight gas-star formation rate (SFR) density relation (Kennicutt-Schmidt relation, "KS", \citealt{Schmidt1959}; \citealt{Kennicutt1998}).
Deviations from this dynamic equilibrium may occur in short starbursting events (associated with major mergers at least in the local Universe, \citealt{Sanders1996}) or due to the cessation of star formation ("quenching", \citealt{Davis2016}). Both these deviations from equilibrium are poorly understood: why do galaxies suddenly ignite the formation of thousands of stars per year? Why do they stop forming stars?

The deviations from the Main Sequence, observed across redshift, and the KS relation might be connected: the merger of gas-rich objects may first result in a burst of star formation, followed by a drop of star formation rate and the subsequent quenching of the galaxy, possibly connected with feedback from active supermassive black holes or environmental effects (\citealt{Hopkins2008}). The results of this process might be the quiescent galaxies, which have been observed up to high redshift ($z \sim 2 - 4$, e.g., \citealt{Cimatti2008, Toft2014,Dudzeviciute2020}, \citep{Glazebrook2017, Valentino2020}. However, what the mechanism responsible for stopping star formation is and how galaxies are maintained quiescent for several Gyrs are still matter of debate. Two alternative physical conditions might be responsible for galaxy quenching: the gas may be either removed from galaxies or prevented from cooling into molecular clouds. Several scenarios have been proposed, which we outline in the following. Outflows driven by supernovae or Active Galactic Nuclei (AGN) can expel gas out of galaxies -- and are the typical quenching mechanisms in simulations \citep{Hopkins2012}. Ram pressure and tidal interactions can strip it in cluster environments \citep{Gunn1972, Werle2022}. The gas might be consumed by continuous star formation and black hole growth, while the normal replenishment from the cosmic web is cut by the accumulation of hot plasma, inducing shocks and heating up the infalling gas (i.e, ``strangulation'', ``starvation'', \citealt{Larson1980, Peng2015}; "halo quenching", \citealt{Dekel2006}). The gas can be prevented from cooling because of dynamical stabilization ("morphological quenching", \citealt{Martig2009}), gravitational heating \citep{Johansson2009}, or AGN heating ("radio-mode feedback", also expelling gas, \citealt{Croton2006}).

To find a way out of this labyrinth, it is key to capture galaxies during or immediately after their quenching, and investigate their molecular gas content and conditions. Galaxies dubbed "K+A", "E+A", or "post-starburst" (post-SB) are thought to be objects on their way towards quenching. Their spectra are characterized by deep Balmer absorption lines, typical of A-stars dominating the continuum emission about 1 Gyr after the last major episode of star formation, and little or no nebular lines associated with ongoing star formation \citep[and references therein]{Dressler1983,Couch1987,French2021}. They have prominent bulges, but their disturbed morphologies are highly suggestive of recent merging activity (e.g., \citealt{Zabludoff1996,Yang2008, Wild2016}). Moreover, many post-SBs lie in the "green valley" of the color-magnitude diagram of galaxies, suggesting once more their nature of intermediate objects between gas rich, blue, star-forming galaxies and red, quiescent systems \citep{Wong2012}. Therefore, this population of galaxies has been a natural choice for sub-millimeter and radio follow-ups in order to constrain their gas content. In the local Universe, a significant fraction of post-SB galaxies have been detected in neutral atomic hydrogen \hi, with intermediate gas fractions similar to gas-poor spirals and gas-rich early-type galaxies \citep{Zwaan2013}. However, \hi\ is not a sensitive tracer of the dense gas from which past and new star formation may occur. Moreover, \hi\ cannot be observed at high redshift, close to the main quenching episode of local massive quiescent galaxies. The far-infrared continuum dust emission has been used as a proxy of the cold gas content of post-starburst and quiescent galaxies up to $z \sim 2$ (e.g., \citealt{Gobat2018, Michalowski2019, Whitaker2021, Gobat2022}). However such studies yield contradictory results, mainly depending on the method used to estimate the dust and in turn gas mass of galaxies. Relatively large gas fractions ($\sim 7\%$) are found when far-infrared spectral energy distributions averaged over large samples are considered \citep{Gobat2018, Magdis2021}, whereas smaller gas fractions ($\sim 0.1 - 1\%$) are found when individual lensed sources are observed \citep{Whitaker2021}. Such discrepancies might be due to uncertainties on the photometric redshift of galaxies entering the stacking, the coarse resolution of the observations and hence blending of quiescent and star-forming sources (which affect the first method), uncertainties on the lensing model and magnification of lensed targets especially in case of extended emission (which affect the second method). Moreover uncertainties about the dust-to-gas conversion factor, dust temperature, and compactness of the emission might also be responsible for such discrepancies \citep{Gobat2022}.

An alternative way to constrain the cold gas content of post-SB and quiescent galaxies is to target carbon monoxide (CO) line transitions, which are a good proxy for the dense molecular phase. If post-SBs are effectively on their route to quiescence, the presence of molecular gas points towards quenching mechanisms preventing gas from forming stars, rather than removing it completely. 
Several studies have detected substantial molecular gas reservoirs (gas fractions $\sim 10\% - 20\%$) in post-SB galaxies up to $z \lesssim 0.1$ and have shown that they have lower star formation efficiency (SFE) than star-forming galaxies, lying below the KS relation \citealt{French2015, Rowlands2015, Alatalo2016, Yesuf2017, French2018, French2021}). From a spatially-resolved analysis, \cite{Brownson2020} finds, in the central regions of local green valley galaxies, lower molecular gas fractions and SFE than in star-forming galaxies and suggest that AGN preventive feedback might inject thermal energy in the ISM, supporting molecular clouds against gravitational collapse, while the prominent bulge stabilizes the ISM against gravitational instabilities, suppressing the SFE \citep{Martig2009}.

To catch quenching in action and understand what its main drivers are, it is crucial to study galaxies close to their main quenching epoch at high redshift. Such studies are challenging though, due to the faint CO emissions of the targets. A handful quiescent galaxies have been detected in CO at $z \sim 0.5 - 2$ \citep{Sargent2015, Rudnik2017, Spilker2018, Hayashi2018, Williams2020} and they typically are massive systems ($\mathrm{M_\star \gtrsim 10^{11} M_\odot}$). Similarly, small samples of massive ($\mathrm{M_\star \gtrsim 10^{11} M_\odot}$) post-SB galaxies have been detected in CO (or have stringent upper limits) at high redshift $z \sim 0.5 - 1.5$ \citep{Suess2017, Bezanson2019, Bezanson2021, Belli2021}.
In this study we expand upon those works, including two post-SB galaxies at $z > 1$, with lower stellar masses ($\mathrm{M_\star \sim 5\times 10^{10} M_\odot}$) than other high-redshift samples from the literature, allowing us to start widening the explored parameter space.

This paper is organized as follows: in Section \ref{sec:data} we describe the sample selection and the data set; in Section \ref{sec:analysis} we discuss how we estimated the observables (CO and dust continuum flux) and derived physical properties (molecular gas mass, stellar mass, star formation rate, stellar age); in Section \ref{sec:results} we discuss our results and scaling relations among observables; in Section \ref{subsec:discussion} we discuss the possible mechanisms halting star formation in our galaxies and report on their environment; finally in Section \ref{sec:conclusions} we conclude and summarize our findings. Throughout the paper we adopt a flat $\Lambda$CDM cosmology with $\Omega_\mathrm{m} = 0.3$, $\Omega_\mathrm{\Lambda} = 0.7$, and $\mathrm{H_0} = 70$ km s$^{-1}$ Mpc$^{-1}$. All magnitudes are AB magnitudes \citep{Oke1974} and we adopt a \cite{Chabrier2003} initial mass function, unless differently stated.

\begin{table}
    \centering
    \caption{Log of the ALMA observations}
    \begin{tabular}{c c c c c}
    \toprule
    \midrule
    ID & Date & t$_\mathrm{exp}$ & Noise R. M. S. & Beam \\
       &      &     (min)        &   mJy/beam     & (arcsec)     \\
    (1) & (2) & (3)              &   (4)          & (5)      \\
    \midrule
    97148  & 09 Jan 2020 & 50 & 0.13 & $1.6 \times 1.3$  \\
    83492  & 09 Jan 2020 & 35 & 0.65 & $1.9 \times 1.3$ \\
    \bottomrule
    \end{tabular}
    \label{tab:log}
    
    \begin{minipage}{8.5cm}
    \textbf{Columns}: (1) Galaxy ID; (2) Date of observations; (3) Integration time on source; (4) Noise r. m. s.; (5) FWHM of the beam.
    \end{minipage}
\end{table}

\begin{table*}
    \centering
    \caption{Measurements for our sample galaxies}
    \begin{tabular}{c c c c c c c c}
    \toprule
    \midrule
    ID & RA & DEC & $z_\mathrm{CO}$ & $z_\mathrm{opt}$ & F$_\mathrm{CO}$ & $\Delta$v & F$_\mathrm{cont}$ \\
       &(deg) & (deg) &  &  & ($\mu$Jy) & km s$^{-1}$ & ($\mu$Jy) \\
    (1) & (2) & (3) & (4) & (5) & (6) & (7) & (8) \\
    \midrule
    97148 & 34.3335000 & -5.0840889 & 1.2728 $\pm$ 0.0007 & 1.273 & $299.4 \pm 78.2$ & 166 & $< 60.1$ \\
    97148 corrected$^1$ & 34.3335000 & -5.0840889 & 1.2728 $\pm$ 0.0007 & 1.273 & $687.7 \pm 78.2$ & 240 & $< 60.1$ \\
    83492 & 34.5469167 & -5.1470694 & 1.1393 $\pm$ 0.0005 & 1.138 & $258.8 \pm 50.9$ & 568 & $< 69.3$ \\
    Companion1 of 83492 & 34.5475633 & -5.1485586 & $1.1375 \pm 0.0002$ & -- & $789.0 \pm 55.5$ & 555 & $48.9 \pm 13.9$ \\
    Companion2 of 83492 & 34.5501383 & -5.1491131 & $1.1379 \pm 0.0001$ & -- & $1054.3 \pm 63.9$ & 439 & $65.3 \pm 13.9$ \\
    \bottomrule
    \end{tabular}
    \label{tab:observations}
    \vspace*{0.2cm}    
    \begin{minipage}{18cm}
    \textbf{Columns}: (1) Galaxy ID; (2) Right ascension of the galaxy center from the $H$-band; (3) Declination of the galaxy center from the $H$-band; (4) Redshift estimated by fitting the CO(3-2) emission line with a Gaussian in our 1D ALMA spectra. The uncertainty that we report is the formal error obtained from the fit; (5) Redshift estimated from the optical spectra \citep{Maltby2016}; (6) CO(3-2) emission line flux; (7) Line velocity width; (8) 5$\sigma$ upper limits on the continuum emission flux.\\
    \textbf{Notes}: $^1$See Section \ref{subsec:CO} for details.
    \end{minipage}
\end{table*}

\section{Data}
\label{sec:data}

In this study we focus on two, post-starburst galaxies at $z > 1$ (ID83492 and ID97148). We aim, by detecting their CO emission, to estimate their molecular gas content and constrain possible physical mechanisms responsible for galaxy quenching. In addition, we include samples from the literature (see Section \ref{subsec:literature}).

\subsection{Sample selection and ancillary data}
\label{subsec:sample}
Our two target galaxies were initially optically selected by \cite{Wild2014} with other 921 candidates applying a principal component analysis (PCA) on the spectral energy distributions (SED) based on the photometry of the Ultra-Deep Survey  that comprises data in the optical $BVRiz$, near-infrared $JHK$, and mid- to far-infrared from Subaru, \textit{Hubble Space Telescope} (\textit{HST}/WFC3), Very Large Telescope (VLT/VIRCAM, and FORS2), \textit{Spitzer} and \textit{Herschel}. \cite{Maltby2016} confirmed the post-SB nature of a subsample of 19 objects with rest-frame optical spectroscopy, showing deep Balmer absorption lines (EW(H$\delta$) $> 5 $\AA) and no or little [OII]3727$\mathrm{\AA}$ emission associated with star formation and low-luminosity AGN. For the present study, we selected two objects with stringent star formation rate limits (SFR $\lesssim 1$ M$_\odot$ yr$^{-1}$) and secure spectroscopic redshift (i.e., marked with flag ``4'' as being a highly reliable redshift, estimated to have $> 99$\% probability of being correct, based on a high signal-to-noise spectrum and supported by obvious and consistent spectral features, \citealt{Garilli2021}).
They are not detected in \textit{Herschel} imaging and neither by the Multi-Band Imaging Photometer for \textit{Spitzer} (MIPS) at 24 $\mu$m or Submillimeter Common-User Bolometer Array 2 (SCUBA-2) at 850$\mu$m probing  the Rayleigh-Jeans tail of the dust continuum emission down to an r.m.s. of 0.9 mJy/beam \citep{Geach2017}.

Both targets have rest-frame optical spectra taken with the VIsible MultiObject Spectrograph (VIMOS) previously mounted on the Very Large Telescope UT3 and available in the ESO archive. In particular, we used the spectrum of ID83492 which is already available among the phase 3 data products released by the VANDELS team\footnote{\url{https://eso.org/rm/api/v1/public/releaseDescriptions/147}}. For ID97148, we reduced the raw data with the VIPGI pipeline \citep{Scodeggio2005} used by the VANDELS team and following the same steps used to reduce the spectrum of ID83492 \citep{Pentericci2018}.

\subsection{ALMA data}
\label{subsec:alma}

We carried out ALMA Band 4 observations for our sample during Cycle 7 (PI: A. Zanella, Project ID: 2019.1.00900.S) with the goal of detecting the CO(3-2) emission line at rest-frame frequency $\nu_\mathrm{rf} = 345.795$ GHz and the underlying continuum, redshifted in the frequency range $\nu_\mathrm{obs} = 150 - 160$ GHz. 

We observed ID83492 for 35 minutes and ID97148 for 50 minutes on source and reached a sensitivity of 0.65 and 0.13 mJy/beam, respectively, over a bandwidth of 150 km s$^{-1}$. The native spectral resolution of the observations is 7.812 MHz ($\sim$ 15 km s$^{-1}$ -- later binned to lower velocity resolutions for our purposes). The imaged beam sizes are $\mathrm{FWHM} = 1\arcsec.9\, \times\, 1\arcsec.3$ for ID83492 and $1\arcsec.6\, \times\, 1\arcsec.3$ for ID97148 (Table \ref{tab:log}).

The data were reduced with the standard ALMA pipeline, based on the CASA software \citep{McMullin2007}. The calibrated data cubes were then converted to \textit{uvfits} format and analyzed with the software GILDAS \citep{Guilloteau2000}.

\subsection{Literature samples}
\label{subsec:literature}

To obtain a more comprehensive understanding of how quenching acts and cover a larger parameter space, we complemented our observations with data from the literature. To have a sample with homogeneously-derived molecular gas masses, we only considered galaxies with CO observations. We excluded from the literature comparison the galaxies with molecular gas mass estimates based on SED fitting and/or dust continuum observations. In particular we considered samples of star-forming (\citealt{Saintonge2011}, \citealt{Tacconi2013}, \citealt{Freundlich2019}), post-SB (\citealt{French2015}, \citealt{Rowlands2015}, \citealt{Suess2017}, \citealt{Bezanson2019}, \citealt{Belli2021}, \citealt{Bezanson2021}), and quiescent (\citealt{Sargent2015}, \citealt{Davis2016}, \citealt{Rudnik2017}, \citealt{Hayashi2018}, \citealt{Spilker2018}, \citealt{Williams2020}) galaxies up to redshift $z \sim 2$. In Appendix \ref{app:literature_data} we briefly describe these literature samples and comment on how the key properties that are relevant for our analysis (i.e., molecular gas mass, SFR) have been derived. We report such properties in Table \ref{tab:literature_data}.

\section{Analysis}
\label{sec:analysis}

\subsection{CO emission}
\label{subsec:CO}

To create the velocity-integrated CO(3-2) line maps for our galaxies we had to determine the optimal spectral range over which to integrate the spectra. We carried out the following iterative procedure, similar to the one described in \cite{Zanella2018}.
Since our targets are expected to be unresolved at the angular resolution of the observations, we modelled their emission in the \textit{uv} plane with point-source profiles in all four sidebands and channel per channel. In this process, we fixed the spatial position to that determined from the optical images (Section \ref{subsec:sample}). We extracted one-dimensional spectra using these models and we looked for positive emission line signal in the resulting spectra. When a
signal was present, we averaged the data over the channels maximizing the detection signal-to-noise ratio (S/N) and we fitted the resulting two-dimensional (channel-averaged) map to obtain the best fitting line spatial position. If this was different from the spatial position of the initial extraction we proceeded to a new spectral extraction at the new position, and iterate the procedure until convergence was reached.

\begin{landscape}
\begin{figure*}[t!]
    \centering
    \vspace{2cm}
    \hspace{-7cm}
    \includegraphics[width=1.3\textwidth]{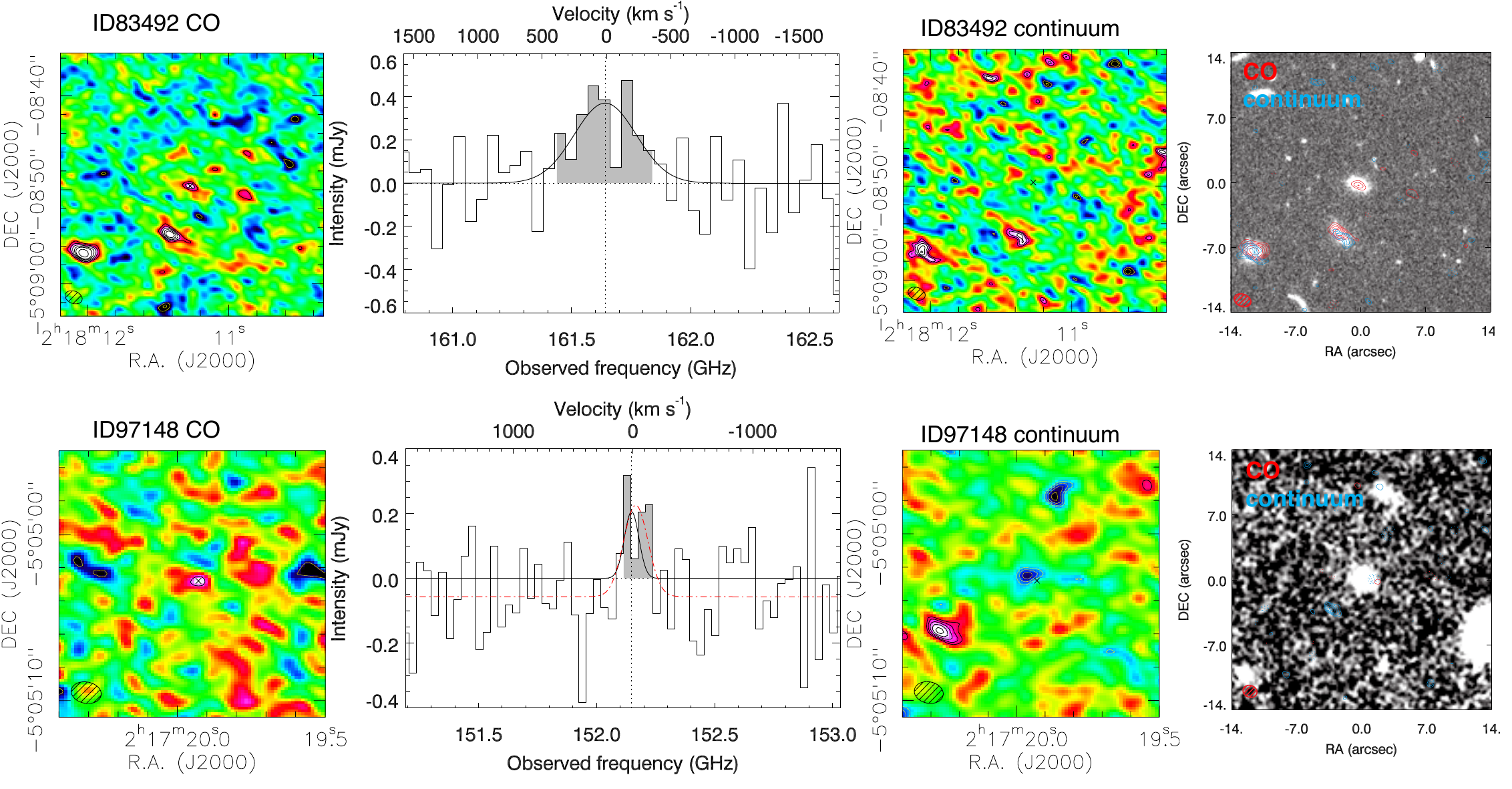}
    \caption{ALMA data of our sample galaxies. \textit{First column}: ALMA 2D maps of the CO(3-2) line. The black and white solid contours indicate respectively positive and negative levels of (9, 7, 6, 5, 4, 3) rms. The beam is reported as the black dashed ellipse. Each stamp has a size of 16" $\times$ 16". The black cross indicates the center of our post-SB targets as estimated from the CO(3-2) emission. Additional serendipitous detections visible in the map of ID83492 are likely companions (see Sections \ref{subsec:CO} and \ref{subsec:merger}). \textit{Second column}: 1D spectra of sources extracted using a PSF to maximize the S/N. The grey shaded areas indicate the 1$\sigma$ velocity range over which we measured the CO(3-2) line flux. For illustrative purposes we also report the Gaussian fit of the emissions: it was not used to estimate the line fluxes, but only as an alternative estimate of the redshift of the galaxies (Section \ref{subsec:CO}). We notice that the continuum of ID97148 is on average below zero despite the fact that the data were not continuum-subtracted. This is due to the negative noise peak which is also visible in the 2D continuum map (third column). The red Gaussian is the fit to the spectrum, allowing for a negative continuum. \textit{Third column}: ALMA 2D maps of the continuum emission. The black and white solid contours indicate respectively positive and negative levels from 4.5$\sigma$ to 2.5$\sigma$, in steps of 0.5$\sigma$. The black cross indicates the center of our post-SB targets as estimated from the CO(3-2) emission. The additional detections visible in both maps are likely companion galaxies (see Sections \ref{subsec:continuum} and \ref{subsec:merger}). \textit{Fourth column}: \textit{HST}/WFC3 F160W rest-frame optical continuum imaging (for ID83492) and VIRCAM/$H$ band imaging (for ID97148). The red contours show the CO(3-2) emission, while the cyan ones show the continuum emission. Contour levels are the same as in the first and third panels (solid for positive levels and dashed for negative levels). The ALMA beam is reported as the red shaded ellipse.}
    \label{fig:maps}
\end{figure*}
\end{landscape}

We securely detected the CO(3-2) emission line at 5$\sigma$ 
and 4$\sigma$ significance for ID83492 and ID97148, respectively (Figure \ref{fig:maps}). As expected, both sources are spatially unresolved. The optimized spatial position for the spectral extraction of ID83492 is consistent with that of the optical peak emission in the astrometrically-calibrated $H$ band

\noindent from \textit{HST}/WFC3, falling within the FWHM of the beam (offset $< 0.25\arcsec$). The spatial position of CO emission of ID97148 instead is offset by $\sim 1.6\arcsec$ with respect to the optical peak emission in the astrometrically-calibrated $H$ band from VLT/VIRCAM\footnote{We calibrated the astrometry of the VIRCAM data against Gaia DR3 by using 30 non-saturated stars in the field of view. Uncertainties on the calibration (estimated as the standard deviation of the $\Delta (RA)$ and $\Delta (DEC)$ offsets) are $\sigma_\mathrm{RA} \sim 0.08\arcsec$ and $\sigma_\mathrm{DEC} \sim 0.06\arcsec$.}. Despite this offset is within the FWHM of the beam ($1.6\arcsec \times 1.3\arcsec$) we notice that the $H$-band photometry show elongation in the direction of the CO detection which could be tracing an ongoing interaction (similarly to the EGS-18045 from \citealt{Belli2021}). Alternatively the CO emission might originate from a nearby dust-obscured satellite which is not detected in the near-infrared imaging (e.g., \citealt{Schreiber2018}). In this case the molecular gas mass that we estimate for ID97148 should be considered as an upper limit. However, the fact that the redshift of
the \cothree\ emission matches the one estimated from the optical spectrum (velocity difference $v_{\rm off} \sim 47\, \mathrm{km\, s^{-1}}$, consistent with null velocity offset within the uncertainties) rules out the possibility of a spurious detection\footnote{If we fit the CO map at the optical position of the galaxy, we retrieve a 3$\sigma$ CO flux upper limit $f_\mathrm{CO} < 235\, \mu\mathrm{m}$, and a corresponding molecular gas mass $\mathrm{M_{H2} < 3.2\times 10^9\, M_\odot}$.}. We also highlight that next to the CO(3-2) detected line there is a negative noise peak which is evident in the continuum map (Figure \ref{fig:maps}) and that could hide part of the emission and artificially offset its peak. To account for the effects of the negative noise peak on the line emission flux, we fitted the 1D spectrum (Figure \ref{fig:maps}) with a Gaussian model, allowing for a negative continuum, and estimated the integrated flux of the emission. We compared it with the integrated flux obtained when forcing the Gaussian to have zero continuum. We find that the flux doubles when we account for the negative continuum. Also the velocity width of the line increases by $\sim 40\%$. This translates into a molecular gas mass and molecular gas fraction that are 3 times larger (Section \ref{subsec:gas_mass}). In the following we  report both molecular gas estimates.

We estimated the redshift from the \cothree\ lines in two ways, both giving consistent results ($\delta z< 0.0004$): by computing the signal-weighted average frequency within the line channels and by fitting the one-dimensional spectrum with a Gaussian function. We compared these redshift estimates with those obtained from optical spectra \citep{Maltby2016} and found that they agree within $1\sigma$, increasing the reliability of the CO detection (Table \ref{tab:observations}). We finally estimated their \cothree\ flux by fitting their average emission line maps in the \textit{uv} plane adopting the Fourier Transform of a two-dimensional PSF model with the GILDAS task \texttt{uv\_fit}. Fluxes and uncertainties determined with the \texttt{uvmodelfit} task provided by CASA give consistent results. The obtained CO(3-2) total fluxes are reported in Table \ref{tab:observations}.\\

We also serendipitously detected bright emission lines and continuum from two additional galaxies in the surroundings of ID83492 (Figure \ref{fig:companions}). The lines are detected at $14\sigma$ and $16.5 \sigma$ and the continuum at $3.5\sigma$ and $4.7\sigma$ respectively (Table \ref{tab:observations}). They both have optical counterparts in \textit{Hubble Space Telescope} Wide Field Camera 3 (\textit{HST}/WFC3) imaging from the 3D-HST program (\citealt{Skelton2014}, Figure \ref{fig:maps}). Their photometric redshift is $z \sim 1.1$, implying that the detected line likely is CO(3-2) at $z = 1.1375 \pm 0.0002$ and $z = 1.1379 \pm 0.0001$, consistent with the spectroscopic redshift of ID83492. Given their velocity difference ($v_{\rm off} \sim 250$ km s$^{-1}$ and 195 km s$^{-1}$) and projected distance from ID83492 ($5.8\arcsec \sim 48\,\mathrm{kpc}$ and $13.7\arcsec \sim 112\,\mathrm{kpc}$), they might be companions that are merging with our target galaxy. In the literature, different criteria about the velocity difference and physical proximity are adopted when defining mergers. Galaxies with spectroscopic velocity differences $v_{\rm off} \lesssim 500$ km s$^{-1}$ and projected distances d$_\mathrm{proj} \lesssim 50$ kpc are typically considered as close pairs (e.g., \citealt{Lin2008, Patton2008, Lotz2011, Mantha2018}) and simulations predict that they merge within less than 1 Gyr (e.g., \citealt{Conselice2006}). Galaxies with larger separation (up to $\sim$ 150 kpc) are still considered pairs \citep{deRavel2009, Lotz2011} and are expected to merge on timescales of $\sim 1.5 - 2.5$ Gyr, depending on their baryonic mass ratio and orbital parameters \citep{Lotz2008, Lotz2010a, Lotz2010b}. Based on these results, ID83492 might merge with two companion galaxies within $< 2.5$ Gyr.
We report their observed and measured properties respectively in Table \ref{tab:observations} and Table \ref{tab:measurements}. We show their CO(3-2) spectra and spectral energy distributions in Figure \ref{fig:companions}. 

\begin{figure*}
    \centering
    \includegraphics[width=\textwidth]{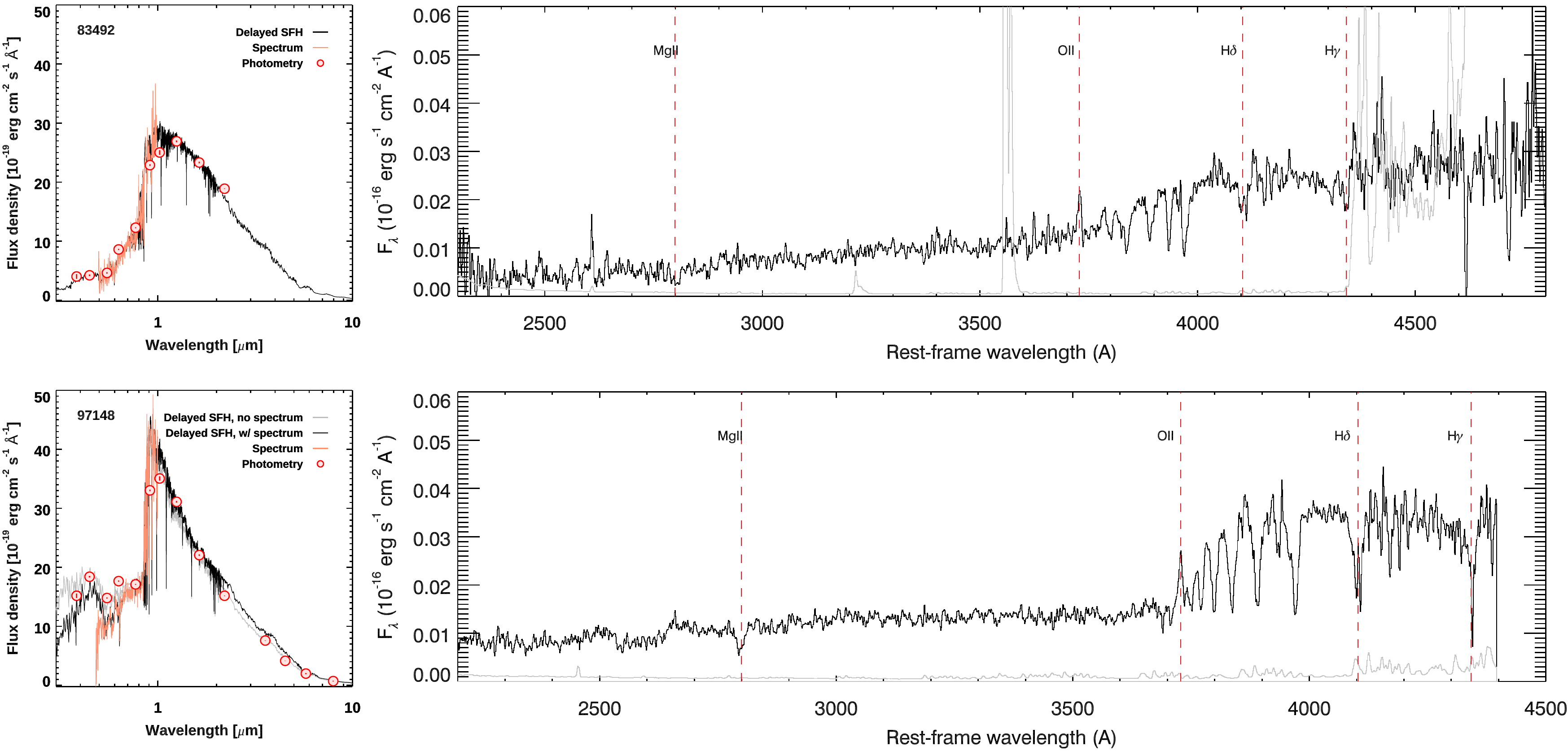}
    \caption{Optical and near-infrared photometry and VIMOS spectroscopy for our target post-SB galaxies. \textit{Left column:} spectral energy distribution modelling. The filled red circles and solid orange line indicate the observed photometry and the optical spectra that we modelled. The black solid line marks the best-fit models at fixed $z_{\rm spec}$ calculated by including the spectrum. For ID97148 the gray solid line shows the best-fit model at fixed $z_{\rm spec}$ calculated by excluding the spectrum. For ID83492 excluding the spectrum produces a comparable fit instead.  \textit{Right column:} VLT/VIMOS spectra. The black curve shows the spectrum, whereas the gray one shows the associated uncertainties. The vertical red dashed lines mark the main absorption or emission lines.}
    \label{fig:sed}
\end{figure*}

\subsection{Continuum emission}
\label{subsec:continuum}
We created averaged continuum maps by integrating the spectral range, after excluding the channels where
the flux is dominated by the CO emission line.
None of our main targets is detected. In Table \ref{tab:observations}, we report the total flux density upper limits that we estimated as the 5$\sigma$ uncertainty obtained when fitting, in the \textit{uv} plane, the Fourier Transform of a two-dimensional PSF model with center fixed at the position of the CO detection.

We detected the continuum emission of the two serendipitous companion galaxies of ID83492 at 3.5$\sigma$ and 4.7$\sigma$ significance. We also serendipitously detected at $4.8\sigma$ the continuum of an additional source nearby ID97148  (Figure \ref{fig:maps}). However this source is not detected in CO(3-2) and it does not have a counterpart in current catalogs including ground-based optical observations \citep{Tanaka2015}. A possible counterpart is found in the catalog by \cite{Mehta2018}, but it has a distance of $\sim 0\arcsec.5$ from ID97148 and a photometric redshift $z \sim 1.1$. The SED fit also returns a very low stellar mass and SFR (log(M$_\star/$M$_\odot$) = 7.7, log(SFR/(M$_\odot$yr$^{-1}$)) = 0.33). Given the angular distance from ID97148, the redshift difference of $\Delta z > 0.15$ and the highly uncertain values retrieved from the SED fit, we do not consider such detection as a companion galaxy, but rather as a spurious detection or lower-redshift and low-mass interloper.

\subsection{Molecular gas mass}
\label{subsec:gas_mass}

We estimated the molecular gas mass of our galaxies from their CO luminosity. We considered a brightness temperature ratio $r_{32} = 0.5$ \citep{Carilli2013} to account for the excitation of the CO(3-2) transition with respect to the thermalized case.  We then converted the estimated CO(1-0) luminosity into molecular gas mass by adopting the standard Milky-Way like conversion factor $\alpha_\mathrm{CO} = 4.4$ M$_\odot$ K km s$^{-1}$ pc$^2$ \citep{Bolatto2013}. This is consistent with most literature studies used to compare with our results (Section \ref{subsec:literature}). We corrected the molecular gas mass from literature samples that adopt a different $\alpha_\mathrm{CO}$ conversion factor to make them consistent with our assumptions. For our sample galaxies, we obtain molecular gas masses M$_\mathrm{H2} = (4.1 \pm 1.1) \times 10^9$ M$_\odot$ for ID97148 and M$_\mathrm{H2} = (9.9 \pm 1.9) \times 10^9$ M$_\odot$ for ID83492 (Table \ref{tab:measurements}). These uncertainties do not account for systematic differences in the (unconstrained) excitation and CO-to-H$_2$ factors in post-SB galaxies, but allow to directly compare our results with the literature on star-forming and quiescent galaxies.

We also estimated an upper limit on the gas mass of our galaxies by using the ALMA continuum which is probing the Rayleigh-Jeans tail of the far-infrared dust emission. The continuum at 850$\mu$m in fact probes a wavelength regime where the dust emission is optically thin, providing direct information on the dust (and thus gas) content \citep{Magdis2012, Scoville2013, Scoville2015}. By considering the continuum flux upper limits estimated as in Section \ref{subsec:continuum} (Table \ref{tab:observations}), a spectral index $\beta = 1.8$ \citep{Magdis2021, Cochrane2022}, and a specific luminosity at 850 GHz rest-frame (L$_{\nu_{850}}$) to gas mass conversion factor $\alpha_{850} = \mathrm{L_{\nu_{850}}/M_{gas}} = 1\times 10^{20}$ erg $\rm s^{-1}$ Hz$^{-1}$ M$_\odot^{-1}$ \citep{Scoville2015}, we obtained M$_\mathrm{gas} \lesssim 1.3 \times 10^{10}$ M$_\odot$ for both galaxies at $5\sigma$. We also stacked the continuum maps of our target galaxies at the position of the CO detections, reaching a 5$\sigma$ upper limit on the flux F$_\mathrm{cont} \lesssim 9.1 \mu$Jy. This implies an upper limit on the average gas mass M$_\mathrm{gas} \lesssim 9.2 \times 10^9$ M$_\odot$. These upper limits are consistent with the molecular gas estimates obtained from CO within the uncertainties related not only to the flux measurements, but also to the conversion factors (e.g., $r_{32}$, $\alpha_\mathrm{CO}$, $\alpha_{850}$).

\subsection{Stellar mass and star formation rate}
\label{subsec:sfr}
We modelled the multi-wavelength photometric catalog in the SXDS field described in \cite{Kubo2018}. The catalog comprises $u$-band data obtained with  CFHT/Megacam, $BVRiz$ with Subaru/Suprime-Cam, $JHK$ from the UKIRT Infrared Deep Sky Survey (UKIDSS, \citealt{Lawrence2007}), and \textit{Spitzer} observations from the UKIDSS Ultra Deep Survey (SpUDS; PI: J. Dunlop). We fit the SEDs with \textsc{Fast++}\footnote{\url{https://github.com/cschreib/fastpp}} \citep{Schreiber2018} using \citet{Bruzual2003} models, the \cite{Chabrier2003} IMF, delayed SFHs ($\mathrm{SFR}(t)\propto te^{-t/\tau}$ with $\rm log(\tau/yr^{-1})\in[6.5,10]$ in log-steps of 0.1), ages in log-steps of 0.1 from a minimum of 100 Myr and a maximum set by the age of the Universe, the \citet{Calzetti2000} dust attenuation curve ($A_{\rm V}\in[0,6]$ mag), and free metallicities including sub- and super-solar estimates ($Z=0.008, 0.02, 0.05$). We fixed the redshifts to the spectroscopic values and modelled the photometry and the spectra simultaneously. In the case of ID83492, the inclusion of the rest-frame optical spectrum does not appreciably affect the best-fit parameters obtained by modelling the photometry only (Figure \ref{fig:sed}). On the contrary, the spectrum of ID97148 drops more steeply than the photometry at short wavelengths. We ascribed this difference to an imperfect flux calibration of the spectrum and compensated for it by boosting its uncertainties by a factor of $3\times$. This more conservative value does not hamper our ability to reproduce the main absorption features in the spectrum and allows us to simultaneously retrieve the overall SED shape at longer wavelengths (Figure \ref{fig:sed}). We note that the photometry adopted here is consistent with an independent extraction by \cite{Mehta2018}, which supports its validity and constraining power. In the case of \ideight, the photometry is also consistent with that from 3D-HST \citep{Skelton2014}. The best-fit parameters and their uncertainties are reported in Table \ref{tab:measurements}. The latter were estimated both analytically following \cite{Avni1976} and via Monte Carlo bootstrapping (see \citealt{Schreiber2018} for details about the exact implementation in \textsc{Fast++}).

We also checked for consistency with SED fitting results reported in the literature \citep{Carnall2019, Wild2020}. Only one of our targets (ID83492) is included in those samples and its stellar mass is reported. Both \cite{Carnall2019} and \cite{Wild2020} fit the SED with the code BAGPIPES, with \cite{Bruzual2003} spectral synthesis models, assuming a \cite{Chabrier2003} IMF. While \cite{Carnall2019} parametrizes the SFH with a double-power-law model, \cite{Wild2020} also allows for a secondary burst of star formation (exponentially declining) not to force all the stellar mass production to happen in a single burst. The stellar mass estimated for ID83492 by \cite{Wild2020} is $\mathrm{log(M_\star/M_\odot)} = 11.19 \pm 0.2$ when fitting the photometric data only and $\mathrm{log(M_\star/M_\odot)} = 11.08 \pm 0.05$ when fitting photometric and spectroscopic data, whereas \cite{Carnall2019} reports $\mathrm{log(M_\star/M_\odot)} = 11.29 \pm 0.2$. Our stellar mass estimate is also consistent with that reported by 3D-HST (\citealt{Skelton2014},  $\mathrm{log(M_\star/M_\odot)}=10.86$ estimated using the code \textsc{Fast++}) and that reported by \cite[$\mathrm{log(M_\star/M_\odot)} = 10.91$]{Mehta2018}. Despite the different assumptions, SFH used, and codes adopted for the SED fit, all these estimates are consistent within them and with ours, within the uncertainties.

Both our target galaxies have detected [OII] emission in the optical spectra (Section \ref{subsec:sample}). We estimate the dust-corrected [OII] luminosity by using the extinction estimated from the best-fit SED and accounting for the fact that emission lines are more attenuated than the stellar continuum ($\mathrm{E(B-V)_{star} = 0.58 E(B-V)_{nebular}}$, \citealt{Calzetti2000}). We convert the dust-corrected [OII] luminosity into SFR by using the calibration by \cite{Kewley2004}. We find SFR = $0.2 \pm 0.1$ M$_\odot$ yr$^{-1}$ for ID97148 and SFR = $0.3 \pm 0.1$ M$_\odot$ yr$^{-1}$ for ID83492, which are consistent, within the uncertainties, with those estimated from the SED fit. In the following we adopt SFR from the SED fit for consistency with the estimates obtained in the literature.

\subsection{Stellar age}
\label{subsec: stellar age}
\begin{figure}
    \centering
    \includegraphics[width=0.4\textwidth]{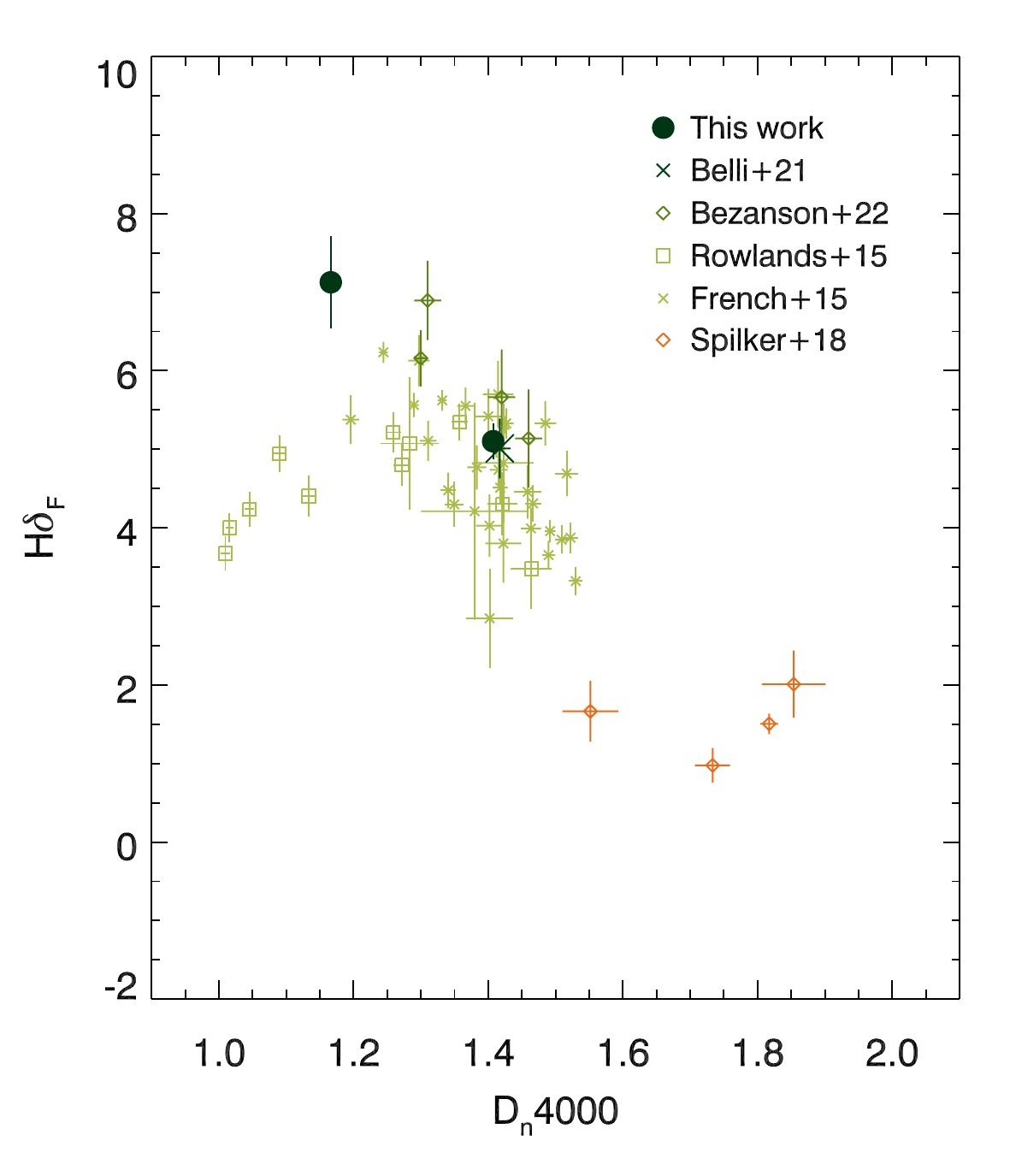}
    \caption{H$\delta_\mathrm{F}$ absorption index as a function of D$n$4000. Our sample galaxies (dark green filled circles) are reported together with literature samples from \citet[light green crosses]{French2015}, \citet[light green squares]{Rowlands2015}, \citet[orange diamonds]{Spilker2018}, \citet[dark green cross, partially hidden behind the symbol indicating one of our sample galaxies]{Belli2021}, \citet[dark green diamonds]{Bezanson2021}. Of the sample of \citet{Bezanson2021} we report only the 4 galaxies that have available measurements for the H$\delta_\mathrm{F}$ index (see Appendix \ref{app:literature_data}).}
    \label{fig:hdf}
\end{figure}

Stellar absorption features in the optical spectra, such as the Balmer lines, \color{black} together with the 4000$\AA$-break (D$_n$4000) are powerful tools to constrain the mean stellar age and the recent star formation history of galaxies \citep{Worthey1994, Kauffmann2003, Gallazzi2005}. From the optical spectra of our galaxies we measure the Lick indices of the high-order Balmer lines H$\delta_\mathrm{F}$, H$\gamma\mathrm{F}$\footnote{Note that we choose the H$\delta_\mathrm{F}$ and H$\gamma_\mathrm{F}$ definitions instead of the wider H$\delta_\mathrm{A}$, H$\gamma_\mathrm{A}$ used in \citep{Gallazzi2014} because the latter cannot be measured for ID97148.}, and the D$_n$4000 index following \cite{Gallazzi2014}. For ID97148 we obtain H$\delta_\mathrm{F} = 7.1 \pm 0.6$, H$\gamma_\mathrm{F} = 7.0 \pm 0.6$, and D$_n4000 = 1.166 \pm 0.010$, whereas for ID83492 we estimate H$\delta_\mathrm{F} = 5.1 \pm 0.2$, H$\gamma_\mathrm{F} = 2.9 \pm 4.6$, and D$_n4000 = 1.407 \pm 0.007$. We add a 10\% error to the error budget of D$_n4000$ to account for uncertainties in spectrophotometric calibration (Figure \ref{fig:hdf}).

We estimated the mean age of the stellar populations of our sample galaxies by adopting the Bayesian approach described in \cite{Gallazzi2005,Gallazzi2014}. Specifically, we compare the observed D$_n4000$, H$\delta_\mathrm{F}$, H$\gamma_\mathrm{F}$ absorption index strengths with those predicted by a large Monte Carlo library of model spectra. We adopted the model library described in \cite{Zibetti2017}, which comprises 500000 synthetic spectra assuming randomly generated star formation histories \citep[][allowing both rising and declining phases, with the addition of random bursts]{Sandage1986} and metal enrichment histories \citep[see][for more details]{Zibetti2017, Zibetti2022}. The base models are \cite{Bruzual2003} Simple Stellar Populations in the 2016 version \citep{Chevallard2016}. Considering only the models with a formation epoch compatible with the age of the Universe at the redshift of our sample galaxies, we construct the posterior probability function (PDF) of the $R$-band light weighted mean stellar age and of the mass-weighted mean stellar age. We take the median of the PDF as our fiducial value and the 16th and 84th percentile range as the uncertainties accounting for model degeneracies. We estimate the uncertainties related to using different sets of indices to measure ages of high-redshift post-SBs to be $\sim 0.02$ dex  (see Appendix \ref{app:literature_data}, point \textit{(viii) High-redshift post-SBs}).

As a result, for ID97148 we estimate an $R$-band light-weighted mean age of $\mathrm{log(age/yr)} = 8.6\pm 0.1$ and for ID83492 $\mathrm{log(age/yr)} = 8.9 \pm 0.1$. The estimated mass-weighted ages are very similar, only 0.03~dex older, but with larger uncertainties of $\sim0.2$dex. \footnote{The model library adopted here differs from the one adopted in \cite{Gallazzi2014}, which assumed exponential SFHs with random bursts and fixed metallicity along each SFH and which is based on the original BC03 SSP models. The new library results in a better coverage of the index-index diagnostic diagrams. As a reference, we note that adopting the \cite{Gallazzi2014} library we would estimate ages older by 0.2~dex and 0.1~dex for ID97148 and ID83492, respectively.} These results are consistent with the ages obtained from SED fitting, within the uncertainties: $\mathrm{log(age/yr)} = 8.4^{+0.5}_{-0.3}$ for ID97148 and $\mathrm{log(age/yr)} = 9.4^{+0.3}_{-0.9}$ for ID83492. In the following we adopt the ages estimated from the stellar population analysis. However, adopting those derived from SED fitting instead, would make our results (e.g., anti-correlation between gas fraction and age, Section \ref{subsec:discussion}) even stronger.

\begin{table*}
    \centering
    \caption{Physical properties of our sample galaxies.}
    \label{tab:measurements}
    \begin{tabular}{c c c c c c c c}
    \toprule
    \midrule
    ID & log(SFR)$^1$ & log(M$_\star$)$^1$ & M$_\mathrm{H2}$ & SFE & t$_\mathrm{d}$ & log(f$_\mathrm{H2}$) & age \\
       &(M$_\odot$ yr$^{-1}$) & (M$_\odot$) & ($10^9$M$_\odot$) & (Gyr$^{-1}$) & (Gyr) & & Gyr \\
    (1) & (2) & (3) & (4) & (5) & (6) & (7) & (8) \\
    \midrule
    97148 & $< -0.14$ & $10.66 \pm 0.13$ & $4.1 \pm 1.1$ & $< 0.20$ & $> 5.7$ & $-1.0 \pm 0.1$ & $8.6^{+0.1}_{-0.1}$ \\
    97148 corrected$^2$ & $< -0.14$ & $10.66 \pm 0.13$ & $13.8 \pm 1.6$ & $< 0.05$ & $> 19$ & $-1.0 \pm 0.1$ & $8.6^{+0.1}_{-0.1}$ \\
    83492 & $-0.28 \pm 0.40$ & $10.79 \pm 0.20$ & $9.9 \pm 1.9$ & 0.05 & 18.9 & $-0.8 \pm 0.1$ & $8.9^{+0.1}_{-0.1}$ \\
    Companion1 of 83492 & $2.47^{+0.24}_{-0.63}$ & $10.57 \pm 0.10$ & $29.5 \pm 2.1$ & $10.00$ & 0.10 & $0.13 \pm 0.1$ & - \\
    Companion2 of 83492 & $2.93^{+0.01}_{-0.63}$ & $10.80 \pm 0.20$ & $31.1 \pm 1.9$ & $27.37$ & 0.04 & $0.31 \pm 0.2$ & - \\
    \bottomrule
    \end{tabular}
    \begin{minipage}{18cm}
    \vspace*{0.2cm}
    \textbf{Columns}: (1) Galaxy ID; (2) For the two post-SBs: SFR averaged over 10 Myr from the SED modelling; for the two companions: SFR averaged over 100 Myr from the SED modelling; (3) stellar mass from the SED modelling; (4) molecular gas mass estimated from \cothree\ adopting $r_{31}=0.5$ and $\alpha_{\rm CO}$ = 4.4 M$_\odot$ K km s$^{-1}$ pc$^2$ (Section \ref{subsec:gas_mass}); (5) star formation efficiency, SFE = SFR/M$_\mathrm{H2}$; (6) depletion time, t$_\mathrm{d} =$ M$_\mathrm{H2}/$SFR (7) gas fraction, f$_\mathrm{H2} = $M$_\mathrm{H2}/$M$_\star$; (8) average $R$-band light-weighted age from stellar population modelling. \\
    \textbf{Notes:} $^1$The uncertainties on the quantities derived from the SED modelling (Section \ref{subsec:sfr}) represent the 95\% confidence interval. $^2$See Section \ref{subsec:CO} for details. \\ Stellar masses and SFRs have all been homogenized to \cite{Chabrier2003} IMF, and molecular gas masses have been homogenized to $\mathrm{\alpha_{CO} = 4.4}$ K km s$^{-1}$ pc$^2$.
    \end{minipage}
\end{table*}

\begin{figure*}
    \centering
    \includegraphics[width=\textwidth]{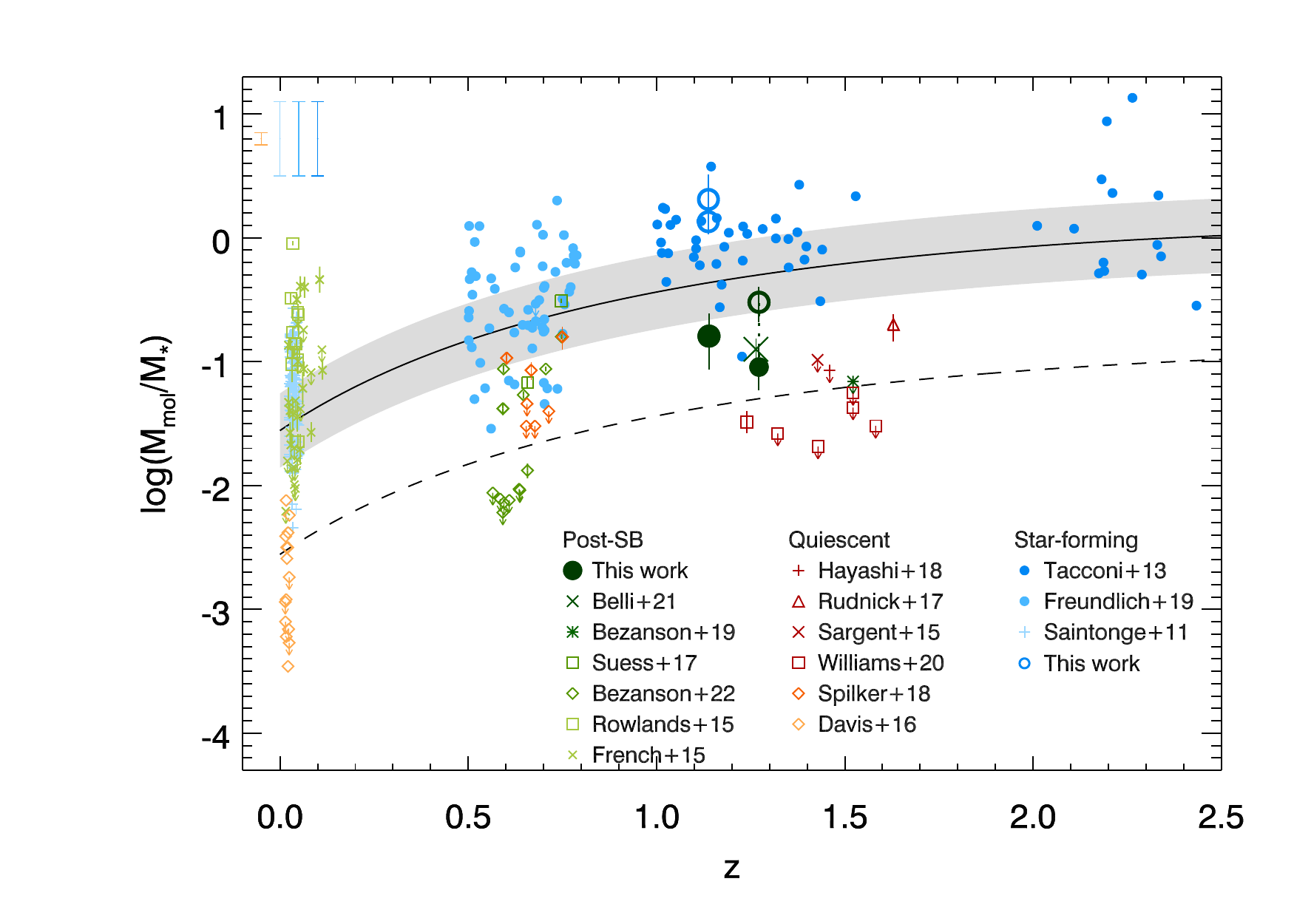}
    \caption{Evolution of the molecular gas fraction with redshift. We show our two post-starburst galaxies (dark green circles, with the filled/empty circle indicating the measurements for ID97148 obtained without or with the correction for the negative continuum, see Section\ref{subsec:CO}), as well as the star forming companions of ID83492 (cyan emtpy circles). We also report star-forming \citep[blue and cyan symbols]{Saintonge2011, Tacconi2013, Freundlich2019}, quiescent \citep[red and orange symbols]{Sargent2015, Davis2016, Rudnik2017, Spilker2018, Hayashi2018, Williams2020}, and post-starburst \citep[green symbols]{French2015, Rowlands2015, Suess2017, Belli2021, Bezanson2019, Bezanson2021} galaxies from the literature. The darkness of symbols correlate with their redshift (lighter colors at lower redshift). Upper limits are indicated with arrows. Typical error bars for the largest statistical samples \citep{Saintonge2011, Davis2016, Tacconi2013, Freundlich2019} are shown as vertical bars in the top left corner. The black curve shows the evolution of the main-sequence for galaxies with stellar masses $9.0 \lesssim \mathrm{log(M_\star/M_\odot)} \lesssim 11.8$ \citep{Tacconi2018}. The gray shaded area shows a 0.2 dex scatter, corresponding to the 1$\sigma$ scatter of the main-sequence. The dashed curve corresponds to a factor 10 vertical shift of the main-sequence.}
    \label{fig:z_evol}
\end{figure*}

\section{Results and scaling relations: linking M$_\mathrm{H2}$, M$_\star$, SFR, and redshift}
\label{sec:results}

\subsection{Molecular gas fraction evolution with redshift}
\label{subsec:gas fraction}

Based on the CO(3-2) detections, we estimate gas fractions M$_\mathrm{mol}$/M$_\star$ of 8\% and 16\% for our sample galaxies. These are comparable to gas fractions of other post-SB galaxies at similar redshift from the literature \citep{Belli2021, Bezanson2019, Suess2017} which have M$_\mathrm{mol}$/M$_\star \sim$ 5 -- 25\%. Such gas fractions are comparable or larger than those measured in quiescent galaxies at high redshift \citep{Spilker2018, Hayashi2018, Rudnik2017, Sargent2015} which are often not detected and have M$_\mathrm{mol}$/M$_\star \lesssim$ 15\%. In Figure \ref{fig:z_evol} we show the molecular gas fraction of galaxies as a function of redshift. We compare different populations: post-SB (our sample, \citealt{Suess2017}, \citealt{Bezanson2019}, \citealt{Belli2021}), quiescent \citep{Sargent2015, Davis2016, Rudnik2017, Hayashi2018, Spilker2018, Williams2020}, and star-forming \citep{Saintonge2011, Tacconi2013, Freundlich2019} galaxies with spectroscopic redshifts in the range $0 < z < 2$. 

The gas fraction of star-forming galaxies increases with redshift, with $z \sim 1 - 1.5$ sources having one order of magnitude higher M$_\mathrm{mol}$/M$_\star$ than local ones in agreement with the evolution of the main-sequence parametrized by \cite{Tacconi2018}. In the local Universe the molecular gas fraction of quiescent galaxies is $\gtrsim 10\times$ lower than that of star-forming systems at the same redshift (Figure \ref{fig:z_evol}). At $z > 0.5$ the CO line of most quiescent galaxies is not detected and therefore $f_\mathrm{H2}$ is an upper limit. Deeper observations are needed to confirm whether high-redshift quiescent galaxies are indeed more gas-rich than local ones. Post-starburst galaxies at $z > 0.5$ on the contrary have been mostly detected in current CO observations and have intermediate $f_\mathrm{H2}$ between star-forming and quiescent systems. In particular, our post-SB galaxies double the sample currently available at $z > 1$ and have intermediate gas fraction between quiescent and star-forming ones (Figure \ref{fig:z_evol}). The other two post-SBs at $z > 1$ from the literature with CO observations are from \cite{Bezanson2019} (not detected in CO) and  \cite{Belli2021} (CO detected). The latter study analyzes three massive galaxies at $1 < z < 1.3$, selected to be detected at $24\mu$m. We only consider the one which is classified as post-SB, namely EGS-17533, which has a gas fraction $f_\mathrm{H2} \sim 13\%$. This is consistent with the results obtained for our post-SB targets ($f_\mathrm{H2} \sim$ 8\% - 16\%), which are not detected at $24\mu$m instead. 

While the decline of gas fraction with redshift is evident for star-forming galaxies, such an evolution is less clear (and possibly absent) for post-SBs . In particular, post-SBs seem to divide in two populations: those with gas fraction $\mathrm{f_{H2} \gtrsim 5\%}$ at all redshifts (from $z \sim 0$ up to $z \sim 1.5$), and those which are not CO detected and have molecular gas fraction upper limits $\mathrm{f_{H2} \lesssim 1\% - 5\%}$, more similar to high-redshift quiescent galaxies (Figure \ref{fig:z_evol}). The latter are, for example, the 8 post-SBs from the sample of \cite{Bezanson2021} which are not CO-detected and have $f_\mathrm{H2} < 1$\%. They are among the most massive (M$_\star > 2.5\times 10^{11}$ M$_\odot$) and oldest (time since quenching $> 0.1$ Gyr, D$_n4000 > 1.3$) post-SBs reported in the literature, suggesting that quenching mechanisms could depend on galaxy mass (see Section \ref{subsec:discussion}).

\begin{figure*}
    \centering
    \includegraphics[width=\textwidth]{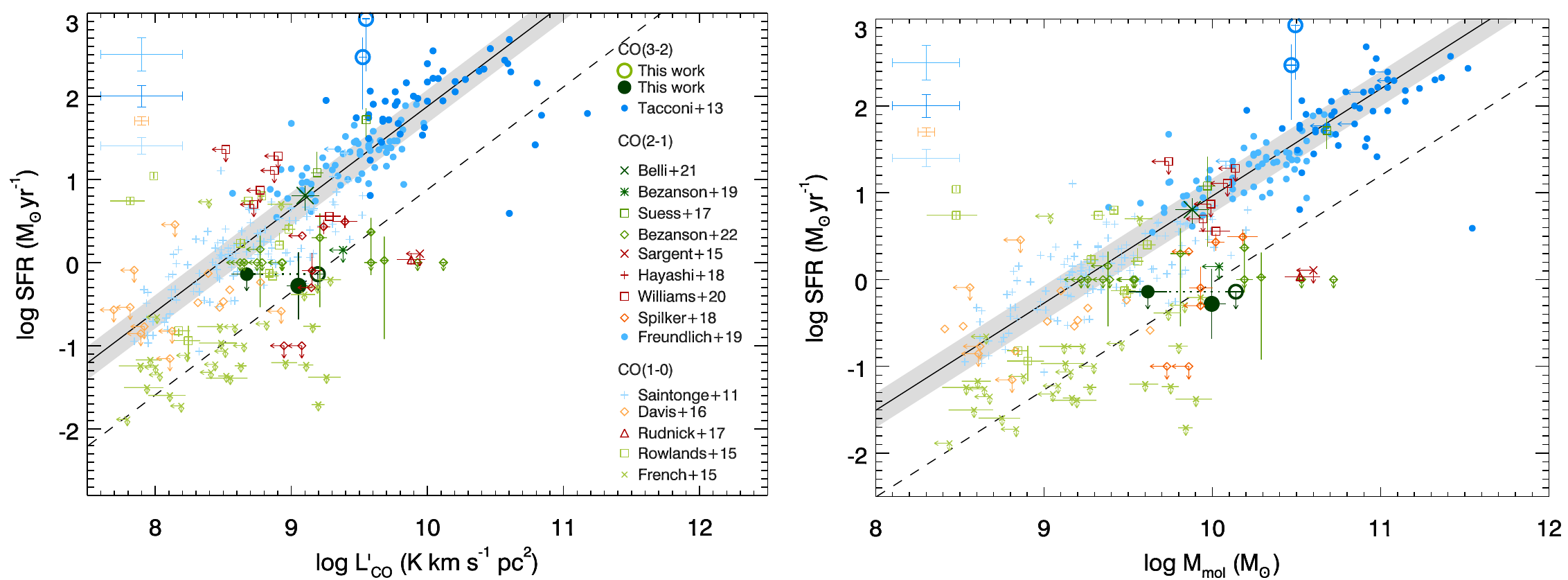}
    \caption{Schmidt-Kennicutt plane. \textit{Left panel}: correlation between SFR and CO luminosity for main-sequence, post-starbursts, and quiescent galaxies both from our work (dark green circles, with the filled/empty circle indicating the measurements for ID97148 obtained without or with the correction for the negative continuum, see Section \ref{subsec:CO}) and the literature \citep[symbols as in Figure \ref{fig:z_evol}]{Saintonge2011, Tacconi2013, Sargent2015, French2015, Rowlands2015, Davis2016, Rudnik2017, Suess2017, Spilker2018, Hayashi2018, Bezanson2019, Freundlich2019, Belli2021, Williams2020, Bezanson2021}. The black line indicates the redshift-independent relation SFR - M$_\mathrm{H2}$ estimated for main-sequence galaxies by \citep{Sargent2014}, re-scaled by considering $\alpha_\mathrm{CO} = 4.4$ M$_\odot$ K km s$^{-1}$ pc$^2$ and $r_\mathrm{21} = 0.8$. The gray shaded area is its scatter. The dashed line corresponds to a factor 10 vertical shift of the main-sequence curve. \textit{Right panel}: correlation between SFR and molecular gas mass. The black line indicates the typical relation for main-sequence galaxies \citep{Sargent2014} and the gray shaded area its scatter. The dashed line corresponds to a factor 10 vertical shift of the main-sequence curve. All symbols are used as in the left panel.}
    \label{fig:SFE}
\end{figure*}

\subsection{Star formation efficiency and Kennicutt-Schmidt plane}
\label{subsec:SFE}

By comparing the SFR and the molecular gas mass of galaxies in the traditional Schmidt-Kennicutt plane \citep{Kennicutt1998} it is possible to assess their star formation efficiency and in turn set constraints on models of galaxy quenching. We show the SFR - M$_\mathrm{H2}$ relation of our sample and literature galaxies in Figure \ref{fig:SFE} (right panel). Star-forming galaxies at all redshifts follow a tight relation with a scatter of $\sim 0.2$ dex, mostly independent of redshift \citep{Sargent2014}. Also the local quiescent galaxies from \cite{Davis2016}, which sample the low-SFR and low-M$_\mathrm{mol}$ part of the parameter space, appear to follow such relation. Quiescent and post-SB systems at $z > 0.5$ instead show a systematic offset from this relation toward lower SFR, but with a large scatter ($\sim 1$ dex). In particular our two post-SBs have one order of magnitude lower molecular gas mass and almost two orders of magnitude lower SFR than star-forming galaxies at similar redshift  (e.g., \citealt{Tacconi2013}). Their star formation efficiency, SFE = SFR/M$_\mathrm{mol} \lesssim 0.1$ Gyr$^{-1}$, is comparable to those of quiescent galaxies at similar redshift.

The molecular gas fraction in different samples has been measured by using different CO line transitions, namely CO(1-0), CO(2-1), and CO(3-2), and converting them to M$_\mathrm{H2}$ by using the excitation factors $r_\mathrm{J1}$ and $\alpha_\mathrm{CO}$ conversion factor, as discussed in Section \ref{subsec:gas_mass}. We investigated whether the use of different tracers affects the offset and scatter of quiescent and post-SB galaxies in the Schmidt-Kennicutt plane by directly comparing their SFR and CO luminosity ($L'_\mathrm{CO}$, Figure \ref{fig:SFE}, left panel). The offset from the relation of star-forming galaxies remains, as well as the scatter. Our sample galaxies have CO luminosity comparable to $z \sim 0$ star-forming galaxies from the COLDGASS survey \citep{Saintonge2011} and $\sim 1$ dex lower than star-forming systems at $z \sim 1.3$ \citep{Tacconi2013}.

Another source of uncertainty is the estimate of the SFR for post-SB and quiescent galaxies as detecting low levels of star formation at high redshift is observationally challenging. In the literature SFR estimated using different tracers have been compared. In particular, \cite{Belli2021} measured the SFR by using three different methods: UV+IR emission, UV-to-IR spectral fit, and dust-corrected [OII] emission. They found that the UV+IR-based SFR estimates are systematically $2 - 6$ times higher than those obtained with the other tracers. This is in agreement with previous findings by \cite{Hayward2014} and \cite{Smercina2018} who pointed out that the relatively old stellar population hosted by post-SBs heats the dust leading to a boost in infrared emission which is unrelated to star formation, hence resulting in higher UV+IR-based SFR estimates than those obtained when properly modelling the UV-to-IR SED. They also pointed out that SFR estimates based on dust-corrected emission lines (e.g., [OII], H$\alpha$, [NeII]+[NeIII]) might be underestimated due to inadequate dust correction and are possibly contaminated by AGN emission. On the other hand, the SFR measurements from SED fitting depend on degeneracies of the parameters (e.g., degeneracies between metallicity, age, and extinction of the stellar population) and on the adopted models themselves. In particular SED fitting results for post-SBs are very sensitive to the choice of star formation history, which can lead to systematic differences in the estimate of SFR up to 0.5 dex (e.g., \citealt{French2018}, \citealt{Wild2020}, \citealt{Suess2022a}). In addition, different SFR proxies trace different timescales with the emission lines probing younger stellar populations than the continuum. Due to these reasons, different SFR values are reported in the literature for the same galaxies. One striking example is SDSSJ0912+1523: \cite{Suess2017} report a $\mathrm{SFR= 2.1 \pm 0.8 M\odot yr^{-1}}$ estimated using the dust-corrected [OII] luminosity; \cite{Bezanson2021} report a $\mathrm{SFR= 0.81^{+1.33}_{-0.76} M\odot yr^{-1}}$ derived from the spectro-photometric fit of UV-to-IR data with a set of non parameteric star formation histories \citep{Suess2022b}; finally \cite{Belli2021} report three different SFR estimates, $\mathrm{SFR < 257 M\odot yr^{-1}}$ derived from the weighted sum of the UV and IR luminosities \citep{Bell2005, Wuyts2008}, $\mathrm{SFR= 52^{+20}_{-20} M\odot yr^{-1}}$ from spectral fit, and $\mathrm{SFR= 4.6 \pm 1.4 M\odot yr^{-1}}$ from the dust-corrected [OII] luminosity including extra dust attenuation toward the HII regions. In the following, we adopt SFR obtained through SED fit whenever available and in Appendix \ref{app:literature_data} we report the assumptions underlying the SFR estimates for the literature comparison samples.

\begin{figure*}
    \centering
    \includegraphics[width=\textwidth]{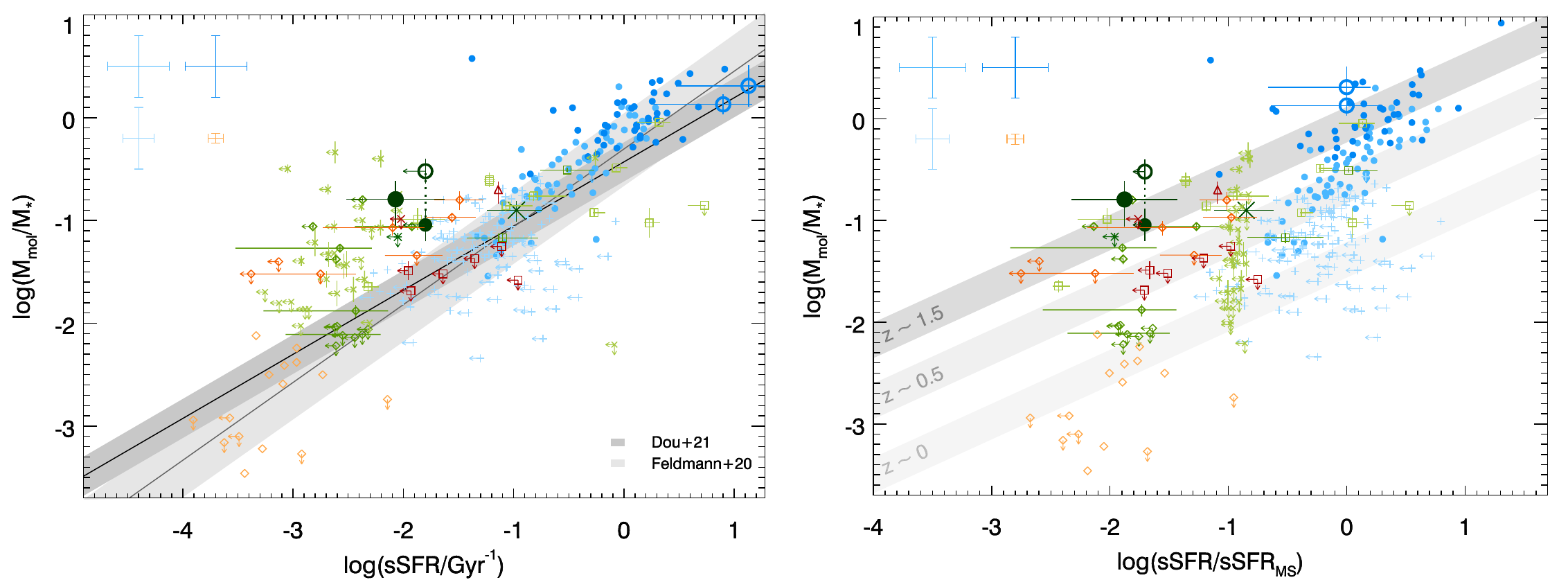}
    \caption{Molecular gas fraction as a function of galaxies specific star formation rate. \textit{Left panel}: Main-sequence, post-starbursts, and quiescent galaxies both from our work (dark green circles, with the filled/empty circle indicating the measurements for ID97148 obtained without or with the correction for the negative continuum, see Section\ref{subsec:CO}) and the literature (symbols as in Figure \ref{fig:z_evol}, \citealt{Saintonge2011, Tacconi2013, Sargent2015, French2015, Rowlands2015, Davis2016, Rudnik2017, Suess2017, Spilker2018, Bezanson2019, Freundlich2019, Williams2020, Belli2021, Bezanson2021}) are reported. The black and gray lines indicate the relations from \citet{Dou2021} and \citet{Feldmann2020}, and the gray shaded areas represent their scatter. \textit{Right panel}: Molecular gas fraction versus galaxies sSFR normalized by the star-forming main-sequence at the redshift and stellar mass of each galaxy from \citet{Sargent2014}. Gray shaded areas indicate the relation from \citet{Tacconi2018}, computed at different redshifts ($z \sim 1.5$ dark gray, $z \sim 0.5$ intermediate shade of gray, $z \sim 0$ light gray) and stellar mass (spanning from M$_\star \sim 10^{10} - 10^{11}$ M$_\odot$ from the upper to the lower envelop of each shaded area). Other symbols are as in the previous panel.}
    \label{fig:sSFR}
\end{figure*}

\subsection{Beyond the KS relation: exploring the full parameter space}
\label{subsec:scaling_rel}

Besides that in the Schmidt-Kennicutt plane, other scaling relations linking galaxies molecular gas reservoirs and star formation activity have been proposed \citep{Lilly2013, Tacconi2013, Scoville2017, Tacconi2018}. In particular, the molecular gas fraction has been compared with the specific star formation rate, sSFR = SFR/M$_\star$ of galaxies, as a function of redshift \citep{Tacconi2020}. Most of such relations though have been investigated considering samples of star-forming galaxies. Only recently samples of post-SB and quiescent galaxies have been adopted to investigate the scatter of such relations at the low-end of the star formation activity \citep{Williams2020, Bezanson2021}. Across all redshifts, star-forming galaxies follow a similar relation (e.g., \citealt{Feldmann2020}, \citealt{Dou2021}), but spanning different sSFR regimes\footnote{We note that most of the star-forming galaxies from \cite{Saintonge2011} that seem to deviate from the \cite{Feldmann2020} and \cite{Dou2021} relations in Figure \ref{fig:sSFR} have sSFR upper limits and are therefore formally consistent with those relations.}. The post-SB galaxies of this sample instead lie outside such relation, having higher molecular gas fraction than expected given their sSFR. The deviation is  still present also when considering the relation between $f_\mathrm{H2}$ and the sSFR, normalized by the sSFR of main-sequence galaxies from \cite{Sargent2014} (Figure \ref{fig:sSFR}, right panel). Our sample post-SBs show similar star formation activity (sSFR/sSFR$_\mathrm{MS}$) as local quiescent galaxies \citep{Davis2016}, but they have two orders of magnitude larger gas fractions. In general, post-SBs span a large range of sSFR/sSFR$_\mathrm{MS} \sim 0.01 - 1$. Their star-formation activity does not seem to correlate with redshift nor with gas fraction (Figure \ref{fig:sSFR}). This is at odds with star-forming galaxies which instead show, on average, a decrease of their gas fraction with redshift and with sSFR/sSFR$_\mathrm{MS}$ \citep{Saintonge2011, Tacconi2013, Freundlich2019}.

Other post-SB galaxies from the literature show similar gas fractions as ours, but their sSFR/sSFR$_\mathrm{MS}$ have a large scatter and can differ up to one order of magnitude (e.g., those by \citealt{French2015}). While the gas fraction of star-forming galaxies declines with sSFR/sSFR$_\mathrm{MS}$ and redshift (in Figure \ref{fig:sSFR} symbols with lighter colors indicate lower-redshift galaxies), such a trend seems to be lacking for post-SBs. 

\begin{figure*}
    \centering
    \includegraphics[width=0.7\textwidth]{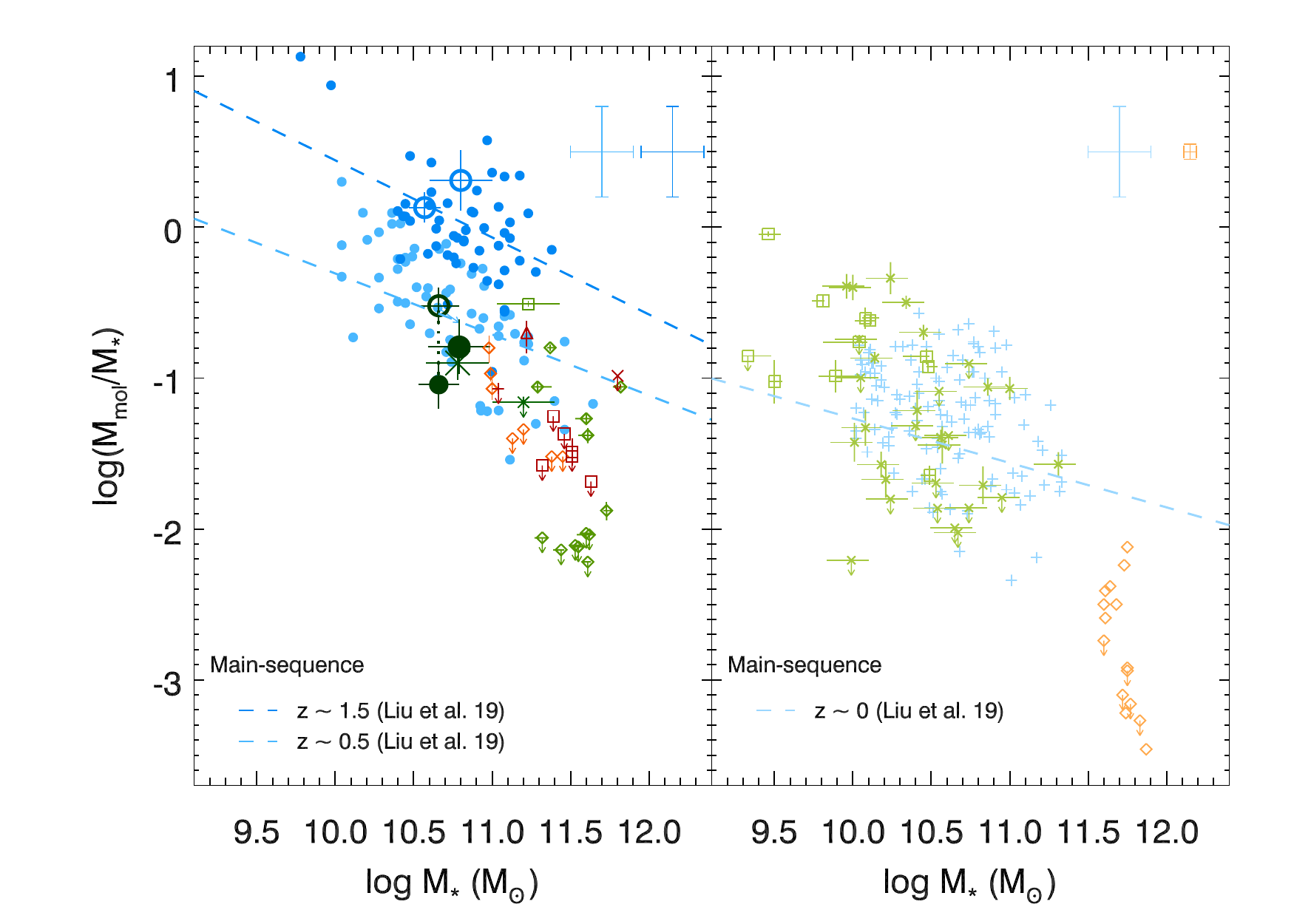}
    \caption{Molecular gas fraction as a function of stellar mass. \textit{Left panel}: we show high-redshift galaxies ($z > 0.5$). \textit{Right panel}: we show low-redshift galaxies ($z < 0.5$). The dashed blue and cyan lines indicate the relation from \citet{Liu2019}, computed at different redshifts. Symbols are as in previous figures.}
    \label{fig:mstar}
\end{figure*}

However, one additional parameter that plays an important role in galaxy evolution is the stellar mass (e.g., \citealt{Peng2010}). In Figure \ref{fig:mstar} we show the gas fraction of galaxies as a function of their stellar mass, divided in two redshfit bins ($z > 0.5$, left panel, and $z < 0.5$ right panel). The gas fraction of all galaxy types (star-forming, post-SB, quiescent) declines with increasing stellar mass, at all redshifts, with quiescent galaxies typically sampling the highest mass end (M$_\star > 10^{11}$ M$_\odot$) and having the lowest gas fraction. In particular, current literature studies of post-SB and quiescent galaxies at high redshift mostly consider massive sources with $\mathrm{M_\star \gtrsim 10^{11} M_\odot}$. Our sample galaxies have $\sim0.3-0.5$ dex lower stellar masses, allowing us to start exploring the gas fraction of post-SBs in this relatively low-mass regime (Figure \ref{fig:mstar}, left panel). More observations of low-mass post-SBs at high redshift are needed to understand whether indeed the gas fraction of post-SBs increases at lower stellar masses (e.g., as for the star-forming population).
  
\begin{figure*}
    \centering
    \includegraphics[width=0.7\textwidth]{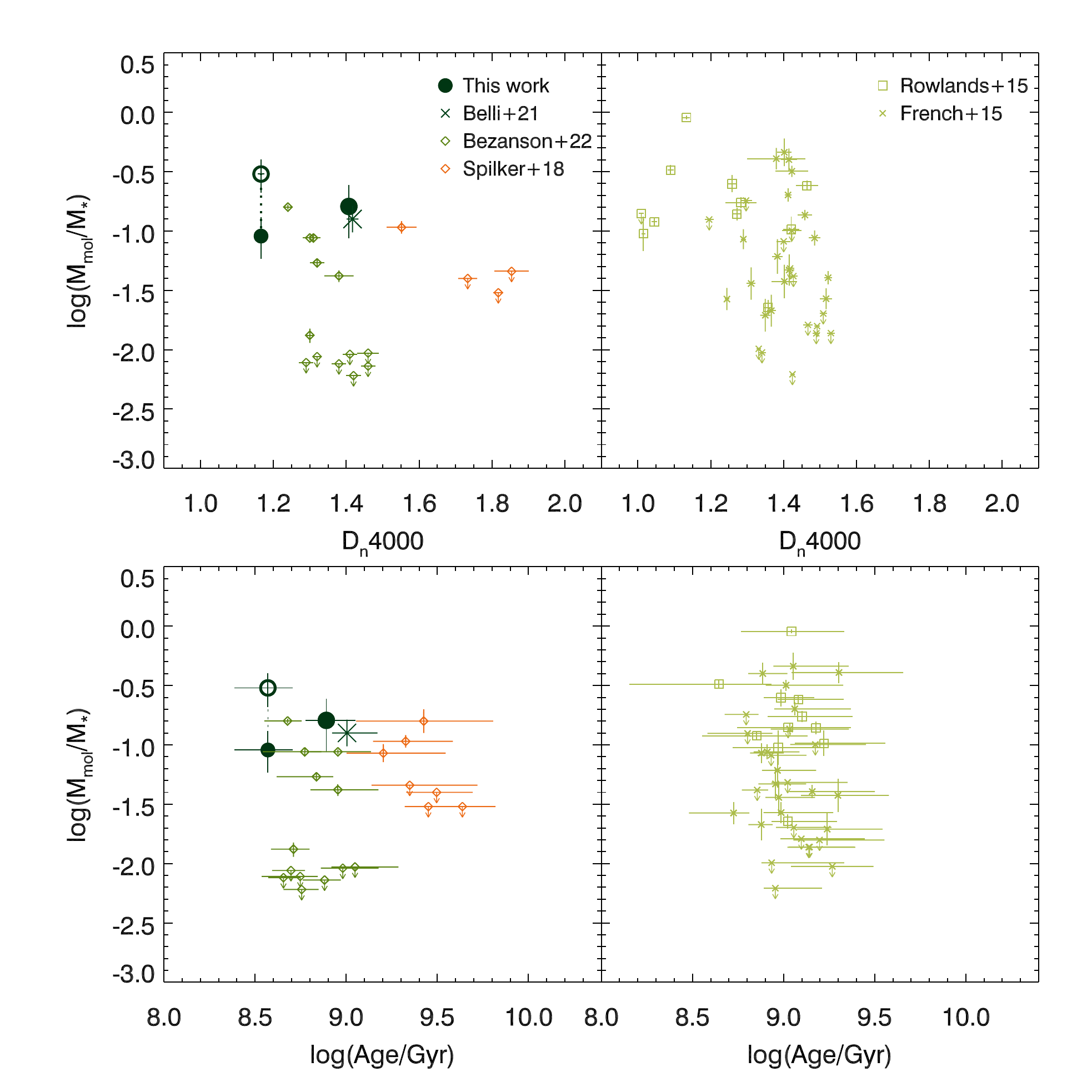}
    \caption{Linking the molecular gas fraction with the properties of the stellar population. \textit{Top panels}: molecular gas fraction as a function of D$n$4000, a proxy for the galaxy age, for our sample galaxies and literature samples \citep{French2015, Rowlands2015, Spilker2018, Bezanson2021, Belli2021} at high redshift ($z > 0.5$, left panel) and low redshift ($z < 0.5$ right panel). \textit{Bottom panels}: molecular gas fraction as a function of $R$-band light-weighted age of the stellar population for the high-redshift (left panel) and low-redshift (right panel) samples. Symbols are as in previous figures.}
    \label{fig:Dn4000}
\end{figure*}

\section{Discussion}
\label{subsec:discussion}

Different pathways have been proposed to explain the physical properties of post-SB galaxies and their contribution to the population of quiescent galaxies. They might be massive galaxies that have rapidly formed the bulk of their stellar content at $z > 2$, consuming most of their molecular gas reservoirs and expelling the remaining gas through outflows, leading them to quiescence; however post-SBs might also be the result of gas-rich major-mergers and/or environment effects that stripped the gas and heated it up preventing new star formation \citep{Wild2016, Pawlik2019}. Both processes might be at play and effective, however it is important to understand what mechanisms are dominating at different cosmic times, as well as possible dependences on the galaxies stellar mass and environment. 

Our post-SB galaxies span the lowest stellar mass range probed by currently available samples (Figure \ref{fig:mstar}) and seem to have retained a significant fraction of molecular gas ($f_\mathrm{H2} \sim$ 8\% - 16\%) detected in CO(3-2), an emission line tracing modestly excited gas. This disfavours scenarios where the gas is consumed and expelled during strong bursts of star formation, which could instead be more common for more massive galaxies (M$_\star > 10^{11}$ M$_\odot$, e.g., \citealt{Williams2000} and part of the sample from \citealt{Bezanson2021} which is not CO-detected). When we complement our sample with literature data, 6 out of 7 (i.e., 85\%) post-SB or quiescent galaxies with stellar mass $\mathrm{M_\star \lesssim 10^{11}\, M_\odot}$ at $z > 0.5$ are detected in CO and have $\mathrm{f_{H2}} \gtrsim 8\%$ (Figure \ref{fig:mstar}). Higher mass galaxies ($\mathrm{M_\star > 10^{11}\, M_\odot}$) at these redshifts instead have a lower CO detection rate: only 8 out of 27 are detected (i.e., 29\%) and have gas fraction consistent with our sample post-SBs ($\mathrm{f_{H2}} \sim 5 - 20\%$), whereas the non-detections have typical upper limits of $\mathrm{f_{H2}} \lesssim 5\%$. The large number of non detections among the high-mass population suggests that multiple mechanisms might be at play for more massive galaxies: in these cases, not only the star formation is suppressed (as for the lower mass galaxies), but also the gas is expelled (e.g., due to AGN or star formation feedback) or stripped during the quenching phase. For the majority of lower mass galaxies instead quenching mechanisms that halt star formation without removing the gas seem to be favoured. Also the fact that post-SBs, especially at low mass and high redshift, have higher gas fraction than expected given their sSFR favours the idea that gas is not expelled before the star formation is quenched or, in other words, that gas depletion is not a direct cause of star formation suppression. This is in agreement with results from \cite{Bezanson2021} probing more massive and lower redshift post-SBs. Mechanisms, such as major mergers, which remove part of the gas and heat up the remaining (e.g., through shocks or by increasing the internal turbulent pressure, \citealt{Smercina2021}) are favoured. Indeed we find that ID83492 is possibly undergoing an early merger phase with two gas-rich and starbursting close companions (Section \ref{subsec:merger}). We do not find clear signatures of mergers for ID97148 instead, but this might simply be due to the lack of high angular resolution observations that prevent us from detecting close pairs and/or due to the fact that low-surface brightness features indicative of past mergers fade quickly ($< 0.5$ Gyr, e.g. \citealt{Pawlik2019}). Studying the CO spectral line energy distribution would clarify whether indeed the remaining gas is excited and high-$J$ CO transitions are favoured in post-SB galaxies. 

Relating the properties of the stellar population (e.g., age) with the gas fraction might further constrain the evolutionary pathways of post-SB galaxies \citep{Bezanson2021}. We investigated whether a relation between the gas fraction and the D$_n$4000 index, a proxy of the age of the stellar population, is in place by complementing our results with the literature (Figure \ref{fig:Dn4000}). When considering galaxies detected in CO at $z > 0.5 - 1.5$ (our sample and \citealt{Spilker2018, Bezanson2021, Belli2021}) we find that galaxies with higher gas fraction preferentially have lower D$_n$4000 (Figure \ref{fig:Dn4000}, top panels) and younger ages (Figure \ref{fig:Dn4000}, bottom panels). We investigated if, after a first episode that drives the quenching of star formation (e.g., major merger), the little ongoing star formation rate (SFR $\sim 0.5 - 1$ M$_\odot$ yr$^{-1}$) can consume the remaining molecular gas reservoir. Even assuming a constant SFR equal to the one we estimated at $z \sim 1$, the gas fraction of our sample galaxies would decrease by a factor two (e.g., for ID83492 it would become $\mathrm{f_{H2} \sim 8\%}$). This is larger than the $\mathrm{f_{H2} \sim 1\% - 0.1\%}$ measured in $z \sim 0$, massive ($\mathrm{M_\star \gtrsim 10^{11.5}\, M_\odot}$) quiescent galaxies \citep{Davis2016}. Similar arguments hold also for ID97148. This indicates that either lower-mass quiescent galaxies have larger molecular gas fractions than massive ($\mathrm{M_\star \gtrsim 10^{11.5}\, M_\odot}$) ones or that additional mechanisms are needed to remove or heat the gas at low redshift.

At lower redshift ($z < 0.5$, \citealt{French2015, Rowlands2015}) a broad anti-correlation of $f_\mathrm{H2}$ with D$_n$4000 is maintained (Figure \ref{fig:Dn4000}, bottom right panel), although the picture becomes less clear, especially when considering the age estimates, possibly indicating that more quenching processes might be at play for local post-SBs. For example at lower redshift some post-SB galaxies might have re-accreted gas (hence show a relatively large gas fraction), while environmental processes (e.g., ram pressure stripping) and AGN feedback might have played an important role in local environments. Larger samples of post-SB and quiescent galaxies with high signal-to-noise optical spectra will be needed to investigate in the future such relations between $f_\mathrm{H2}$ and stellar populations.

Our sample galaxies are not detected in dust-continuum. However, assuming a standard dust-to-gas conversion factor (Section \ref{subsec:gas_mass}) we obtain M$_\mathrm{H2}$ upper limits consistent with the CO-based moelcular gas mass estimates. This suggests that exotic gas-to-dust ratios (e.g., \citealt{Whitaker2021}) are not needed to explain the molecular gas fractions observed in post-SB galaxies. However deeper submillimeter observations are needed to detect the dust-continuum underlying the CO(3-2) emission and properly measure the gas-to-dust ratio in post-SBs. Moreover, our results seem to confirm that the less massive post-SB galaxies retain substantial molecular gas reservoirs, possibly in agreement with studies targeting the dust emission of post-SB and quiescent galaxies and stacking statistical samples (e.g., \citealt{Gobat2018, Magdis2021}, Bl\'anquez-Sese et al., submitted). Having deep observations of post-SBs to detect individually their dust continuum will be crucial to confirm these results.

\subsection{The merger scenario}
\label{subsec:merger}
It has been suggested several times in previous studies that post-starburst galaxies represent a transition phase between gas-rich mergers and gas-poor quiescent galaxies. A large fraction of local post-SBs in the field show major merger signatures (e.g., strong tidal features, disturbed kinematics, nearby companions) in \textit{HST} imaging \citep{Zabludoff1996, Yang2006, Chandar2021} and in high-resolution ALMA observations \citep{Smercina2021}. \cite{Wilkinson2022} find that the merger fraction in local post-SBs ranges between 19\% and 42\%, depending on the merger identification criteria, but merger signatures fade quickly ($< 200$ Myr) and therefore such fractions might actually be higher. Indeed, when considering deep \textit{HST} and SDSS images, \cite{Sazonova2021} find that 88\% of their local post-SBs show disturbed morphologies. They also point out that most disturbances are low-surface brightness features which require high sensitivity and fine spatial resolution to be identified, and that fade in short timescales, suggesting that all their post-SBs might have had a merger origin. Similar results are also found by \cite{Verrico2022}, who also show that post-SBs are $3.6^{+2.9}_{-1.3}$ ($2.1^{+1.9}_{0.73}$) times more likely to show disturbances than quiescent (star-forming) galaxies. In the cosmological hydrodynamical EAGLE simulations, between 30\% and 50\% of post-SBs have undergone a merger just before the onset of their post-starburst episode \citep{Davis2018, Pawlik2019}. Similarly, idealised simulations of galaxy mergers have found that a brief ($< 600$ Myr) post-starburst phase can follow a gas-rich merger which triggers a central starburst phase, succeeded by the rapid quenching of star formation \citep{Wild2009, Bekki2005, Snyder2011}.
Studying the morphology of high-redshift post-SBs is still challenging, due to the need of high-resolution and deep observations. \cite{Setton2022} have studied the morphology of 145 post-SBs at $z\sim 0.7$ and found no correlation between their size and time since quenching. This finding disfavours the formation of post-SBs purely through mergers which would instead ignite a central burst of star formation, that in turn would make the galaxy more compact and shut off star formation. On the contrary, other works found that post-SBs are in general smaller than the average population of quiescent galaxies (e.g., \citealt{Whitaker2012, Yano2016, Almaini2017, Wu2018}). Therefore, evidence for mergers being at the origin of post-SBs, especially at high redshift, is still inconclusive.

To assess whether our sample galaxies are undergoing a merger we investigated their environment. In particular, we serendipitously found two CO(3-2) emitters located respectively at $5\arcsec.8$ and $13\arcsec.7$ away from ID83492. Their redshift, estimated from the CO emission, is $\Delta z < 0.0018$ ($v_\mathrm{off} < 250$ km s$^{-1}$) different from that of our post-SB target. This suggests that ID83492 might be in an early (multiple) merger stage. Given the stellar masses of the companions, this is a major merger (mass ratios 1:1 and 1:1.5). ID83492 and its two companions might be part of a larger galaxy group or be the parents of a very massive galaxy to be formed by $z \sim 0$ through the merger of these sources. In the current \textit{HST} and ALMA data we do not detect tidal features or morphological asymmetries indicating an ongoing merger, although deeper and higher resolution observations might be needed to detect those, especially if the merger is in its early phase. The two companion galaxies are gas-rich, star-forming sources in a starburst phase (sSFR $> 10$, Figure \ref{fig:sSFR}), lying above main-sequence galaxies in the KS plane (Figure \ref{fig:SFE}).

On the contrary ID97148 does not show any clear companion, a part from a continuum-detected source that might rather be a lower redshift interloper (Section \ref{subsec:continuum}). Observations with better angular resolution are needed to understand whether ID97148 is undergoing a late merger phase with two nuclei which are not resolved with the current angular resolution (pair closer than $\sim 1\arcsec.5 \sim 12.5$ kpc at $z \sim 1.2$).
The fact that at least one of our post-SBs is undergoing a merger suggests that interactions might be a major cause for the suppression of star formation. The large molecular gas reservoir measured for ID83492 might be due to the fact that while the merger episode halts the star formation it does not remove the galaxy ISM entirely, but leaves a large ($\sim 15\%$) fraction of the molecular gas. Possibly due to the high excitation and turbulence, the leftover gas does not form new stars and hence the galaxy becomes a post-SB with little or no ongoing star formation and still a considerable reservoir of gas. Another possible scenario is the fact that the post-SB is a quiescent galaxy which was replenished with molecular gas, stripped from the star-forming companions, during the merger phase. If the galaxy rejuvenates (i.e., goes through a new spark of star formation) the correlation between molecular gas fraction with D$_n$4000 and age is still expected, whereas it is not if the accreted gas does not produce new star formation (i.e., the galaxy does not rejuvenate). This shows the importance of combining measurements of molecular gas tracers with excellent optical spectra which allow us to accurately determine absorption indices and hence stellar population ages. Future studies with larger samples of post-SBs will confirm whether indeed mergers at high redshift are key ingredients to undertsand the nature of post-SB galaxies. 

\section{Summary and conclusions}
\label{sec:conclusions}

We have investigated the molecular gas content (traced by ALMA CO(3-2) observations) of two post-starburst galaxies at redshift $z = 1.1 - 1.3$ and stellar masses M$_\star \sim 5\times 10^{10}$ M$_\odot$ and complement our study with literature samples. We find that:

\begin{itemize}
\item our post-SB galaxies retain a significant fraction of molecular gas ($f_\mathrm{H2} \sim$ 8\% - 16\%), which is modestly excited. Such fractions are consistent with other samples of post-SB galaxies from the literature and are less than 10 times lower than the gas fraction of star-forming galaxies at similar redshift (Figure \ref{fig:z_evol}). Such gas fractions however are larger than those found for local quiescent galaxies (e.g., \citealt{Davis2016});
  
\item our post-SBs have a 10 times lower star formation efficiency than star-forming galaxies. This result does not depend on the CO transition that is probed (up to CO(3-2)), indicating that post-SBs have in general less CO-emitting gas than star-forming galaxies (Figure \ref{fig:SFE});

\item post-SBs are outliers of the sSFR - $f_\mathrm{H2}$ relation of star-forming galaxies as they have larger molecular gas fraction than expected given their sSFR. In particular, their molecular gas fraction seems to be quite constant with redshift, while that of star-forming and quiescent galaxies decreases with redshift (Figures \ref{fig:z_evol} and \ref{fig:sSFR});

\item our post-SBs probe the lower mass end of current post-SB and quiescent samples at high redshift ($z > 0.5$). They retain a significant molecular gas fraction ($f_\mathrm{H2} \sim$ 8\% - 16\%, similar to other post-SB and quiescent galaxies at similar redshift, especially with $\mathrm{M_\star < 10^{11}\, M_\odot}$), hinting at major mergers as a possible mechanism at play at $z > 0.5$, which halt star formation by increasing the turbulence and excitation of the molecular gas rather than expelling it. This scenario is supported by the fact that post-SBs have higher gas fraction than expected given their sSFR (indicating that gas removal is not the cause of star formation suppression). Moreover, one of our targets is undergoing a major merger with two nearby companion galaxies which are actively star forming;

\item the gas fraction of post-SBs seems to decrease with the age of their stellar population (Figure \ref{fig:Dn4000}), as suggested also by previous literature works studying lower-redshift post-SBs \citep{French2015, Bezanson2021}. This is in agreement with the scenario that a perturber event (e.g., major merger) halts star formation without completely removing the gas. The leftover cold molecular gas is then exhausted by the low remaining level of star formation and/or expelled (or heated up) at a later stage.
\end{itemize}

Having in the future larger samples of high-redshift post-SBs with CO observations will allow us to confirm the results reported in this study. In particular, by targeting lower mass post-SBs we can investigate whether a trend between the $f_\mathrm{H2}$ and stellar mass is indeed in place and in turn constrain the role of different quenching mechanisms in different M$_\star$ regimes. In addition, sampling the CO spectral line energy distribution of post-SB galaxies will allow us to understand whether indeed the molecular gas is more excited than in star-forming galaxies. Also having higher-resolution observations will help to constrain the size of the CO emitting region, its morphology and deblend possible close pairs to clarify whether indeed mergers are ubiquitous in the post-SB population and in turn constrain their role in the suppression of star formation at cosmic noon.

\section*{Acknowledgements}

We thank the referee whose valuable comments helped to clarify the text and improve the paper. We warmly thank Paolo Franzetti for his help with the reduction of the VLT/VIMOS data and Masayuki Tanaka and David Bl\'anquez-Sese for sharing their multi-wavelength catalog and literature compilation. We also thank Paolo Cassata for useful discussions. GEM acknowledges financial support from the Villum Young Investigator grant 37440 and 13160. The Cosmic Dawn Center (DAWN) is funded by the Danish National Research Foundation under grant No. 140. This paper makes use of the following ALMA data: 2019.1.00900.S. ALMA is a partnership of ESO (representing its member states), NSF (USA), and NINS (Japan), together with NRC (Canada), MOST and ASIAA (Taiwan), and KASI (Republic of Korea), in cooperation with the Republic of Chile. The Joint ALMA Observatory is operated by ESO, AUI/NRAO and NAOJ.

\section*{Data Availability}

The data used in this study are publicly available from telescope archives. Software and derived data generated for this research can be made available upon reasonable request made to the corresponding author. 



\bibliographystyle{mnras}
\bibliography{bibliography} 

\begin{thebibliography}{}
\makeatletter
\relax
\def\mn@urlcharsother{\let\do\@makeother \do\$\do\&\do\#\do\^\do\_\do\%\do\~}
\def\mn@doi{\begingroup\mn@urlcharsother \@ifnextchar [ {\mn@doi@}
  {\mn@doi@[]}}
\def\mn@doi@[#1]#2{\def\@tempa{#1}\ifx\@tempa\@empty \href
  {http://dx.doi.org/#2} {doi:#2}\else \href {http://dx.doi.org/#2} {#1}\fi
  \endgroup}
\def\mn@eprint#1#2{\mn@eprint@#1:#2::\@nil}
\def\mn@eprint@arXiv#1{\href {http://arxiv.org/abs/#1} {{\tt arXiv:#1}}}
\def\mn@eprint@dblp#1{\href {http://dblp.uni-trier.de/rec/bibtex/#1.xml}
  {dblp:#1}}
\def\mn@eprint@#1:#2:#3:#4\@nil{\def\@tempa {#1}\def\@tempb {#2}\def\@tempc
  {#3}\ifx \@tempc \@empty \let \@tempc \@tempb \let \@tempb \@tempa \fi \ifx
  \@tempb \@empty \def\@tempb {arXiv}\fi \@ifundefined
  {mn@eprint@\@tempb}{\@tempb:\@tempc}{\expandafter \expandafter \csname
  mn@eprint@\@tempb\endcsname \expandafter{\@tempc}}}

\bibitem[\protect\citeauthoryear{{Alatalo} et~al.,}{{Alatalo}
  et~al.}{2016}]{Alatalo2016}
{Alatalo} K.,  et~al., 2016, \mn@doi [\apj] {10.3847/0004-637X/827/2/106},
  \href {https://ui.adsabs.harvard.edu/abs/2016ApJ...827..106A} {827, 106}

\bibitem[\protect\citeauthoryear{{Almaini} et~al.,}{{Almaini}
  et~al.}{2017}]{Almaini2017}
{Almaini} O.,  et~al., 2017, \mn@doi [\mnras] {10.1093/mnras/stx1957}, \href
  {https://ui.adsabs.harvard.edu/abs/2017MNRAS.472.1401A} {472, 1401}

\bibitem[\protect\citeauthoryear{{Avni}}{{Avni}}{1976}]{Avni1976}
{Avni} Y.,  1976, \mn@doi [\apj] {10.1086/154870}, \href
  {https://ui.adsabs.harvard.edu/abs/1976ApJ...210..642A} {210, 642}

\bibitem[\protect\citeauthoryear{{Bekki}, {Couch}, {Shioya}  \&
  {Vazdekis}}{{Bekki} et~al.}{2005}]{Bekki2005}
{Bekki} K.,  {Couch} W.~J.,  {Shioya} Y.,   {Vazdekis} A.,  2005, \mn@doi
  [\mnras] {10.1111/j.1365-2966.2005.08932.x}, \href
  {https://ui.adsabs.harvard.edu/abs/2005MNRAS.359..949B} {359, 949}

\bibitem[\protect\citeauthoryear{{Bell} et~al.,}{{Bell}
  et~al.}{2005}]{Bell2005}
{Bell} E.~F.,  et~al., 2005, \mn@doi [\apj] {10.1086/429552}, \href
  {https://ui.adsabs.harvard.edu/abs/2005ApJ...625...23B} {625, 23}

\bibitem[\protect\citeauthoryear{{Belli} et~al.,}{{Belli}
  et~al.}{2021}]{Belli2021}
{Belli} S.,  et~al., 2021, \mn@doi [\apjl] {10.3847/2041-8213/abe6a6}, \href
  {https://ui.adsabs.harvard.edu/abs/2021ApJ...909L..11B} {909, L11}

\bibitem[\protect\citeauthoryear{{Bezanson}, {Spilker}, {Williams}, {Whitaker},
  {Narayanan}, {Weiner}  \& {Franx}}{{Bezanson} et~al.}{2019}]{Bezanson2019}
{Bezanson} R.,  {Spilker} J.,  {Williams} C.~C.,  {Whitaker} K.~E.,
  {Narayanan} D.,  {Weiner} B.,   {Franx} M.,  2019, \mn@doi [\apjl]
  {10.3847/2041-8213/ab0c9c}, \href
  {https://ui.adsabs.harvard.edu/abs/2019ApJ...873L..19B} {873, L19}

\bibitem[\protect\citeauthoryear{{Bezanson} et~al.,}{{Bezanson}
  et~al.}{2022}]{Bezanson2021}
{Bezanson} R.,  et~al., 2022, \mn@doi [\apj] {10.3847/1538-4357/ac3dfa}, \href
  {https://ui.adsabs.harvard.edu/abs/2022ApJ...925..153B} {925, 153}

\bibitem[\protect\citeauthoryear{{Bolatto}, {Wolfire}  \& {Leroy}}{{Bolatto}
  et~al.}{2013}]{Bolatto2013}
{Bolatto} A.~D.,  {Wolfire} M.,   {Leroy} A.~K.,  2013, \mn@doi [\araa]
  {10.1146/annurev-astro-082812-140944}, \href
  {http://adsabs.harvard.edu/abs/2013ARA%26A..51..207B} {51, 207}

\bibitem[\protect\citeauthoryear{{Brownson}, {Belfiore}, {Maiolino}, {Lin}  \&
  {Carniani}}{{Brownson} et~al.}{2020}]{Brownson2020}
{Brownson} S.,  {Belfiore} F.,  {Maiolino} R.,  {Lin} L.,   {Carniani} S.,
  2020, \mn@doi [\mnras] {10.1093/mnrasl/slaa128}, \href
  {https://ui.adsabs.harvard.edu/abs/2020MNRAS.498L..66B} {498, L66}

\bibitem[\protect\citeauthoryear{{Bruzual} \& {Charlot}}{{Bruzual} \&
  {Charlot}}{2003}]{Bruzual2003}
{Bruzual} G.,  {Charlot} S.,  2003, \mn@doi [\mnras]
  {10.1046/j.1365-8711.2003.06897.x}, \href
  {http://adsabs.harvard.edu/abs/2003MNRAS.344.1000B} {344, 1000}

\bibitem[\protect\citeauthoryear{{Calzetti}, {Armus}, {Bohlin}, {Kinney},
  {Koornneef}  \& {Storchi-Bergmann}}{{Calzetti} et~al.}{2000}]{Calzetti2000}
{Calzetti} D.,  {Armus} L.,  {Bohlin} R.~C.,  {Kinney} A.~L.,  {Koornneef} J.,
   {Storchi-Bergmann} T.,  2000, \mn@doi [\apj] {10.1086/308692}, \href
  {http://adsabs.harvard.edu/abs/2000ApJ...533..682C} {533, 682}

\bibitem[\protect\citeauthoryear{{Carilli} \& {Walter}}{{Carilli} \&
  {Walter}}{2013}]{Carilli2013}
{Carilli} C.~L.,  {Walter} F.,  2013, \mn@doi [\araa]
  {10.1146/annurev-astro-082812-140953}, \href
  {http://adsabs.harvard.edu/abs/2013ARA%26A..51..105C} {51, 105}

\bibitem[\protect\citeauthoryear{{Carnall} et~al.,}{{Carnall}
  et~al.}{2019}]{Carnall2019}
{Carnall} A.~C.,  et~al., 2019, \mn@doi [\mnras] {10.1093/mnras/stz2544}, \href
  {https://ui.adsabs.harvard.edu/abs/2019MNRAS.490..417C} {490, 417}

\bibitem[\protect\citeauthoryear{{Chabrier}}{{Chabrier}}{2003}]{Chabrier2003}
{Chabrier} G.,  2003, \mn@doi [\pasp] {10.1086/376392}, \href
  {http://adsabs.harvard.edu/abs/2003PASP..115..763C} {115, 763}

\bibitem[\protect\citeauthoryear{{Chandar}, {Mok}, {French}, {Smercina}  \&
  {Smith}}{{Chandar} et~al.}{2021}]{Chandar2021}
{Chandar} R.,  {Mok} A.,  {French} K.~D.,  {Smercina} A.,   {Smith} J.-D.~T.,
  2021, \mn@doi [\apj] {10.3847/1538-4357/ac0c19}, \href
  {https://ui.adsabs.harvard.edu/abs/2021ApJ...920..105C} {920, 105}

\bibitem[\protect\citeauthoryear{{Chevallard} \& {Charlot}}{{Chevallard} \&
  {Charlot}}{2016}]{Chevallard2016}
{Chevallard} J.,  {Charlot} S.,  2016, \mn@doi [\mnras]
  {10.1093/mnras/stw1756}, \href
  {https://ui.adsabs.harvard.edu/abs/2016MNRAS.462.1415C} {462, 1415}

\bibitem[\protect\citeauthoryear{{Cimatti} et~al.,}{{Cimatti}
  et~al.}{2008}]{Cimatti2008}
{Cimatti} A.,  et~al., 2008, \mn@doi [\aap] {10.1051/0004-6361:20078739}, \href
  {http://adsabs.harvard.edu/abs/2008A%26A...482...21C} {482, 21}

\bibitem[\protect\citeauthoryear{{Cochrane}, {Hayward}  \&
  {Angl{\'e}s-Alc{\'a}zar}}{{Cochrane} et~al.}{2022}]{Cochrane2022}
{Cochrane} R.~K.,  {Hayward} C.~C.,   {Angl{\'e}s-Alc{\'a}zar} D.,  2022,
  \mn@doi [\apjl] {10.3847/2041-8213/ac951d}, \href
  {https://ui.adsabs.harvard.edu/abs/2022ApJ...939L..27C} {939, L27}

\bibitem[\protect\citeauthoryear{{Conselice}}{{Conselice}}{2006}]{Conselice2006}
{Conselice} C.~J.,  2006, \mn@doi [\apj] {10.1086/499067}, \href
  {https://ui.adsabs.harvard.edu/abs/2006ApJ...638..686C} {638, 686}

\bibitem[\protect\citeauthoryear{{Couch} \& {Sharples}}{{Couch} \&
  {Sharples}}{1987}]{Couch1987}
{Couch} W.~J.,  {Sharples} R.~M.,  1987, \mn@doi [\mnras]
  {10.1093/mnras/229.3.423}, \href
  {https://ui.adsabs.harvard.edu/abs/1987MNRAS.229..423C} {229, 423}

\bibitem[\protect\citeauthoryear{{Croton} et~al.,}{{Croton}
  et~al.}{2006}]{Croton2006}
{Croton} D.~J.,  et~al., 2006, \mn@doi [\mnras]
  {10.1111/j.1365-2966.2005.09675.x}, \href
  {https://ui.adsabs.harvard.edu/abs/2006MNRAS.365...11C} {365, 11}

\bibitem[\protect\citeauthoryear{{Daddi} et~al.,}{{Daddi}
  et~al.}{2007}]{Daddi2007}
{Daddi} E.,  et~al., 2007, \mn@doi [\apj] {10.1086/521818}, \href
  {http://adsabs.harvard.edu/abs/2007ApJ...670..156D} {670, 156}

\bibitem[\protect\citeauthoryear{{Davis} et~al.,}{{Davis}
  et~al.}{2014}]{Davis2014}
{Davis} T.~A.,  et~al., 2014, \mn@doi [\mnras] {10.1093/mnras/stu570}, \href
  {https://ui.adsabs.harvard.edu/abs/2014MNRAS.444.3427D} {444, 3427}

\bibitem[\protect\citeauthoryear{{Davis}, {Greene}, {Ma}, {Pandya},
  {Blakeslee}, {McConnell}  \& {Thomas}}{{Davis} et~al.}{2016}]{Davis2016}
{Davis} T.~A.,  {Greene} J.,  {Ma} C.-P.,  {Pandya} V.,  {Blakeslee} J.~P.,
  {McConnell} N.,   {Thomas} J.,  2016, \mn@doi [\mnras]
  {10.1093/mnras/stv2313}, \href
  {https://ui.adsabs.harvard.edu/abs/2016MNRAS.455..214D} {455, 214}

\bibitem[\protect\citeauthoryear{{Davis}, {van de Voort}, {Rowlands},
  {McAlpine}, {Wild}  \& {Crain}}{{Davis} et~al.}{2019}]{Davis2018}
{Davis} T.~A.,  {van de Voort} F.,  {Rowlands} K.,  {McAlpine} S.,  {Wild} V.,
   {Crain} R.~A.,  2019, \mn@doi [\mnras] {10.1093/mnras/stz180}, \href
  {https://ui.adsabs.harvard.edu/abs/2019MNRAS.484.2447D} {484, 2447}

\bibitem[\protect\citeauthoryear{{Dekel} \& {Birnboim}}{{Dekel} \&
  {Birnboim}}{2006}]{Dekel2006}
{Dekel} A.,  {Birnboim} Y.,  2006, \mn@doi [\mnras]
  {10.1111/j.1365-2966.2006.10145.x}, \href
  {https://ui.adsabs.harvard.edu/abs/2006MNRAS.368....2D} {368, 2}

\bibitem[\protect\citeauthoryear{{Dou} et~al.,}{{Dou} et~al.}{2021}]{Dou2021}
{Dou} J.,  et~al., 2021, \mn@doi [\apj] {10.3847/1538-4357/abd17c}, \href
  {https://ui.adsabs.harvard.edu/abs/2021ApJ...907..114D} {907, 114}

\bibitem[\protect\citeauthoryear{{Dressler} \& {Sandage}}{{Dressler} \&
  {Sandage}}{1983}]{Dressler1983}
{Dressler} A.,  {Sandage} A.,  1983, \mn@doi [\apj] {10.1086/160712}, \href
  {https://ui.adsabs.harvard.edu/abs/1983ApJ...265..664D} {265, 664}

\bibitem[\protect\citeauthoryear{{Dudzevi{\v{c}}i{\={u}}t{\.{e}}}
  et~al.,}{{Dudzevi{\v{c}}i{\={u}}t{\.{e}}} et~al.}{2020}]{Dudzeviciute2020}
{Dudzevi{\v{c}}i{\={u}}t{\.{e}}} U.,  et~al., 2020, \mn@doi [\mnras]
  {10.1093/mnras/staa769}, \href
  {https://ui.adsabs.harvard.edu/abs/2020MNRAS.494.3828D} {494, 3828}

\bibitem[\protect\citeauthoryear{{Feldmann}}{{Feldmann}}{2020}]{Feldmann2020}
{Feldmann} R.,  2020, \mn@doi [Communications Physics]
  {10.1038/s42005-020-00493-0}, \href
  {https://ui.adsabs.harvard.edu/abs/2020CmPhy...3..226F} {3, 226}

\bibitem[\protect\citeauthoryear{{French}}{{French}}{2021}]{French2021}
{French} K.~D.,  2021, \mn@doi [\pasp] {10.1088/1538-3873/ac0a59}, \href
  {https://ui.adsabs.harvard.edu/abs/2021PASP..133g2001F} {133, 072001}

\bibitem[\protect\citeauthoryear{{French}, {Yang}, {Zabludoff}, {Narayanan},
  {Shirley}, {Walter}, {Smith}  \& {Tremonti}}{{French}
  et~al.}{2015}]{French2015}
{French} K.~D.,  {Yang} Y.,  {Zabludoff} A.,  {Narayanan} D.,  {Shirley} Y.,
  {Walter} F.,  {Smith} J.-D.,   {Tremonti} C.~A.,  2015, \mn@doi [\apj]
  {10.1088/0004-637X/801/1/1}, \href
  {https://ui.adsabs.harvard.edu/abs/2015ApJ...801....1F} {801, 1}

\bibitem[\protect\citeauthoryear{{French}, {Yang}, {Zabludoff}  \&
  {Tremonti}}{{French} et~al.}{2018}]{French2018}
{French} K.~D.,  {Yang} Y.,  {Zabludoff} A.~I.,   {Tremonti} C.~A.,  2018,
  \mn@doi [\apj] {10.3847/1538-4357/aacb2d}, \href
  {https://ui.adsabs.harvard.edu/abs/2018ApJ...862....2F} {862, 2}

\bibitem[\protect\citeauthoryear{{Freundlich} et~al.,}{{Freundlich}
  et~al.}{2019}]{Freundlich2019}
{Freundlich} J.,  et~al., 2019, \mn@doi [\aap] {10.1051/0004-6361/201732223},
  \href {https://ui.adsabs.harvard.edu/abs/2019A&A...622A.105F} {622, A105}

\bibitem[\protect\citeauthoryear{{Gallazzi}, {Charlot}, {Brinchmann}, {White}
  \& {Tremonti}}{{Gallazzi} et~al.}{2005}]{Gallazzi2005}
{Gallazzi} A.,  {Charlot} S.,  {Brinchmann} J.,  {White} S. D.~M.,   {Tremonti}
  C.~A.,  2005, \mn@doi [\mnras] {10.1111/j.1365-2966.2005.09321.x}, \href
  {https://ui.adsabs.harvard.edu/abs/2005MNRAS.362...41G} {362, 41}

\bibitem[\protect\citeauthoryear{{Gallazzi}, {Bell}, {Zibetti}, {Brinchmann}
  \& {Kelson}}{{Gallazzi} et~al.}{2014}]{Gallazzi2014}
{Gallazzi} A.,  {Bell} E.~F.,  {Zibetti} S.,  {Brinchmann} J.,   {Kelson}
  D.~D.,  2014, \mn@doi [\apj] {10.1088/0004-637X/788/1/72}, \href
  {https://ui.adsabs.harvard.edu/abs/2014ApJ...788...72G} {788, 72}

\bibitem[\protect\citeauthoryear{{Gallazzi}, {Pasquali}, {Zibetti}  \&
  {Barbera}}{{Gallazzi} et~al.}{2021}]{Gallazzi2021}
{Gallazzi} A.~R.,  {Pasquali} A.,  {Zibetti} S.,   {Barbera} F.~L.,  2021,
  \mn@doi [\mnras] {10.1093/mnras/stab265}, \href
  {https://ui.adsabs.harvard.edu/abs/2021MNRAS.502.4457G} {502, 4457}

\bibitem[\protect\citeauthoryear{{Garilli} et~al.,}{{Garilli}
  et~al.}{2021}]{Garilli2021}
{Garilli} B.,  et~al., 2021, \mn@doi [\aap] {10.1051/0004-6361/202040059},
  \href {https://ui.adsabs.harvard.edu/abs/2021A&A...647A.150G} {647, A150}

\bibitem[\protect\citeauthoryear{{Geach} et~al.,}{{Geach}
  et~al.}{2017}]{Geach2017}
{Geach} J.~E.,  et~al., 2017, \mn@doi [\mnras] {10.1093/mnras/stw2721}, \href
  {https://ui.adsabs.harvard.edu/abs/2017MNRAS.465.1789G} {465, 1789}

\bibitem[\protect\citeauthoryear{{Glazebrook} et~al.,}{{Glazebrook}
  et~al.}{2017}]{Glazebrook2017}
{Glazebrook} K.,  et~al., 2017, \mn@doi [\nat] {10.1038/nature21680}, \href
  {https://ui.adsabs.harvard.edu/abs/2017Natur.544...71G} {544, 71}

\bibitem[\protect\citeauthoryear{{Gobat} et~al.,}{{Gobat}
  et~al.}{2018}]{Gobat2018}
{Gobat} R.,  et~al., 2018, \mn@doi [Nature Astronomy]
  {10.1038/s41550-017-0352-5}, \href
  {https://ui.adsabs.harvard.edu/abs/2018NatAs...2..239G} {2, 239}

\bibitem[\protect\citeauthoryear{{Gobat}, {D'Eugenio}, {Liu}, {Caminha},
  {Daddi}  \& {Bl{\'a}nquez}}{{Gobat} et~al.}{2022}]{Gobat2022}
{Gobat} R.,  {D'Eugenio} C.,  {Liu} D.,  {Caminha} G.~B.,  {Daddi} E.,
  {Bl{\'a}nquez} D.,  2022, \mn@doi [\aap] {10.1051/0004-6361/202244995}, \href
  {https://ui.adsabs.harvard.edu/abs/2022A&A...668L...4G} {668, L4}

\bibitem[\protect\citeauthoryear{{Guilloteau} \& {Lucas}}{{Guilloteau} \&
  {Lucas}}{2000}]{Guilloteau2000}
{Guilloteau} S.,  {Lucas} R.,  2000, in {Mangum} J.~G.,  {Radford} S.~J.~E.,
  eds,  Astronomical Society of the Pacific Conference Series Vol. 217, Imaging
  at Radio through Submillimeter Wavelengths. p.~299

\bibitem[\protect\citeauthoryear{{Gunn} \& {Gott}}{{Gunn} \&
  {Gott}}{1972}]{Gunn1972}
{Gunn} J.~E.,  {Gott} J.~Richard I.,  1972, \mn@doi [\apj] {10.1086/151605},
  \href {https://ui.adsabs.harvard.edu/abs/1972ApJ...176....1G} {176, 1}

\bibitem[\protect\citeauthoryear{{Hayashi} et~al.,}{{Hayashi}
  et~al.}{2018}]{Hayashi2018}
{Hayashi} M.,  et~al., 2018, \mn@doi [\apj] {10.3847/1538-4357/aab3e7}, \href
  {https://ui.adsabs.harvard.edu/abs/2018ApJ...856..118H} {856, 118}

\bibitem[\protect\citeauthoryear{{Hayward} et~al.,}{{Hayward}
  et~al.}{2014}]{Hayward2014}
{Hayward} C.~C.,  et~al., 2014, \mn@doi [\mnras] {10.1093/mnras/stu1843}, \href
  {https://ui.adsabs.harvard.edu/abs/2014MNRAS.445.1598H} {445, 1598}

\bibitem[\protect\citeauthoryear{{Hopkins}, {Cox}, {Kere{\v{s}}}  \&
  {Hernquist}}{{Hopkins} et~al.}{2008}]{Hopkins2008}
{Hopkins} P.~F.,  {Cox} T.~J.,  {Kere{\v{s}}} D.,   {Hernquist} L.,  2008,
  \mn@doi [\apjs] {10.1086/524363}, \href
  {https://ui.adsabs.harvard.edu/abs/2008ApJS..175..390H} {175, 390}

\bibitem[\protect\citeauthoryear{{Hopkins}, {Kere{\v s}}, {Murray}, {Quataert}
  \& {Hernquist}}{{Hopkins} et~al.}{2012}]{Hopkins2012}
{Hopkins} P.~F.,  {Kere{\v s}} D.,  {Murray} N.,  {Quataert} E.,   {Hernquist}
  L.,  2012, \mn@doi [\mnras] {10.1111/j.1365-2966.2012.21981.x}, \href
  {http://adsabs.harvard.edu/abs/2012MNRAS.427..968H} {427, 968}

\bibitem[\protect\citeauthoryear{{Johansson}, {Naab}  \&
  {Ostriker}}{{Johansson} et~al.}{2009}]{Johansson2009}
{Johansson} P.~H.,  {Naab} T.,   {Ostriker} J.~P.,  2009, \mn@doi [\apjl]
  {10.1088/0004-637X/697/1/L38}, \href
  {https://ui.adsabs.harvard.edu/abs/2009ApJ...697L..38J} {697, L38}

\bibitem[\protect\citeauthoryear{{Kauffmann} et~al.,}{{Kauffmann}
  et~al.}{2003}]{Kauffmann2003}
{Kauffmann} G.,  et~al., 2003, \mn@doi [\mnras]
  {10.1046/j.1365-8711.2003.06291.x}, \href
  {http://adsabs.harvard.edu/abs/2003MNRAS.341...33K} {341, 33}

\bibitem[\protect\citeauthoryear{{Kennicutt}}{{Kennicutt}}{1998}]{Kennicutt1998}
{Kennicutt} Jr. R.~C.,  1998, \mn@doi [\apj] {10.1086/305588}, \href
  {http://adsabs.harvard.edu/abs/1998ApJ...498..541K} {498, 541}

\bibitem[\protect\citeauthoryear{{Kewley}, {Geller}  \& {Jansen}}{{Kewley}
  et~al.}{2004}]{Kewley2004}
{Kewley} L.~J.,  {Geller} M.~J.,   {Jansen} R.~A.,  2004, \mn@doi [\aj]
  {10.1086/382723}, \href {http://adsabs.harvard.edu/abs/2004AJ....127.2002K}
  {127, 2002}

\bibitem[\protect\citeauthoryear{{Kubo}, {Tanaka}, {Yabe}, {Toft}, {Stockmann}
  \& {G{\'o}mez-Guijarro}}{{Kubo} et~al.}{2018}]{Kubo2018}
{Kubo} M.,  {Tanaka} M.,  {Yabe} K.,  {Toft} S.,  {Stockmann} M.,
  {G{\'o}mez-Guijarro} C.,  2018, \mn@doi [\apj] {10.3847/1538-4357/aae3e8},
  \href {https://ui.adsabs.harvard.edu/abs/2018ApJ...867....1K} {867, 1}

\bibitem[\protect\citeauthoryear{{Larson}, {Tinsley}  \& {Caldwell}}{{Larson}
  et~al.}{1980}]{Larson1980}
{Larson} R.~B.,  {Tinsley} B.~M.,   {Caldwell} C.~N.,  1980, \mn@doi [\apj]
  {10.1086/157917}, \href
  {https://ui.adsabs.harvard.edu/abs/1980ApJ...237..692L} {237, 692}

\bibitem[\protect\citeauthoryear{{Lawrence} et~al.,}{{Lawrence}
  et~al.}{2007}]{Lawrence2007}
{Lawrence} A.,  et~al., 2007, \mn@doi [\mnras]
  {10.1111/j.1365-2966.2007.12040.x}, \href
  {https://ui.adsabs.harvard.edu/abs/2007MNRAS.379.1599L} {379, 1599}

\bibitem[\protect\citeauthoryear{{Lilly}, {Carollo}, {Pipino}, {Renzini}  \&
  {Peng}}{{Lilly} et~al.}{2013}]{Lilly2013}
{Lilly} S.~J.,  {Carollo} C.~M.,  {Pipino} A.,  {Renzini} A.,   {Peng} Y.,
  2013, \mn@doi [\apj] {10.1088/0004-637X/772/2/119}, \href
  {https://ui.adsabs.harvard.edu/abs/2013ApJ...772..119L} {772, 119}

\bibitem[\protect\citeauthoryear{{Lin} et~al.,}{{Lin} et~al.}{2008}]{Lin2008}
{Lin} L.,  et~al., 2008, \mn@doi [\apj] {10.1086/587928}, \href
  {https://ui.adsabs.harvard.edu/abs/2008ApJ...681..232L} {681, 232}

\bibitem[\protect\citeauthoryear{{Liu} et~al.,}{{Liu} et~al.}{2019}]{Liu2019}
{Liu} D.,  et~al., 2019, \mn@doi [\apj] {10.3847/1538-4357/ab578d}, \href
  {https://ui.adsabs.harvard.edu/abs/2019ApJ...887..235L} {887, 235}

\bibitem[\protect\citeauthoryear{{Lotz} et~al.,}{{Lotz}
  et~al.}{2008}]{Lotz2008}
{Lotz} J.~M.,  et~al., 2008, \mn@doi [\apj] {10.1086/523659}, \href
  {https://ui.adsabs.harvard.edu/abs/2008ApJ...672..177L} {672, 177}

\bibitem[\protect\citeauthoryear{{Lotz}, {Jonsson}, {Cox}  \& {Primack}}{{Lotz}
  et~al.}{2010a}]{Lotz2010b}
{Lotz} J.~M.,  {Jonsson} P.,  {Cox} T.~J.,   {Primack} J.~R.,  2010a, \mn@doi
  [\mnras] {10.1111/j.1365-2966.2010.16268.x}, \href
  {https://ui.adsabs.harvard.edu/abs/2010MNRAS.404..575L} {404, 575}

\bibitem[\protect\citeauthoryear{{Lotz}, {Jonsson}, {Cox}  \& {Primack}}{{Lotz}
  et~al.}{2010b}]{Lotz2010a}
{Lotz} J.~M.,  {Jonsson} P.,  {Cox} T.~J.,   {Primack} J.~R.,  2010b, \mn@doi
  [\mnras] {10.1111/j.1365-2966.2010.16269.x}, \href
  {https://ui.adsabs.harvard.edu/abs/2010MNRAS.404..590L} {404, 590}

\bibitem[\protect\citeauthoryear{{Lotz}, {Jonsson}, {Cox}, {Croton}, {Primack},
  {Somerville}  \& {Stewart}}{{Lotz} et~al.}{2011}]{Lotz2011}
{Lotz} J.~M.,  {Jonsson} P.,  {Cox} T.~J.,  {Croton} D.,  {Primack} J.~R.,
  {Somerville} R.~S.,   {Stewart} K.,  2011, \mn@doi [\apj]
  {10.1088/0004-637X/742/2/103}, \href
  {http://adsabs.harvard.edu/abs/2011ApJ...742..103L} {742, 103}

\bibitem[\protect\citeauthoryear{{Magdis} et~al.,}{{Magdis}
  et~al.}{2012}]{Magdis2012}
{Magdis} G.~E.,  et~al., 2012, \mn@doi [\apj] {10.1088/0004-637X/760/1/6},
  \href {http://adsabs.harvard.edu/abs/2012ApJ...760....6M} {760, 6}

\bibitem[\protect\citeauthoryear{{Magdis} et~al.,}{{Magdis}
  et~al.}{2021}]{Magdis2021}
{Magdis} G.~E.,  et~al., 2021, \mn@doi [\aap] {10.1051/0004-6361/202039280},
  \href {https://ui.adsabs.harvard.edu/abs/2021A&A...647A..33M} {647, A33}

\bibitem[\protect\citeauthoryear{{Maltby} et~al.,}{{Maltby}
  et~al.}{2016}]{Maltby2016}
{Maltby} D.~T.,  et~al., 2016, \mn@doi [\mnras] {10.1093/mnrasl/slw057}, \href
  {https://ui.adsabs.harvard.edu/abs/2016MNRAS.459L.114M} {459, L114}

\bibitem[\protect\citeauthoryear{{Mantha} et~al.,}{{Mantha}
  et~al.}{2018}]{Mantha2018}
{Mantha} K.~B.,  et~al., 2018, \mn@doi [\mnras] {10.1093/mnras/stx3260}, \href
  {http://adsabs.harvard.edu/abs/2018MNRAS.475.1549M} {475, 1549}

\bibitem[\protect\citeauthoryear{{Martig}, {Bournaud}, {Teyssier}  \&
  {Dekel}}{{Martig} et~al.}{2009}]{Martig2009}
{Martig} M.,  {Bournaud} F.,  {Teyssier} R.,   {Dekel} A.,  2009, \mn@doi
  [\apj] {10.1088/0004-637X/707/1/250}, \href
  {http://adsabs.harvard.edu/abs/2009ApJ...707..250M} {707, 250}

\bibitem[\protect\citeauthoryear{{McMullin}, {Waters}, {Schiebel}, {Young}  \&
  {Golap}}{{McMullin} et~al.}{2007}]{McMullin2007}
{McMullin} J.~P.,  {Waters} B.,  {Schiebel} D.,  {Young} W.,   {Golap} K.,
  2007, in {Shaw} R.~A.,  {Hill} F.,   {Bell} D.~J.,  eds,  Astronomical
  Society of the Pacific Conference Series Vol. 376, Astronomical Data Analysis
  Software and Systems XVI. p.~127

\bibitem[\protect\citeauthoryear{{Mehta} et~al.,}{{Mehta}
  et~al.}{2018}]{Mehta2018}
{Mehta} V.,  et~al., 2018, \mn@doi [\apjs] {10.3847/1538-4365/aab60c}, \href
  {https://ui.adsabs.harvard.edu/abs/2018ApJS..235...36M} {235, 36}

\bibitem[\protect\citeauthoryear{{Micha{\l}owski} et~al.,}{{Micha{\l}owski}
  et~al.}{2019}]{Michalowski2019}
{Micha{\l}owski} M.~J.,  et~al., 2019, \mn@doi [\aap]
  {10.1051/0004-6361/201936055}, \href
  {https://ui.adsabs.harvard.edu/abs/2019A&A...632A..43M} {632, A43}

\bibitem[\protect\citeauthoryear{{Oke}}{{Oke}}{1974}]{Oke1974}
{Oke} J.~B.,  1974, \mn@doi [\apjl] {10.1086/181461}, \href
  {http://adsabs.harvard.edu/abs/1974ApJ...189L..47O} {189, L47}

\bibitem[\protect\citeauthoryear{{Patton} \& {Atfield}}{{Patton} \&
  {Atfield}}{2008}]{Patton2008}
{Patton} D.~R.,  {Atfield} J.~E.,  2008, \mn@doi [\apj] {10.1086/590542}, \href
  {https://ui.adsabs.harvard.edu/abs/2008ApJ...685..235P} {685, 235}

\bibitem[\protect\citeauthoryear{{Pawlik}, {McAlpine}, {Trayford}, {Wild},
  {Bower}, {Crain}, {Schaller}  \& {Schaye}}{{Pawlik}
  et~al.}{2019}]{Pawlik2019}
{Pawlik} M.~M.,  {McAlpine} S.,  {Trayford} J.~W.,  {Wild} V.,  {Bower} R.,
  {Crain} R.~A.,  {Schaller} M.,   {Schaye} J.,  2019, \mn@doi [Nature
  Astronomy] {10.1038/s41550-019-0725-z}, \href
  {https://ui.adsabs.harvard.edu/abs/2019NatAs...3..440P} {3, 440}

\bibitem[\protect\citeauthoryear{{Peng} et~al.,}{{Peng}
  et~al.}{2010}]{Peng2010}
{Peng} Y.-j.,  et~al., 2010, \mn@doi [\apj] {10.1088/0004-637X/721/1/193},
  \href {http://adsabs.harvard.edu/abs/2010ApJ...721..193P} {721, 193}

\bibitem[\protect\citeauthoryear{{Peng}, {Maiolino}  \& {Cochrane}}{{Peng}
  et~al.}{2015}]{Peng2015}
{Peng} Y.,  {Maiolino} R.,   {Cochrane} R.,  2015, \mn@doi [\nat]
  {10.1038/nature14439}, \href
  {https://ui.adsabs.harvard.edu/abs/2015Natur.521..192P} {521, 192}

\bibitem[\protect\citeauthoryear{{Pentericci} et~al.,}{{Pentericci}
  et~al.}{2018}]{Pentericci2018}
{Pentericci} L.,  et~al., 2018, \mn@doi [\aap] {10.1051/0004-6361/201833047},
  \href {https://ui.adsabs.harvard.edu/abs/2018A&A...616A.174P} {616, A174}

\bibitem[\protect\citeauthoryear{{Rowlands}, {Wild}, {Nesvadba}, {Sibthorpe},
  {Mortier}, {Lehnert}  \& {da Cunha}}{{Rowlands} et~al.}{2015}]{Rowlands2015}
{Rowlands} K.,  {Wild} V.,  {Nesvadba} N.,  {Sibthorpe} B.,  {Mortier} A.,
  {Lehnert} M.,   {da Cunha} E.,  2015, \mn@doi [\mnras]
  {10.1093/mnras/stu2714}, \href
  {https://ui.adsabs.harvard.edu/abs/2015MNRAS.448..258R} {448, 258}

\bibitem[\protect\citeauthoryear{{Rudnick} et~al.,}{{Rudnick}
  et~al.}{2017}]{Rudnik2017}
{Rudnick} G.,  et~al., 2017, \mn@doi [\apj] {10.3847/1538-4357/aa87b2}, \href
  {https://ui.adsabs.harvard.edu/abs/2017ApJ...849...27R} {849, 27}

\bibitem[\protect\citeauthoryear{{Saintonge} et~al.,}{{Saintonge}
  et~al.}{2011}]{Saintonge2011}
{Saintonge} A.,  et~al., 2011, \mn@doi [\mnras]
  {10.1111/j.1365-2966.2011.18677.x}, \href
  {https://ui.adsabs.harvard.edu/abs/2011MNRAS.415...32S} {415, 32}

\bibitem[\protect\citeauthoryear{{Sandage}}{{Sandage}}{1986}]{Sandage1986}
{Sandage} A.,  1986, \aap, \href
  {https://ui.adsabs.harvard.edu/abs/1986A&A...161...89S} {161, 89}

\bibitem[\protect\citeauthoryear{{Sanders} \& {Mirabel}}{{Sanders} \&
  {Mirabel}}{1996}]{Sanders1996}
{Sanders} D.~B.,  {Mirabel} I.~F.,  1996, \mn@doi [\araa]
  {10.1146/annurev.astro.34.1.749}, \href
  {http://adsabs.harvard.edu/abs/1996ARA%26A..34..749S} {34, 749}

\bibitem[\protect\citeauthoryear{{Sargent} et~al.,}{{Sargent}
  et~al.}{2014}]{Sargent2014}
{Sargent} M.~T.,  et~al., 2014, \mn@doi [\apj] {10.1088/0004-637X/793/1/19},
  \href {http://adsabs.harvard.edu/abs/2014ApJ...793...19S} {793, 19}

\bibitem[\protect\citeauthoryear{{Sargent} et~al.,}{{Sargent}
  et~al.}{2015}]{Sargent2015}
{Sargent} M.~T.,  et~al., 2015, \mn@doi [\apjl] {10.1088/2041-8205/806/1/L20},
  \href {https://ui.adsabs.harvard.edu/abs/2015ApJ...806L..20S} {806, L20}

\bibitem[\protect\citeauthoryear{{Sazonova} et~al.,}{{Sazonova}
  et~al.}{2021}]{Sazonova2021}
{Sazonova} E.,  et~al., 2021, \mn@doi [\apj] {10.3847/1538-4357/ac0f7f}, \href
  {https://ui.adsabs.harvard.edu/abs/2021ApJ...919..134S} {919, 134}

\bibitem[\protect\citeauthoryear{{Schmidt}}{{Schmidt}}{1959}]{Schmidt1959}
{Schmidt} M.,  1959, \mn@doi [\apj] {10.1086/146614}, \href
  {https://ui.adsabs.harvard.edu/abs/1959ApJ...129..243S} {129, 243}

\bibitem[\protect\citeauthoryear{{Schreiber} et~al.,}{{Schreiber}
  et~al.}{2018}]{Schreiber2018}
{Schreiber} C.,  et~al., 2018, \mn@doi [\aap] {10.1051/0004-6361/201833070},
  \href {https://ui.adsabs.harvard.edu/abs/2018A&A...618A..85S} {618, A85}

\bibitem[\protect\citeauthoryear{{Scodeggio} et~al.,}{{Scodeggio}
  et~al.}{2005}]{Scodeggio2005}
{Scodeggio} M.,  et~al., 2005, \mn@doi [\pasp] {10.1086/496937}, \href
  {https://ui.adsabs.harvard.edu/abs/2005PASP..117.1284S} {117, 1284}

\bibitem[\protect\citeauthoryear{{Scoville} \& {Murchikova}}{{Scoville} \&
  {Murchikova}}{2013}]{Scoville2013}
{Scoville} N.,  {Murchikova} L.,  2013, \mn@doi [\apj]
  {10.1088/0004-637X/779/1/75}, \href
  {https://ui.adsabs.harvard.edu/abs/2013ApJ...779...75S} {779, 75}

\bibitem[\protect\citeauthoryear{{Scoville}, {Faisst}, {Capak}, {Kakazu}, {Li}
  \& {Steinhardt}}{{Scoville} et~al.}{2015}]{Scoville2015}
{Scoville} N.,  {Faisst} A.,  {Capak} P.,  {Kakazu} Y.,  {Li} G.,
  {Steinhardt} C.,  2015, \mn@doi [\apj] {10.1088/0004-637X/800/2/108}, \href
  {http://adsabs.harvard.edu/abs/2015ApJ...800..108S} {800, 108}

\bibitem[\protect\citeauthoryear{{Scoville} et~al.,}{{Scoville}
  et~al.}{2017}]{Scoville2017}
{Scoville} N.,  et~al., 2017, \mn@doi [\apj] {10.3847/1538-4357/aa61a0}, \href
  {http://adsabs.harvard.edu/abs/2017ApJ...837..150S} {837, 150}

\bibitem[\protect\citeauthoryear{{Setton} et~al.,}{{Setton}
  et~al.}{2022}]{Setton2022}
{Setton} D.~J.,  et~al., 2022, \mn@doi [\apj] {10.3847/1538-4357/ac6096}, \href
  {https://ui.adsabs.harvard.edu/abs/2022ApJ...931...51S} {931, 51}

\bibitem[\protect\citeauthoryear{{Skelton} et~al.,}{{Skelton}
  et~al.}{2014}]{Skelton2014}
{Skelton} R.~E.,  et~al., 2014, \mn@doi [\apjs] {10.1088/0067-0049/214/2/24},
  \href {https://ui.adsabs.harvard.edu/abs/2014ApJS..214...24S} {214, 24}

\bibitem[\protect\citeauthoryear{{Smercina} et~al.,}{{Smercina}
  et~al.}{2018}]{Smercina2018}
{Smercina} A.,  et~al., 2018, \mn@doi [\apj] {10.3847/1538-4357/aaafcd}, \href
  {https://ui.adsabs.harvard.edu/abs/2018ApJ...855...51S} {855, 51}

\bibitem[\protect\citeauthoryear{{Smercina} et~al.,}{{Smercina}
  et~al.}{2022}]{Smercina2021}
{Smercina} A.,  et~al., 2022, \mn@doi [\apj] {10.3847/1538-4357/ac5d5f}, \href
  {https://ui.adsabs.harvard.edu/abs/2022ApJ...929..154S} {929, 154}

\bibitem[\protect\citeauthoryear{{Snyder}, {Cox}, {Hayward}, {Hernquist}  \&
  {Jonsson}}{{Snyder} et~al.}{2011}]{Snyder2011}
{Snyder} G.~F.,  {Cox} T.~J.,  {Hayward} C.~C.,  {Hernquist} L.,   {Jonsson}
  P.,  2011, \mn@doi [\apj] {10.1088/0004-637X/741/2/77}, \href
  {https://ui.adsabs.harvard.edu/abs/2011ApJ...741...77S} {741, 77}

\bibitem[\protect\citeauthoryear{{Speagle}, {Steinhardt}, {Capak}  \&
  {Silverman}}{{Speagle} et~al.}{2014}]{Speagle2014}
{Speagle} J.~S.,  {Steinhardt} C.~L.,  {Capak} P.~L.,   {Silverman} J.~D.,
  2014, \mn@doi [\apjs] {10.1088/0067-0049/214/2/15}, \href
  {https://ui.adsabs.harvard.edu/abs/2014ApJS..214...15S} {214, 15}

\bibitem[\protect\citeauthoryear{{Spilker} et~al.,}{{Spilker}
  et~al.}{2018}]{Spilker2018}
{Spilker} J.,  et~al., 2018, \mn@doi [\apj] {10.3847/1538-4357/aac438}, \href
  {https://ui.adsabs.harvard.edu/abs/2018ApJ...860..103S} {860, 103}

\bibitem[\protect\citeauthoryear{{Suess}, {Bezanson}, {Spilker}, {Kriek},
  {Greene}, {Feldmann}, {Hunt}  \& {Narayanan}}{{Suess}
  et~al.}{2017}]{Suess2017}
{Suess} K.~A.,  {Bezanson} R.,  {Spilker} J.~S.,  {Kriek} M.,  {Greene} J.~E.,
  {Feldmann} R.,  {Hunt} Q.,   {Narayanan} D.,  2017, \mn@doi [\apjl]
  {10.3847/2041-8213/aa85dc}, \href
  {https://ui.adsabs.harvard.edu/abs/2017ApJ...846L..14S} {846, L14}

\bibitem[\protect\citeauthoryear{{Suess} et~al.,}{{Suess}
  et~al.}{2022a}]{Suess2022a}
{Suess} K.~A.,  et~al., 2022a, \mn@doi [\apj] {10.3847/1538-4357/ac404a}, \href
  {https://ui.adsabs.harvard.edu/abs/2022ApJ...926...89S} {926, 89}

\bibitem[\protect\citeauthoryear{{Suess} et~al.,}{{Suess}
  et~al.}{2022b}]{Suess2022b}
{Suess} K.~A.,  et~al., 2022b, \mn@doi [\apj] {10.3847/1538-4357/ac82b0}, \href
  {https://ui.adsabs.harvard.edu/abs/2022ApJ...935..146S} {935, 146}

\bibitem[\protect\citeauthoryear{{Tacconi} et~al.,}{{Tacconi}
  et~al.}{2013}]{Tacconi2013}
{Tacconi} L.~J.,  et~al., 2013, \mn@doi [\apj] {10.1088/0004-637X/768/1/74},
  \href {http://adsabs.harvard.edu/abs/2013ApJ...768...74T} {768, 74}

\bibitem[\protect\citeauthoryear{{Tacconi} et~al.,}{{Tacconi}
  et~al.}{2018}]{Tacconi2018}
{Tacconi} L.~J.,  et~al., 2018, \mn@doi [\apj] {10.3847/1538-4357/aaa4b4},
  \href {https://ui.adsabs.harvard.edu/abs/2018ApJ...853..179T} {853, 179}

\bibitem[\protect\citeauthoryear{{Tacconi}, {Genzel}  \& {Sternberg}}{{Tacconi}
  et~al.}{2020}]{Tacconi2020}
{Tacconi} L.~J.,  {Genzel} R.,   {Sternberg} A.,  2020, \mn@doi [\araa]
  {10.1146/annurev-astro-082812-141034}, \href
  {https://ui.adsabs.harvard.edu/abs/2020ARA&A..58..157T} {58, 157}

\bibitem[\protect\citeauthoryear{{Tanaka}}{{Tanaka}}{2015}]{Tanaka2015}
{Tanaka} M.,  2015, \mn@doi [\apj] {10.1088/0004-637X/801/1/20}, \href
  {https://ui.adsabs.harvard.edu/abs/2015ApJ...801...20T} {801, 20}

\bibitem[\protect\citeauthoryear{{Toft} et~al.,}{{Toft}
  et~al.}{2014}]{Toft2014}
{Toft} S.,  et~al., 2014, \mn@doi [\apj] {10.1088/0004-637X/782/2/68}, \href
  {https://ui.adsabs.harvard.edu/abs/2014ApJ...782...68T} {782, 68}

\bibitem[\protect\citeauthoryear{{Valentino} et~al.,}{{Valentino}
  et~al.}{2020}]{Valentino2020}
{Valentino} F.,  et~al., 2020, \mn@doi [\apj] {10.3847/1538-4357/ab64dc}, \href
  {https://ui.adsabs.harvard.edu/abs/2020ApJ...889...93V} {889, 93}

\bibitem[\protect\citeauthoryear{{Verrico} et~al.,}{{Verrico}
  et~al.}{2022}]{Verrico2022}
{Verrico} M.,  et~al., 2022, \mn@doi [arXiv e-prints]
  {10.48550/arXiv.2211.16532}, \href
  {https://ui.adsabs.harvard.edu/abs/2022arXiv221116532V} {p. arXiv:2211.16532}

\bibitem[\protect\citeauthoryear{{Werle} et~al.,}{{Werle}
  et~al.}{2022}]{Werle2022}
{Werle} A.,  et~al., 2022, \mn@doi [\apj] {10.3847/1538-4357/ac5f06}, \href
  {https://ui.adsabs.harvard.edu/abs/2022ApJ...930...43W} {930, 43}

\bibitem[\protect\citeauthoryear{{Whitaker}, {Kriek}, {van Dokkum}, {Bezanson},
  {Brammer}, {Franx}  \& {Labb{\'e}}}{{Whitaker} et~al.}{2012}]{Whitaker2012}
{Whitaker} K.~E.,  {Kriek} M.,  {van Dokkum} P.~G.,  {Bezanson} R.,  {Brammer}
  G.,  {Franx} M.,   {Labb{\'e}} I.,  2012, \mn@doi [\apj]
  {10.1088/0004-637X/745/2/179}, \href
  {http://adsabs.harvard.edu/abs/2012ApJ...745..179W} {745, 179}

\bibitem[\protect\citeauthoryear{{Whitaker} et~al.,}{{Whitaker}
  et~al.}{2021}]{Whitaker2021}
{Whitaker} K.~E.,  et~al., 2021, \mn@doi [\apjl] {10.3847/2041-8213/ac399f},
  \href {https://ui.adsabs.harvard.edu/abs/2021ApJ...922L..30W} {922, L30}

\bibitem[\protect\citeauthoryear{{Wild}, {Walcher}, {Johansson}, {Tresse},
  {Charlot}, {Pollo}, {Le F{\`e}vre}  \& {de Ravel}}{{Wild}
  et~al.}{2009}]{Wild2009}
{Wild} V.,  {Walcher} C.~J.,  {Johansson} P.~H.,  {Tresse} L.,  {Charlot} S.,
  {Pollo} A.,  {Le F{\`e}vre} O.,   {de Ravel} L.,  2009, \mn@doi [\mnras]
  {10.1111/j.1365-2966.2009.14537.x}, \href
  {https://ui.adsabs.harvard.edu/abs/2009MNRAS.395..144W} {395, 144}

\bibitem[\protect\citeauthoryear{{Wild} et~al.,}{{Wild}
  et~al.}{2014}]{Wild2014}
{Wild} V.,  et~al., 2014, \mn@doi [\mnras] {10.1093/mnras/stu212}, \href
  {https://ui.adsabs.harvard.edu/abs/2014MNRAS.440.1880W} {440, 1880}

\bibitem[\protect\citeauthoryear{{Wild}, {Almaini}, {Dunlop}, {Simpson},
  {Rowlands}, {Bowler}, {Maltby}  \& {McLure}}{{Wild} et~al.}{2016}]{Wild2016}
{Wild} V.,  {Almaini} O.,  {Dunlop} J.,  {Simpson} C.,  {Rowlands} K.,
  {Bowler} R.,  {Maltby} D.,   {McLure} R.,  2016, \mn@doi [\mnras]
  {10.1093/mnras/stw1996}, \href
  {https://ui.adsabs.harvard.edu/abs/2016MNRAS.463..832W} {463, 832}

\bibitem[\protect\citeauthoryear{{Wild} et~al.,}{{Wild}
  et~al.}{2020}]{Wild2020}
{Wild} V.,  et~al., 2020, \mn@doi [\mnras] {10.1093/mnras/staa674}, \href
  {https://ui.adsabs.harvard.edu/abs/2020MNRAS.494..529W} {494, 529}

\bibitem[\protect\citeauthoryear{{Wilkinson}, {Ellison}, {Bottrell}, {Bickley},
  {Gwyn}, {Cuillandre}  \& {Wild}}{{Wilkinson} et~al.}{2022}]{Wilkinson2022}
{Wilkinson} S.,  {Ellison} S.~L.,  {Bottrell} C.,  {Bickley} R.~W.,  {Gwyn} S.,
   {Cuillandre} J.-C.,   {Wild} V.,  2022, \mn@doi [\mnras]
  {10.1093/mnras/stac1962}, \href
  {https://ui.adsabs.harvard.edu/abs/2022MNRAS.516.4354W} {516, 4354}

\bibitem[\protect\citeauthoryear{{Williams} et~al.,}{{Williams}
  et~al.}{2000}]{Williams2000}
{Williams} R.~E.,  et~al., 2000, \mn@doi [\aj] {10.1086/316854}, \href
  {http://adsabs.harvard.edu/abs/2000AJ....120.2735W} {120, 2735}

\bibitem[\protect\citeauthoryear{{Williams} et~al.,}{{Williams}
  et~al.}{2021}]{Williams2020}
{Williams} C.~C.,  et~al., 2021, \mn@doi [\apj] {10.3847/1538-4357/abcbf6},
  \href {https://ui.adsabs.harvard.edu/abs/2021ApJ...908...54W} {908, 54}

\bibitem[\protect\citeauthoryear{{Wong} et~al.,}{{Wong}
  et~al.}{2012}]{Wong2012}
{Wong} O.~I.,  et~al., 2012, \mn@doi [\mnras]
  {10.1111/j.1365-2966.2011.20159.x}, \href
  {https://ui.adsabs.harvard.edu/abs/2012MNRAS.420.1684W} {420, 1684}

\bibitem[\protect\citeauthoryear{{Worthey}, {Faber}, {Gonzalez}  \&
  {Burstein}}{{Worthey} et~al.}{1994}]{Worthey1994}
{Worthey} G.,  {Faber} S.~M.,  {Gonzalez} J.~J.,   {Burstein} D.,  1994,
  \mn@doi [\apjs] {10.1086/192087}, \href
  {https://ui.adsabs.harvard.edu/abs/1994ApJS...94..687W} {94, 687}

\bibitem[\protect\citeauthoryear{{Wu} et~al.,}{{Wu} et~al.}{2018}]{Wu2018}
{Wu} P.-F.,  et~al., 2018, \mn@doi [\apj] {10.3847/1538-4357/aae822}, \href
  {https://ui.adsabs.harvard.edu/abs/2018ApJ...868...37W} {868, 37}

\bibitem[\protect\citeauthoryear{{Wuyts}, {Labb{\'e}}, {F{\"o}rster Schreiber},
  {Franx}, {Rudnick}, {Brammer}  \& {van Dokkum}}{{Wuyts}
  et~al.}{2008}]{Wuyts2008}
{Wuyts} S.,  {Labb{\'e}} I.,  {F{\"o}rster Schreiber} N.~M.,  {Franx} M.,
  {Rudnick} G.,  {Brammer} G.~B.,   {van Dokkum} P.~G.,  2008, \mn@doi [\apj]
  {10.1086/588749}, \href
  {https://ui.adsabs.harvard.edu/abs/2008ApJ...682..985W} {682, 985}

\bibitem[\protect\citeauthoryear{{Yang}, {Tremonti}, {Zabludoff}  \&
  {Zaritsky}}{{Yang} et~al.}{2006}]{Yang2006}
{Yang} Y.,  {Tremonti} C.~A.,  {Zabludoff} A.~I.,   {Zaritsky} D.,  2006,
  \mn@doi [\apjl] {10.1086/506909}, \href
  {https://ui.adsabs.harvard.edu/abs/2006ApJ...646L..33Y} {646, L33}

\bibitem[\protect\citeauthoryear{{Yang}, {Zabludoff}, {Zaritsky}  \&
  {Mihos}}{{Yang} et~al.}{2008}]{Yang2008}
{Yang} Y.,  {Zabludoff} A.~I.,  {Zaritsky} D.,   {Mihos} J.~C.,  2008, \mn@doi
  [\apj] {10.1086/591656}, \href
  {https://ui.adsabs.harvard.edu/abs/2008ApJ...688..945Y} {688, 945}

\bibitem[\protect\citeauthoryear{{Yano}, {Kriek}, {van der Wel}  \&
  {Whitaker}}{{Yano} et~al.}{2016}]{Yano2016}
{Yano} M.,  {Kriek} M.,  {van der Wel} A.,   {Whitaker} K.~E.,  2016, \mn@doi
  [\apjl] {10.3847/2041-8205/817/2/L21}, \href
  {https://ui.adsabs.harvard.edu/abs/2016ApJ...817L..21Y} {817, L21}

\bibitem[\protect\citeauthoryear{{Yesuf}, {French}, {Faber}  \& {Koo}}{{Yesuf}
  et~al.}{2017}]{Yesuf2017}
{Yesuf} H.~M.,  {French} K.~D.,  {Faber} S.~M.,   {Koo} D.~C.,  2017, \mn@doi
  [\mnras] {10.1093/mnras/stx1046}, \href
  {https://ui.adsabs.harvard.edu/abs/2017MNRAS.469.3015Y} {469, 3015}

\bibitem[\protect\citeauthoryear{{Zabludoff}, {Zaritsky}, {Lin}, {Tucker},
  {Hashimoto}, {Shectman}, {Oemler}  \& {Kirshner}}{{Zabludoff}
  et~al.}{1996}]{Zabludoff1996}
{Zabludoff} A.~I.,  {Zaritsky} D.,  {Lin} H.,  {Tucker} D.,  {Hashimoto} Y.,
  {Shectman} S.~A.,  {Oemler} A.,   {Kirshner} R.~P.,  1996, \mn@doi [\apj]
  {10.1086/177495}, \href
  {https://ui.adsabs.harvard.edu/abs/1996ApJ...466..104Z} {466, 104}

\bibitem[\protect\citeauthoryear{{Zanella} et~al.,}{{Zanella}
  et~al.}{2018}]{Zanella2018}
{Zanella} A.,  et~al., 2018, \mn@doi [\mnras] {10.1093/mnras/sty2394}, \href
  {https://ui.adsabs.harvard.edu/abs/2018MNRAS.481.1976Z} {481, 1976}

\bibitem[\protect\citeauthoryear{{Zibetti} \& {Gallazzi}}{{Zibetti} \&
  {Gallazzi}}{2022}]{Zibetti2022}
{Zibetti} S.,  {Gallazzi} A.~R.,  2022, \mn@doi [\mnras]
  {10.1093/mnras/stac370}, \href
  {https://ui.adsabs.harvard.edu/abs/2022MNRAS.512.1415Z} {512, 1415}

\bibitem[\protect\citeauthoryear{{Zibetti} et~al.,}{{Zibetti}
  et~al.}{2017}]{Zibetti2017}
{Zibetti} S.,  et~al., 2017, \mn@doi [\mnras] {10.1093/mnras/stx251}, \href
  {https://ui.adsabs.harvard.edu/abs/2017MNRAS.468.1902Z} {468, 1902}

\bibitem[\protect\citeauthoryear{{Zwaan}, {Kuntschner}, {Pracy}  \&
  {Couch}}{{Zwaan} et~al.}{2013}]{Zwaan2013}
{Zwaan} M.~A.,  {Kuntschner} H.,  {Pracy} M.~B.,   {Couch} W.~J.,  2013,
  \mn@doi [\mnras] {10.1093/mnras/stt496}, \href
  {https://ui.adsabs.harvard.edu/abs/2013MNRAS.432..492Z} {432, 492}

\bibitem[\protect\citeauthoryear{{de Ravel} et~al.,}{{de Ravel}
  et~al.}{2009}]{deRavel2009}
{de Ravel} L.,  et~al., 2009, \mn@doi [\aap] {10.1051/0004-6361/200810569},
  \href {https://ui.adsabs.harvard.edu/abs/2009A&A...498..379D} {498, 379}

\makeatother
\end{thebibliography}


\newpage
\appendix

\section{Literature data}
\label{app:literature_data}

We complemented our observations with literature samples. We describe such ancillary datasets and the methods used to derive the parameters that are relevant for our analysis (redshift, SFR, molecular gas mass, stellar mass).
To properly compare different samples, we homogenized the IMF to \citealt{Chabrier2003} adopting the following conversions: SFR$_\mathrm{Chabrier} = $SFR$_\mathrm{Salpeter}/0.63$ and M$_\mathrm{*,Chabrier} = $M$_\mathrm{*,Salpeter}/0.61$.

\textit{(i) Local star-forming galaxies} \citep{Saintonge2011}. Sample of galaxies from the COLD GASS survey with measurements of the CO(1-0) emission line from the IRAM 30m telescope. The galaxies have redshift $0.025 < z < 0.05$ and stellar masses $10.0 < \mathrm{log(M_\star/M_\odot)} < 11.5$. The molecular gas masses have been estimated considering an $\alpha_\mathrm{CO} = 3.2$ M$_\odot$ K km s$^{-1}$ pc$^2$ conversion factor (throught the paper and in Table \ref{tab:literature_data} we homogenized this sample to the other literature ones by converting to $\alpha_\mathrm{CO} = 4.4$ M$_\odot$ K km s$^{-1}$ pc$^2$). SFRs have been derived through SED fitting assuming exponentially declining SFHs without bursts.

\textit{(ii) Intermediate-redshift star-forming galaxies} \citep{Freundlich2019}. Sample of 61 galaxies from the PHIBSS survey with measurements of the CO(2-1) emission from IRAM NOEMA. The galaxies have redshift $0.5 < z < 0.8$ and stellar masses $10.0 < \mathrm{log(M_\star/M_\odot)} < 11.8$. The molecular gas masses have been estimated considering an excitation factor $r_{21} = 0.77$ and a CO-to-H2 conversion factor that depends on metallicity and has an average value of $\alpha_\mathrm{CO} = 4.0$ M$_\odot$ K km s$^{-1}$ pc$^2$ (in Figures and in Table \ref{tab:literature_data} we homogenized this sample to the other literature ones by converting to $\alpha_\mathrm{CO} = 4.4$ M$_\odot$ K km s$^{-1}$ pc$^2$). SFRs have been derived by combining UV and IR luminosities to account for both unobscured and dust-obscured star formation \citep{Kennicutt1998}.

\textit{(iii) High-redshift star-forming galaxies} \citep{Tacconi2013}. Sample of 52 galaxies from the PHIBSS survey with measurements of the CO(3-2) emission from IRAM NOEMA. The galaxies have redshift $1 < z < 2.5$ and stellar masses $10.4 < \mathrm{log(M_\star/M_\odot)} < 11.5$. The molecular gas masses have been estimated considering an excitation factor $r_{31} = 0.5$ and an $\alpha_\mathrm{CO} = 4.4$ M$_\odot$ K km s$^{-1}$ pc$^2$ conversion factor. SFRs have been derived by combining UV and IR luminosities \citep{Kennicutt1998}.

\textit{(iv) Local quiescent galaxies} \citep{Davis2016}. Sample of 15 passive galaxies from the MASSIVE survey with measurement of the CO(1-0) emission from IRAM 30m telescope. The galaxies have distances $< 108$ Mpc and stellar masses $\mathrm{log(M_\star/M_\odot)} > 11.5$. The molecular gas masses have been estimated considering a conversion factor X$_\mathrm{CO} = 3\times 10^{20}$. SFRs have been derived from \textit{WISE} 22$\mu$m as in \cite{Davis2014}.

\textit{(v) Intermediate-redshift quiescent galaxies} \citep{Spilker2018}. Sample of 8 passive galaxies with measurements of the CO(2-1) emission from ALMA, at redshift $z \sim 0.7$ and stellar mass $\mathrm{log(M_\star/M_\odot)} \sim 11$. The molecular gas masses have been estimated considering an excitation factor $r_{21} = 0.8$ and a conversion factor $\alpha_\mathrm{CO} = 4.4$. SFRs have been derived by combining UV and IR luminosities \citep{Kennicutt1998}. The D$n$4000 and H$\gamma_\mathrm{F}$ indices are from the LEGA-C DR3 catalog. We estimated the $R$-band light-weighted age adopting the same procedure and model assumptions as for our sample (Section \ref{subsec: stellar age}).

\textit{(vi) High-redshift quiescent galaxies} \citep{Sargent2015, Rudnik2017, Hayashi2018, Williams2020}. The galaxy by \cite{Sargent2015} and \cite{Hayashi2018} have CO(2-1) measurements at $z = 1.43$ from IRAM NOEMA and $z = 1.46$ from ALMA respectively, and have stellar masses $\mathrm{log(M_\star/M_\odot) \sim 11}$. The molecular gas masses have been computed by considering an excitation factor $r_{21} = 0.8 - 1$ and a conversion factor $\alpha_\mathrm{CO} = 4.4$ M$_\odot$ K km s$^{-1}$ pc$^2$. \cite{Sargent2015} estimated the SFRs from SED fitting assuming an exponentially declining SFH, whereas \cite{Hayashi2018} from the dust-corrected [OII] emission line. 
\\
The galaxy by \cite{Rudnik2017} instead is detected in CO(1-0) emission from VLA at redshift $z = 1.62$, with stellar mass $\mathrm{log(M_\star/M_\odot) \sim 11}$. The molecular gas mass has been computed by considering a conversion factor $\alpha_\mathrm{CO} = 4.4$ M$_\odot$ K km s$^{-1}$ pc$^2$. The SFR has been estimated from SED modelling, but the used SFH is not reported.
\\
The sample by \cite{Williams2020} is made of massive quiescent galaxies ($\mathrm{log(M_\star/M_\odot) > 11.3}$), with CO(2-1) emission from ALMA at redshift $z \sim 1.5$. The molecular gas masses have been computed by considering an excitation factor $r_{21} = 0.8$ and a conversion factor $\alpha_\mathrm{CO} = 4.4$ M$_\odot$ K km s$^{-1}$ pc$^2$. The SFR is estimated from UV+IR, as well as from SED fitting assuming an exponentially declyning SFH.

\textit{(vii) Low-redshift post-SBs} \citep{French2015, Rowlands2015}. The galaxies in both samples have CO(1-0) observed with IRAM. They have redshift $z < 0.1$ and stellar masses in the range $\mathrm{log(M_\star/M_\odot) \sim 9.5 - 11.3}$. The molecular gas masses have been computed considering a conversion factor $\alpha_\mathrm{CO} = 4.0$ M$_\odot$ K km s$^{-1}$ pc$^2$ \citep{French2015} and $\alpha_\mathrm{CO} = 4.4$ M$_\odot$ K km s$^{-1}$ pc$^2$ \citep{Rowlands2015}. In Figures and in Table \ref{tab:literature_data} we homogenized the sample of \cite{French2015} to the other literature ones by converting to $\alpha_\mathrm{CO} = 4.4$ M$_\odot$ K km s$^{-1}$ pc$^2$. \cite{French2015} estimates the SFR in two ways, from the H$\alpha$ emission and from the D$_n$4000 break. \cite{Rowlands2015} estimates the SFR from SED fitting assuming  exponentially declining SFHs with additional superimposed random bursts (i.e., stochastic SFHs). We used the D$n$4000, H$\delta_\mathrm{F}$, and H$\gamma_\mathrm{F}$ indices reported in the MPA/JHU catalog\footnote{Link to the catalog: \href{https://wwwmpa.mpa-garching.mpg.de/SDSS/DR7/raw_data.html}{link}}. We have estimated the $R$-band light-weighted ages of these galaxies by using only the spectral features adopted also for the high-redshift samples (D$_n$4000, H$\delta$, H$\gamma$), for consistency with our sample galaxies. For the sample by \cite{French2015} these ages are systematically 0.2 dex younger than those estimated using five spectral indices (D$_n$4000, H$\beta$, H$\delta_\mathrm{A}$+H$\gamma_\mathrm{A}$, [MgFe]', and [Mg$_2$Fe]) as in \cite{Gallazzi2021} and the library of SFHs (exponential + burst) and metallicities (fixed along the SFH) of \cite{Gallazzi2005, Gallazzi2014}. They are instead 0.17 dex younger than the ages estimated using the new SFH library and varying metallicities as in \cite{Zibetti2017}. For the sample by \cite{Rowlands2015} the systematic differences are slightly smaller, although there is a fairly large scatter.

\textit{(viii) Intermediate-redshift post-SBs} \citep{Suess2017, Bezanson2021}. Sample of 2 post-SB galaxies with CO(2-1) emission observed by ALMA. They have a redshift $z \sim 0.7$ and stellar mass $\mathrm{log(M_\star/M_\odot) \sim 11}$. The molecular gas mass has been computed by considering an excitation factor $r_{21} = 1$ and a conversion factor $\alpha_\mathrm{CO} = 4.0$ M$_\odot$ K km s$^{-1}$ pc$^2$ (in Figures and in Table \ref{tab:literature_data} we homogenized this sample to the other literature ones by converting to $\alpha_\mathrm{CO} = 4.4$ M$_\odot$ K km s$^{-1}$ pc$^2$). The SFR has been estimated from the dust-corrected [OII] luminosity \citep{Kennicutt1998}.
\\
The sample from \cite{Bezanson2021} is made of galaxies at redshift $z \sim 1.5$, with CO(2-1) observed with ALMA and stellar masses $\mathrm{log(M_\star/M_\odot)} \sim 11.3$. The molecular gas masses have estimated considering an excitation factor $r_{21} = 1$ and a $\alpha_{CO} = 4.0$ M$_\odot$ K km s$^{-1}$ pc$^2$ conversion factor (in Figures and in Table \ref{tab:literature_data} we homogenized this sample to the other literature ones by converting to $\alpha_\mathrm{CO} = 4.4$ M$_\odot$ K km s$^{-1}$ pc$^2$). The SFRs have been estimated through the joint photometric and spectroscopic modelling, assuming a custom set of non parametric SFHs. As mentioned in \cite{Bezanson2021} (their Section 2.1), when making comparisons to scaling relations, we set a floor value to the SFR = 1 M$_\odot$ yr$^{-1}$ and consider those estimates as upper limits. We estimated the ages of all the 13 sample galaxies by fitting the D$_n$4000 and H$\delta_\mathrm{A}$ indices that are reported in \citet[Table 1]{Suess2022b}, adopting the same procedure described in our manuscript (\ref{subsec: stellar age}). In the following we refer to this as ``method 1''.
As a check, for the 4 galaxies that are also in the SDSS DR7 catalogue and have measurements of indices from the MPA/JHU catalogs, we have also estimated the ages by fitting 
(2) the D$_n$4000 and H$\delta_\mathrm{A}$ indices (as reported in the MPA/JHU catalogue); (3) the D$_n$4000, H$\delta_\mathrm{F}$, and H$\gamma_\mathrm{F}$ (as reported in the MPA/JHU catalogue). The comparison of the results obtained from methods 1 and 2 allows us to estimate the uncertainties related to using indices calculated by different groups to estimate ages. The comparison of results obtained using method 2 and 3 instead quantifies the uncertainty related to estimating ages using the H$\delta_\mathrm{A}$ index (as reported by \citealt{Suess2022b}) instead of the $H\delta_\mathrm{F}$ and H$\gamma_\mathrm{F}$ as for our post-SBs. In all cases we find very small differences ($\sim 0.02$ dex).

\textit{(ix) High-redshift post-SBs} \citep{Bezanson2019, Belli2021}. The post-starburst galaxy from \cite{Bezanson2019} has CO(2-1) observations from ALMA at $z \sim 1.5$ and it has a stellar mass $\mathrm{log(M_\star/M_\odot)} \sim 11.2$. The molecular gas mass has been computed by considering an excitation correction $r_{21} = 1$ and a $\alpha_{CO} = 4.4$ M$_\odot$ K km s$^{-1}$ pc$^2$ conversion factor. The SFR has been estimated by combining UV and IR luminosities \citep{Kennicutt1998}.
\\
The sample from \cite{Belli2021} instead is made of 3 post-starburst galaxies with CO(3-2) and CO(2-1) observations from IRAM NOEMA. They have a redshift $1 < z < 1.3$ and stellar masses $\mathrm{10.8 < log(M_\star/M_\odot} < 11.3$. The molecular gas mass has been computed by considering excitation corrections $r_{31} = 0.5$ and $r_{21} = 0.77$ and a conversion factor $\alpha_\mathrm{CO} = 4.4$ M$_\odot$ K km s$^{-1}$ pc$^2$. SFRs have been estimated through SED modelling assuming a nonparametric SFH consisting of seven independent age bins logarithmically spaced. In this work we only consider EGS-17533 that is, among their three sample galaxies, the only one classified as post-SB, having a spectrum dominated by Balmer absorption lines and a SFR that dropped by an order of magnitude over the last $\sim$ 500 Myr \citep{Belli2021}. We estimated the D$n$4000, H$\gamma_\mathrm{F}$ indices, and $R$-band light-weighted age adopting the same procedure as for our sample (Section \ref{subsec: stellar age}).

\newpage
\onecolumn
    \begin{longtable}{c c c c c c}
    \toprule
    \midrule
    ID & $z$ & log(M$_\star$) & log(L'$\mathrm{CO})$ & log(M$_\mathrm{H2}$) & SFR \\
       &      &     (M$_\odot$)        &    (K km s$^{-1}$ pc$^2$)    & (M$_\odot$)  &  (M$_\odot$ yr$^{-1}$)   \\
    (1) & (2) & (3)    &   (4)    & (5)  & (6)    \\
    \midrule
    \multicolumn{6}{c}{Local star-forming galaxies \citep{Saintonge2011}} \\
       1 & 0.0395 & 10.0 & 7.9 & 8.6 & 0.4 \\ 
       2 & 0.0384 & 10.0 & 8.3 & 9.1 & 2.3 \\ 
       3 & 0.0375 & 10.0 & 8.4 & 9.1 & 1.4 \\ 
       4 & 0.0292 & 10.1 & 8.0 & 8.8 & 0.2 \\ 
       5 & 0.0484 & 10.8 & 8.7 & 9.4 & 1.2 \\ 
       6 & 0.0427 & 10.7 & 8.7 & 9.4 & 3.0 \\ 
       7 & 0.0380 & 10.1 & 8.3 & 9.0 & 0.7 \\ 
       8 & 0.0453 & 10.6 & 9.3 & 10.0 & 10. \\ 
       9 & 0.0438 & 10.7 & 9.4 & 10.1 & 9.6 \\ 
      10 & 0.0427 & 10.9 & 9.5 & 10.2 & 6.3 \\ 
      11 & 0.0258 & 10.2 & 8.0 & 8.8 & 0.6 \\ 
      12 & 0.0324 & 10.2 & 7.9 & 8.6 & 0.4 \\ 
      13 & 0.0491 & 11.3 & 9.1 & 9.8 & 1.1 \\ 
      14 & 0.0417 & 10.7 & 8.8 & 9.5 & 1.4 \\ 
      15 & 0.0265 & 10.4 & 8.9 & 9.7 & 4.5 \\ 
      16 & 0.0262 & 10.4 & 8.4 & 9.1 & 0.9 \\ 
      17 & 0.0414 & 10.5 & 8.7 & 9.4 & 1.5 \\ 
      18 & 0.0416 & 10.5 & 8.6 & 9.4 & 4.3 \\ 
      19 & 0.0264 & 10.5 & 8.1 & 8.8 & 0.3 \\ 
      20 & 0.0296 & 10.3 & 8.8 & 9.5 & 2.4 \\ 
      21 & 0.0464 & 10.8 & 9.1 & 9.9 & 1.2 \\ 
      22 & 0.0345 & 10.1 & 8.1 & 8.8 & 0.4 \\ 
      23 & 0.0293 & 10.4 & 8.6 & 9.3 & 0.7 \\ 
      24 & 0.0344 & 10.1 & 8.5 & 9.2 & 1.1 \\ 
      25 & 0.0340 & 10.8 & 8.5 & 9.2 & 1.4 \\ 
      26 & 0.0462 & 10.8 & 8.7 & 9.5 & 1.8 \\ 
      27 & 0.0292 & 10.0 & 7.9 & 8.6 & 0.3 \\ 
      28 & 0.0328 & 10.4 & 7.9 & 8.6 & 0.1 \\ 
      29 & 0.0289 & 10.6 & 8.8 & 9.5 & 1.7 \\ 
      30 & 0.0368 & 10.5 & 8.6 & 9.3 & 2.3 \\ 
      31 & 0.0296 & 10.1 & 8.6 & 9.3 & 1.6 \\ 
      32 & 0.0334 & 10.6 & 8.6 & 9.4 & 0.7 \\ 
      33 & 0.0383 & 10.5 & 9.1 & 9.9 & 4.2 \\ 
      34 & 0.0323 & 10.5 & 8.4 & 9.1 & 0.3 \\ 
      35 & 0.0318 & 10.0 & 8.4 & 9.2 & 1.1 \\ 
      36 & 0.0489 & 10.8 & 8.8 & 9.6 & 5.3 \\ 
      37 & 0.0318 & 11.0 & 8.5 & 9.2 & 1.3 \\   
      38 & 0.0285 & 10.3 & 8.5 & 9.3 & 0.8 \\ 
      39 & 0.0296 & 10.1 & 8.4 & 9.2 & 1.7 \\ 
      40 & 0.0299 & 10.2 & 8.4 & 9.2 & 0.6 \\ 
      41 & 0.0385 & 10.3 & 8.9 & 9.6 & 3.3 \\ 
      42 & 0.0411 & 10.5 & 8.8 & 9.5 & 1.9 \\ 
      43 & 0.0299 & 10.6 & 8.4 & 9.2 & 0.2 \\ 
      44 & 0.0346 & 10.5 & 8.8 & 9.5 & 0.9 \\ 
      45 & 0.0444 & 11.2 & 9.0 & 9.7 & 1.1 \\ 
      46 & 0.0317 & 10.9 & 9.5 & 10.2 & 3.7 \\ 
      47 & 0.0329 & 10.2 & 8.7 & 9.5 & 1.7 \\ 
      48 & 0.0448 & 11.1 & 8.6 & 9.4 & 1.3 \\ 
      49 & 0.0455 & 10.6 & 9.1 & 9.8 & 4.6 \\ 
      50 & 0.0468 & 11.1 & 9.0 & 9.8 & 1.4 \\ 
      51 & 0.0339 & 10.0 & 8.1 & 8.8 & 1.2 \\ 
      52 & 0.0316 & 10.3 & 7.9 & 8.6 & 0.1 \\ 
      53 & 0.0266 & 10.4 & 8.7 & 9.5 & 1.4 \\ 
      54 & 0.0324 & 10.6 & 7.8 & 8.5 & 0.1 \\ 
      55 & 0.0348 & 10.1 & 8.2 & 9.0 & 1.9 \\ 
      56 & 0.0394 & 11.1 & 9.3 & 10.0 & 3.6 \\ 
      57 & 0.0340 & 10.3 & 8.8 & 9.5 & 1.0 \\ 
      58 & 0.0340 & 11.0 & 7.9 & 8.7 & 0.1 \\ 
      59 & 0.0432 & 10.8 & 8.7 & 9.5 & 2.2 \\ 
      60 & 0.0496 & 11.3 & 8.8 & 9.6 & 0.9 \\ 
      61 & 0.0494 & 10.9 & 8.5 & 9.3 & 0.3 \\ 
      62 & 0.0341 & 10.8 & 9.3 & 10.0 & 2.3 \\ 
      63 & 0.0470 & 10.7 & 9.1 & 9.8 & 2.4 \\ 
      64 & 0.0270 & 10.2 & 8.4 & 9.2 & 1.2 \\ 
      65 & 0.0366 & 11.0 & 8.7 & 9.4 & 1.1 \\ 
      66 & 0.0260 & 10.4 & 8.0 & 8.8 & 0.1 \\ 
      67 & 0.0429 & 10.7 & 9.1 & 9.8 & 4.3 \\ 
      68 & 0.0492 & 10.9 & 8.9 & 9.7 & 4.4 \\ 
      69 & 0.0434 & 10.6 & 8.6 & 9.4 & 1.8 \\ 
      70 & 0.0431 & 11.0 & 8.7 & 9.4 & 0.8 \\ 
      71 & 0.0338 & 10.1 & 8.4 & 9.1 & 1.0 \\ 
      72 & 0.0305 & 10.1 & 8.5 & 9.2 & 1.2 \\ 
      73 & 0.0478 & 10.1 & 8.4 & 9.2 & 12. \\        
      74 & 0.0487 & 10.7 & 9.3 & 10.0 & 7.0 \\ 
      75 & 0.0376 & 11.0 & 9.2 & 9.9 & 2.7 \\ 
      76 & 0.0483 & 11.2 & 8.8 & 9.5 & 1.2 \\ 
      77 & 0.0350 & 10.9 & 9.3 & 10.0 & 5.3 \\ 
      78 & 0.0447 & 10.6 & 8.4 & 9.2 & 1.8 \\ 
      79 & 0.0343 & 10.7 & 8.1 & 8.8 & 0.4 \\ 
      80 & 0.0411 & 10.3 & 8.4 & 9.2 & 1.2 \\ 
      81 & 0.0462 & 11.3 & 8.9 & 9.7 & 1.3 \\ 
      82 & 0.0269 & 10.7 & 8.1 & 8.9 & 1.4 \\ 
      83 & 0.0258 & 10.7 & 8.8 & 9.6 & 0.7 \\ 
      84 & 0.0354 & 10.7 & 8.7 & 9.5 & 2.2 \\ 
      85 & 0.0267 & 10.5 & 7.9 & 8.6 & 0.2 \\ 
      86 & 0.0269 & 10.9 & 9.4 & 10.1 & 1.8 \\ 
      87 & 0.0293 & 10.8 & 8.7 & 9.5 & 0.5 \\ 
      88 & 0.0405 & 10.4 & 8.6 & 9.4 & 4.5 \\ 
      89 & 0.0308 & 10.6 & 8.9 & 9.7 & 1.0 \\ 
      90 & 0.0276 & 10.1 & 8.5 & 9.3 & 0.6 \\ 
      91 & 0.0464 & 10.7 & 9.1 & 9.8 & 3.5 \\ 
      92 & 0.0345 & 10.1 & 8.4 & 9.2 & 1.4 \\ 
      93 & 0.0367 & 10.8 & 8.4 & 9.2 & 0.6 \\ 
      94 & 0.0351 & 10.4 & 9.1 & 9.9 & 4.4 \\ 
      95 & 0.0368 & 10.7 & 9.3 & 10.0 & 4.9 \\ 
      96 & 0.0347 & 10.4 & 8.6 & 9.3 & 2.4 \\ 
      97 & 0.0435 & 10.8 & 9.0 & 9.7 & 3.1 \\ 
      98 & 0.0312 & 10.5 & 8.5 & 9.3 & 1.1 \\ 
      99 & 0.0419 & 11.2 & 9.3 & 10.1 & 20. \\ 
     100 & 0.0318 & 10.5 & 8.4 & 9.2 & 0.8 \\ 
     101 & 0.0460 & 11.0 & 8.5 & 9.3 & 1.2 \\ 
     102 & 0.0428 & 11.1 & 8.3 & 9.0 & 0.5 \\ 
     103 & 0.0273 & 10.4 & 8.1 & 8.8 & 0.2 \\ 
     104 & 0.0355 & 10.2 & 8.3 & 9.1 & 0.3 \\ 
     105 & 0.0270 & 10.3 & 8.3 & 9.0 & 0.4 \\ 
     106 & 0.0275 & 10.0 & 8.5 & 9.3 & 1.5 \\ 
     107 & 0.0275 & 10.2 & 8.1 & 8.9 & 0.3 \\ 
     108 & 0.0258 & 10.0 & 7.9 & 8.7 & 0.1 \\ 
     109 & 0.0262 & 10.4 & 8.3 & 9.0 & 0.0 \\        
     110 & 0.0257 & 10.5 & 8.1 & 8.8 & 0.7 \\ 
     111 & 0.0277 & 10.8 & 9.3 & 10.0 & 2.8 \\ 
     112 & 0.0271 & 10.4 & 8.5 & 9.2 & 2.5 \\ 
     113 & 0.0341 & 10.9 & 9.0 & 9.8 & 0.5 \\ 
     114 & 0.0394 & 10.1 & 8.5 & 9.3 & 0.3 \\ 
     115 & 0.0398 & 11.0 & 9.0 & 9.8 & 0.5 \\ 
     116 & 0.0298 & 10.7 & 8.9 & 9.7 & 1.3 \\ 
     117 & 0.0395 & 10.0 & 8.1 & 8.9 & 0.9 \\ 
     118 & 0.0256 & 10.1 & 8.5 & 9.2 & 1.0 \\ 
     119 & 0.0264 & 10.5 & 8.2 & 9.0 & 0.4 \\ 
     120 & 0.0260 & 10.0 & 8.4 & 9.2 & 1.1 \\ 
     121 & 0.0411 & 10.6 & 8.0 & 8.8 & 0.1 \\ 
     122 & 0.0428 & 10.2 & 8.3 & 9.1 & 1.8 \\ 
     123 & 0.0479 & 10.7 & 8.7 & 9.4 & 4.3 \\ 
     124 & 0.0363 & 10.6 & 8.9 & 9.7 & 3.1 \\ 
     125 & 0.0380 & 10.2 & 8.2 & 8.9 & 0.9 \\ 
     \midrule
    \multicolumn{6}{c}{Intermediate-redshift star-forming galaxies \citep{Freundlich2019}} \\
       1 & 0.7000 & 11.4 & 9.9 & 10.7 & 47.0 \\ 
       2 & 0.6227 & 10.9 & 9.0 & 9.7 & 47.0 \\ 
       3 & 0.7028 & 10.9 & 9.8 & 10.6 & 39.0 \\ 
       4 & 0.5297 & 10.3 & 9.6 & 10.5 & 25.0 \\ 
       5 & 0.5020 & 11.0 & 9.7 & 10.5 & 18.0 \\ 
       6 & 0.6223 & 11.2 & 9.6 & 10.4 & 21.0 \\ 
       7 & 0.7028 & 10.7 & 9.2 & 10.0 & 18.0 \\ 
       8 & 0.7026 & 11.4 & 9.3 & 10.1 & 20.0 \\ 
       9 & 0.7506 & 11.2 & 9.7 & 10.5 & 28.0 \\ 
      10 & 0.7007 & 11.6 & 9.7 & 10.5 & 23.0 \\ 
      11 & 0.6967 & 11.0 & 9.5 & 10.3 & 24.0 \\ 
      12 & 0.6077 & 11.3 & 9.5 & 10.2 & 11.0 \\ 
      13 & 0.6793 & 10.9 & 9.4 & 10.2 & 26.0 \\ 
      14 & 0.5165 & 11.2 & 9.2 & 10.0 & 14.0 \\ 
      15 & 0.7021 & 11.0 & 9.6 & 10.3 & 22.0 \\ 
      16 & 0.5172 & 10.2 & 9.4 & 10.2 & 28.0 \\ 
      17 & 0.6248 & 10.7 & 9.8 & 10.6 & 23.0 \\ 
      18 & 0.7503 & 10.3 & 9.1 & 9.9 & 13.0 \\ 
      19 & 0.5024 & 10.1 & 9.4 & 10.3 & 29.0 \\ 
      20 & 0.6885 & 10.4 & 9.1 & 9.9 & 8.8 \\ 
      21 & 0.5015 & 10.7 & 9.1 & 9.9 & 4.1 \\ 
      22 & 0.6081 & 10.9 & 9.6 & 10.3 & 13.0 \\ 
      23 & 0.6976 & 10.4 & 9.4 & 10.2 & 21.0 \\ 
      24 & 0.6985 & 10.4 & 9.6 & 10.4 & 29.0 \\ 
      25 & 0.7007 & 10.5 & 9.4 & 10.2 & 10.0 \\ 
      26 & 0.6590 & 11.1 & 9.8 & 10.5 & 51.0 \\ 
      27 & 0.6702 & 11.2 & 9.7 & 10.5 & 29.0 \\ 
      28 & 0.5093 & 11.2 & 9.6 & 10.3 & 37.0 \\ 
      29 & 0.7541 & 10.3 & 9.5 & 10.4 & 28.0 \\ 
      30 & 0.5090 & 10.3 & 9.3 & 10.1 & 11.0 \\ 
      31 & 0.7683 & 10.7 & 9.5 & 10.3 & 19.0 \\ 
      32 & 0.6593 & 11.1 & 9.6 & 10.4 & 14.0 \\ 
      33 & 0.7560 & 10.2 & 8.9 & 9.7 & 13.0 \\ 
      34 & 0.5010 & 10.4 & 9.0 & 9.8 & 7.4 \\ 
      35 & 0.7315 & 10.9 & 9.9 & 10.7 & 79.0 \\ 
      36 & 0.7359 & 10.0 & 9.4 & 10.3 & 9.9 \\ 
      37 & 0.6702 & 10.7 & 9.1 & 9.8 & 9.3 \\     
      38 & 0.5705 & 10.7 & 9.5 & 10.3 & 25.0 \\ 
      39 & 0.5447 & 11.0 & 9.0 & 9.8 & 9.1 \\ 
      40 & 0.7099 & 10.9 & 8.9 & 9.7 & 5.9 \\ 
      41 & 0.7369 & 10.9 & 9.0 & 9.7 & 13.0 \\ 
      42 & 0.6445 & 10.6 & 9.1 & 9.9 & 6.6 \\ 
      43 & 0.7610 & 10.4 & 9.4 & 10.2 & 44.0 \\ 
      44 & 0.6790 & 10.6 & $< 9.3$ & $<10.1$ & 23.0 \\ 
      45 & 0.7800 & 10.6 & 9.7 & 10.5 & 29.0 \\ 
      46 & 0.7790 & 10.4 & 9.5 & 10.3 & 21.0 \\ 
      47 & 0.7720 & 10.5 & 9.3 & 10.1 & 14.0 \\ 
      48 & 0.6382 & 10.0 & 9.0 & 9.9 & 11.0 \\ 
      49 & 0.5110 & 10.5 & 9.3 & 10.1 & 8.5 \\ 
      50 & 0.7784 & 10.2 & 9.3 & 10.1 & 13.0 \\ 
      51 & 0.7880 & 10.5 & 9.5 & 10.3 & 22.0 \\ 
      52 & 0.6825 & 10.3 & 9.7 & 10.5 & 23.0 \\ 
      53 & 0.5950 & 10.8 & 9.5 & 10.3 & 8.9 \\ 
      54 & 0.5035 & 10.2 & 9.1 & 9.9 & 5.5 \\ 
      55 & 0.7837 & 10.3 & 9.3 & 10.2 & 32.0 \\ 
      56 & 0.6380 & 10.7 & 9.8 & 10.6 & 76.0 \\ 
      57 & 0.5561 & 10.1 & 8.5 & 9.4 & 6.8 \\ 
      58 & 0.5320 & 10.6 & 8.8 & 9.6 & 3.5 \\ 
      59 & 0.5605 & 11.1 & 8.8 & 9.6 & 7.6 \\ 
      60 & 0.5609 & 10.0 & 8.8 & 9.7 & 6.7 \\ 
      61 & 0.5200 & 10.8 & 9.8 & 10.6 & 8.7 \\ 
    \midrule
    \multicolumn{6}{c}{High-redshift star-forming galaxies \citep{Tacconi2013}} \\
       1 & 1.0230 & 10.6 & 9.8 & 10.8 & 100.0 \\ 
       2 & 1.0170 & 10.9 & 10.2 & 11.1 & 150.0 \\ 
       3 & 1.0260 & 10.9 & 9.6 & 10.6 & 53.0 \\ 
       4 & 1.1600 & 10.7 & 9.9 & 10.8 & 94.0 \\ 
       5 & 1.2820 & 10.4 & 9.5 & 10.5 & 6.4 \\ 
       6 & 1.1190 & 11.0 & 10.2 & 11.1 & -- \\ 
       7 & 1.3740 & 10.6 & 9.7 & 10.7 & 113.0 \\ 
       8 & 1.3790 & 10.6 & 10.0 & 11.0 & 163.0 \\ 
       9 & 1.3510 & 10.7 & 9.5 & 10.5 & 115.0 \\ 
      10 & 1.3980 & 10.7 & 9.7 & 10.7 & 78.0 \\ 
      11 & 1.0020 & 10.3 & 9.5 & 10.5 & 31.0 \\ 
      12 & 1.1590 & 10.4 & 9.2 & 10.2 & 89.0 \\ 
      13 & 1.2300 & 11.2 & 10.3 & 11.3 & 200.0 \\ 
      14 & 1.1450 & 10.9 & 10.6 & 11.5 & 3.9 \\ 
      15 & 1.1920 & 10.4 & 9.5 & 10.5 & 60.0 \\ 
      16 & 1.0140 & 11.0 & 9.9 & 10.9 & 42.0 \\ 
      17 & 1.1730 & 11.0 & 9.7 & 10.6 & 52.0 \\ 
      18 & 1.0120 & 11.0 & 10.1 & 11.0 & 201.0 \\ 
      19 & 1.5290 & 11.0 & 10.4 & 11.4 & 373.0 \\ 
      20 & 1.0990 & 10.9 & 9.8 & 10.7 & -- \\ 
      21 & 1.1800 & 11.1 & 10.1 & 11.0 & 88.0 \\ 
      22 & 1.0370 & 10.8 & 10.0 & 10.9 & 351.0 \\ 
      23 & 1.0520 & 10.6 & 9.7 & 10.7 & 35.0 \\ 
      24 & 1.0310 & 10.6 & 9.5 & 10.5 & 55.0 \\ 
      25 & 1.2270 & 11.0 & 9.0 & 10.0 & -- \\ 
      26 & 1.1050 & 10.8 & 9.8 & 10.8 & -- \\ 
      27 & 1.2290 & 10.7 & 9.5 & 10.5 & -- \\ 
      28 & 1.1050 & 10.8 & 9.7 & 10.7 & 47.0 \\ 
      29 & 1.3500 & 10.6 & 9.6 & 10.6 & 87.0 \\ 
      30 & 1.2410 & 11.1 & 10.2 & 11.1 & 113.0 \\ 
      31 & 1.3170 & 10.9 & 10.0 & 10.9 & 148.0 \\ 
      32 & 1.3170 & 10.4 & 9.6 & 10.6 & 28.0 \\ 
      33 & 1.1680 & 11.0 & 9.5 & 10.5 & -- \\ 
      34 & 1.4400 & 10.8 & 9.7 & 10.7 & 86.0 \\ 
      35 & 1.4350 & 10.7 & 9.2 & 10.2 & -- \\ 
      36 & 1.1150 & 11.1 & 10.0 & 10.9 & 87.0 \\ 
      37 & 1.3930 & 10.5 & 9.4 & 10.4 & 55.0 \\      
      38 & 2.1960 & 9.97 & 10.6 & 10.9 & 480.0 \\ 
      39 & 2.1870 & 10.7 & 9.6 & 10.5 & 80.0 \\ 
      40 & -- & $< 10.0$ & 10.5 & $< 11.0$ & 246.0 \\ 
      41 & 2.1820 & 10.4 & 10.0 & 10.9 & 246.0 \\ 
      42 & 2.2680 & $< 9.7$ & 10.8 & $< 10.6$ & 59.0 \\ 
      43 & 2.3300 & 10.7 & 9.7 & 10.6 & -- \\ 
      44 & -99.00 & $< 9.9$ & 10.8 & $< 10.8$ & 144.0 \\ 
      45 & 2.3330 & 11.1 & 10.5 & 11.5 & 271.0 \\ 
      46 & 2.2890 & 11.2 & 10.0 & 10.9 & 164.0 \\ 
      47 & 2.3400 & 11.3 & 10.2 & 11.2 & 159.0 \\ 
      48 & -- & $< 10.0$ & 10.8 & $< 11.0$ & -0 \\ 
      49 & 2.4340 & 11.0 & 9.5 & 10.5 & 17.0 \\ 
      50 & 2.1890 & 10.8 & 9.6 & 10.6 & 50.0 \\ 
      51 & 2.2110 & 11.0 & 10.4 & 11.3 & 212.0 \\ 
      52 & 2.1090 & 10.4 & 9.5 & 10.5 & -- \\ 
      53 & 2.1750 & 11.0 & 9.8 & 10.7 & 145.0 \\ 
      54 & 2.0110 & 10.8 & 10.7 & 10.9 & 26.0 \\ 
      55 & -- & $<10.1$ & 10.6 & $< 11.0$ & 196.0 \\ 
      56 & -- & $< 9.9$ & 11.1 & $< 10.8$ & 62.0 \\ 
      57 & 2.2640 & 9.77 & 9.9 & 10.9 & 34.0 \\ 
    \midrule
    \multicolumn{6}{c}{Local quiescent galaxies \citep{Davis2016}} \\
        1 & 0.0183 & 11.6 & 8.3 & 9.0 & 0.3 \\ 
       2 & 0.0176 & 11.6 & 8.4 & 9.1 & 0.2 \\ 
       3 & 0.0162 & 11.7 & $< 8.1$ & 8.8 & $< 0.15$ \\ 
       4 & 0.0213 & 11.6 & 8.5 & 9.2 & 0.4 \\ 
       5 & 0.0230 & 11.7 & $< 7.9$ & $< 8.6$ & 0.1 \\ 
       6 & 0.0165 & 11.7 & $< 7.8$ & 8.5 & 0.2 \\ 
       7 & 0.0207 & 11.6 & 8.3 & 9.1 & 0.7 \\ 
       8 & 0.0149 & 11.6 & 8.5 & 9.2 & 0.3 \\ 
       9 & 0.0139 & 11.7 & $< 8.1$ & $< 8.8$ & 0.0 \\ 
      10 & 0.0211 & 11.8 & $< 7.7$ & 8.4 & 0.2 \\ 
      11 & 0.0247 & 11.7 & 8.7 & 9.4 & 0.5 \\ 
      12 & 0.0245 & 11.6 & $< 8.1$ & $< 8.9$ & 2.8 \\ 
      13 & 0.0145 & 11.7 & $< 7.9$ & $< 8.6$ & $< 0.17$ \\ 
      14 & 0.0163 & 11.7 & $< 8.9$ & 9.6 & 0.2 \\ 
      15 & 0.0243 & 11.8 & $< 7.8$ & $< 8.6$ & 0.8 \\ 
    \midrule
    \multicolumn{6}{c}{Intermediate-redshift quiescent galaxies \citep{Spilker2018}} \\
      110509 & 0.6670 & 11.0 & 9.1 & 9.9 & 0.8 \\ 
      130284 & 0.6020 & 10.9 & 9.2 & 10.0 & 2.7 \\ 
      132776 & 0.7500 & 10.9 & 9.3 & 10.0 & 3.1 \\ 
      138718 & 0.6560 & 11.2 & $< 9.0$ & $< 9.8$ & 2.1 \\ 
      169076 & 0.6770 & 11.4 & $< 9.1$ & $< 9.9$ & 0.5 \\ 
      210210 & 0.6540 & 11.3 & $< 9.0$ & $< 9.8$ & $< 0.1$ \\ 
      211409 & 0.7140 & 11.1 & $< 8.9$ & $< 9.7$ & $< 0.1$ \\ 
    \midrule
    \multicolumn{6}{c}{High-redshift quiescent galaxies \citep{Sargent2015}} \\
      pBzK-217431 & 1.4277 & 11.8 & $< 9.9$ & $< 10.6$ & 0.03 \\ 
    \midrule
    \multicolumn{6}{c}{High-redshift quiescent galaxies \citep{Hayashi2018}} \\
      Stack & 1.4600 & 11.0 & -Na & $< 9.9$ & 1.0 \\ 
    \midrule
    \multicolumn{6}{c}{High-redshift quiescent galaxies \citep{Rudnik2017}} \\
      30169 & 1.6290 & 11.2 & 9.8 & 10.5 & 12.0 \\ 
    \midrule
    \multicolumn{6}{c}{High-redshift quiescent galaxies \citep{Williams2020}} \\
    22260  & 1.240   & 11.5 & $< 1.3$ & $< 1.0$  & 3.6  \\
    20866  & 1.522   & 11.5 & $< 0.9$ & $< 1.1$  & 12.8 \\
    34879  & 1.322   & 11.3 & $< 0.5$ & $< 0.7$  & 22.9  \\
    34265  & 1.582   & 11.5 & $< 0.8$ & $< 1.0$  & 7.4  \\
    21434  & 1.522   & 11.4 & $< 0.9$ & $< 1.1$  & 19.1 \\
    307881 & 1.429   & 11.6 & $< 0.7$ & $< 0.9$  & 5.0 \\    
    \midrule
    \multicolumn{6}{c}{Low-redshift post-SBs \citep{French2015}} \\
    EAH01  & 0.0478  & 10.5 & 2.1     & 2.8      & 0.06 \\
    EAH02  & 0.0541  & 10.0 & 1.9     & 2.6      & 5.03 \\
    EAH03  & 0.0449  & 10.3 & 2.2     & 2.8      & 0.02 \\
    EAH04  & 0.0269  & 10.2 & 1.0     & 1.6      & 0.07 \\
    EAH05  & 0.0653  & 10.0 & 2.0     & 2.6      & 0.06 \\
    EAH06  & 0.0441  & 10.5 & $< 1.4$ & $< 2.0$  & 5.37 \\
    EAH07  & 0.038   & 10.7 & $< 1.0$ & $< 1.6$  & 0.22 \\
    EAH08  & 0.0604  & 10.4 & 1.6     & 2.2      & 0.04 \\
    EAH09  & 0.0291  & 10.2 & 0.9     & 1.5      & 0.06 \\
    EAH10  & 0.1053  & 10.2 & 2.3     & 2.9      & 0.04 \\
    EAH11  & 0.0542  & 10.6 & $< 1.6$ & $< 2.2$  & 0.17 \\
    EAH12  & 0.0826  & 10.6 & $< 1.8$ & $< 2.4$  & 0.18 \\
    EAH13  & 0.1129  & 11.0 & 2.3     & 2.9      & 0.62 \\
    EAH14  & 0.0622  & 10.0 & $< 1.7$ & $< 2.3$  & 0.10 \\
    EAH15  & 0.0411  & 10.4 & $< 1.4$ & $< 2.0$  & 0.06 \\
    EAH16  & 0.1113  & 10.7 & $< 2.2$ & $< 2.8$  & 0.51 \\
    EAH17  & 0.0481  & 10.1 & $< 1.4$ & $< 2.0$  & 0.05 \\ 
    EAS01  & 0.0196  & 10.2 & $< 0.8$ & $< 1.4$  & 0.01 \\
    EAS02  & 0.0231  & 10.1 & 1.1     & 1.8      & 0.03 \\
    EAS03  & 0.0628  & 10.9 & 2.2     & 2.8      & 0.17 \\
    EAS04  & 0.0153  & 10.0 & $< 0.1$ & $< 0.7$  & 8.01 \\
    EAS05  & 0.0467  & 10.6 & 1.5     & 2.1      & 0.11 \\
    EAS06  & 0.0227  & 10.1 & 1.6     & 2.3      & 0.06 \\
    EAS07  & 0.0325  & 10.5 & $< 1.0$ & $< 1.6$  & 0.04 \\
    EAS08  & 0.0408  & 10.7 & $< 1.0$ & $< 1.6$  & 0.06 \\
    EAS09  & 0.027   & 10.6 & 1.5     & 2.2      & 0.04 \\
    EAS10  & 0.0381  & 10.5 & $< 1.2$ & $< 1.8$  & 0.02 \\
    EAS11  & 0.0395  & 10.7 & $< 1.2$ & $< 1.8$  & 0.08 \\
    EAS12  & 0.0336  & 10.0 & 0.9     & 1.6      & 0.03 \\
    EAS13  & 0.0462  & 11.0 & $< 1.5$ & $< 2.1$  & 0.09 \\
    EAS14  & 0.0826  & 11.3 & 2.1     & 2.7      & 0.41 \\
    EAS15  & 0.0533  & 10.8 & 1.5     & 2.1      & 0.17 \\
    \midrule
    \multicolumn{6}{c}{Low-redshift post-SBs \citep{Rowlands2015}} \\
    PSB1   & 0.039   & 9.3  & $< 8.0$ & $< -0.5$  & 10.96 \\
    PSB2   & 0.036   & 10.5 & 8.9     & 0.6       & 1.62 \\
    PSB3   & 0.032   & 9.5  & 7.8     & -0.5      & 5.50 \\
    PSB4   & 0.029   & 9.8  & 8.7     & 0.3       & 5.50 \\
    PSB5   & 0.034   & 9.5  & 8.8     & 0.4       & 6.31 \\
    PSB6   & 0.043   & 10.5 & 9.0     & 0.6       & 2.51 \\
    PSB7   & 0.047   & 10.1 & 8.8     & 0.5       & 0.74 \\
    PSB8   & 0.046   & 9.9  & 8.2     & -0.1      & 0.11 \\
    PSB9   & 0.049   & 10.1 & 8.9     & 0.5       & 0.69 \\
    PSB10  & 0.034   & 10.0 & 8.6     & 0.3       & 1.70 \\
    PSB11  & 0.048   & 10.5 & 8.2     & -0.2      & 0.15 \\
    \midrule
    \multicolumn{6}{c}{Intermediate-redshift post-SBs \citep{Suess2017}} \\
      SDSSJ0912+1523 & 0.7470 & 11.2 & 9.9 & 10.7 & 52.0 \\ 
      SDSSJ2202-0033 & 0.6570 & 11.1 & 9.1 & 10.0 & 12.0 \\ 
    \midrule    
    \multicolumn{6}{c}{Intermediate-redshift post-SBs \citep{Bezanson2021}} \\
    SDSS J1448+1010 & 0.6462 & 11.6 & 9.7     & 10.3    & 1.06 \\
    SDSS J0753+2403 & 0.5652 & 11.3 & $< 8.6$ & $< 9.3$ & 0.10 \\
    SDSS J1053+2342 & 0.6370 & 11.6 & $< 8.9$ & $< 9.6$ & 0.29 \\
    SDSS J0027+0129 & 0.5851 & 11.5 & $< 8.8$ & $< 9.4$ & 1.44 \\
    SDSS J2202-0033 & 0.6573 & 11.7 & 9.2     & 9.9     & 1.99 \\
    SDSS J2258+2313 & 0.7058 & 11.8 &  10.1   & 10.8    & 0.94 \\
    SDSS J0233+0052 & 0.5918 & 11.6 & $< 8.7$ & $< 9.4$ & 0.01 \\
    SDSS J0046-0147 & 0.6088 & 11.6 & $< 8.8$ & $< 9.4$ & 0.14 \\
    SDSS J1109-0040 & 0.5935 & 11.3 & 9.6     & 10.2    & 2.33 \\
    SDSS J1203+1807 & 0.5946 & 11.4 & $< 8.7$ & $< 9.3$ & 0.02 \\
    SDSS J1007+2330 & 0.6353 & 11.6 & $< 8.9$ & $< 9.6$ & 0.89 \\
    SDSS J0912+1523 & 0.7473 & 11.4 & 9.9     & 10.6    & 0.81 \\
    SDSS J1302+1043 & 0.5921 & 11.6 & 9.6     & 10.2    & 0.26 \\
    \midrule    
    \multicolumn{6}{c}{High-redshift post-SBs \citep{Bezanson2019}} \\
      C21434 & 1.5220 & 11.2 & $< 9.3$ & $< 10.0$ & $< 1.4$ \\ 
    \midrule
    \multicolumn{6}{c}{High-redshift post-SBs \citep{Belli2021}} \\
      EGS-17533 & 1.2640 & 10.7 & 9.1 & 9.8 & 6.4 \\ 
    \bottomrule
    \caption{Compilation of literature data used in this paper. The full table is available online.}
    \label{tab:literature_data}
    \end{longtable}
    \begin{minipage}{17cm}
    \textbf{Columns}: (1) Galaxy ID; (2) Redshift; (3) Stellar mass; (4) CO luminosity; (5) Molecular gas mass; (6) Star formation rate.
    \end{minipage}

\clearpage
\section{Companion galaxies}
\label{app:companions}

\begin{figure*}
    \centering
    \includegraphics[width=\textwidth]{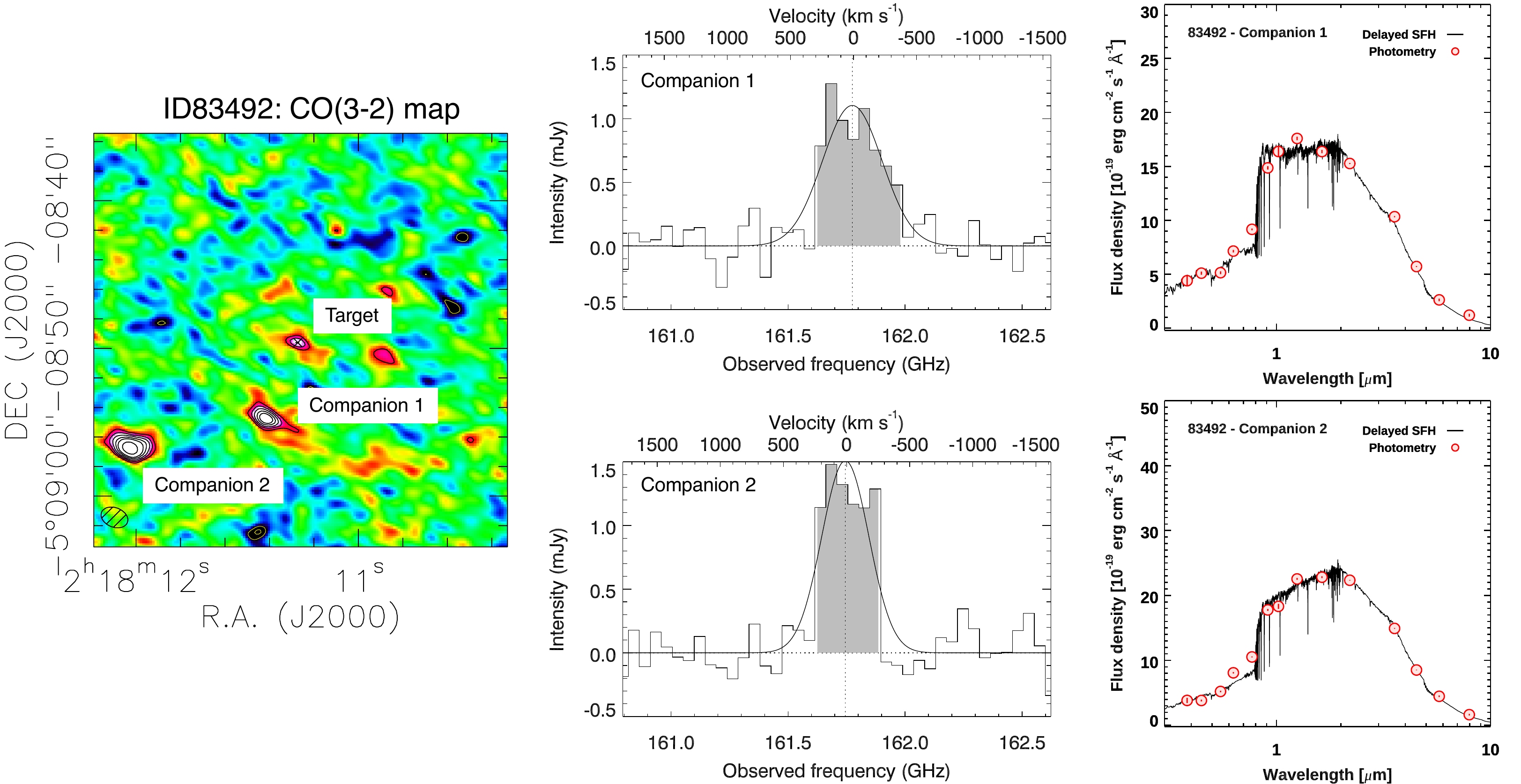}
    \caption{Available photometric and spectroscopic data of the galaxies detected in CO(3-2) that are likely companions of our post-starburst target ID83492. \textit{Left panel:} CO(3-2) ALMA map showing the target and the two CO-detected companions. \textit{Central panels:} one-dimensional spectrum extracted at the location of each companion, showing the CO(3-2) emission. \textit{Right panels:} SED modelling. The filled red circles indicate the observed photometry. The black line shows the best-fit model.}
    \label{fig:companions}
  \end{figure*}

We detected the CO(3-2) emission of two galaxies nearby our post-starburst target ID83492 (Section \ref{subsec:CO}). In Figure \ref{fig:companions} we show their submillimeter spectra and spectral energy distributions.

\bsp	
\label{lastpage}
\end{document}